\documentclass[]{interact}

\usepackage{epstopdf}
\usepackage[caption=false]{subfig}
\usepackage[numbers,sort&compress]{natbib}
\bibpunct[, ]{[}{]}{,}{n}{,}{,}
\renewcommand\bibfont{\fontsize{10}{12}\selectfont}

\usepackage{hyperref}
\usepackage{color}
\usepackage{braket}
\usepackage{dsfont}
\usepackage{xcolor}
\usepackage{mathrsfs}

\newcommand\mystackrel[2]{\overset{\scriptstyle  #1\mathstrut}{#2}}

\theoremstyle{plain}

\theoremstyle{definition}

\theoremstyle{remark}

\DeclareMathOperator{\rank}{rank}
\DeclareMathOperator{\tr}{tr}

\begin{document}

\articletype{INVITED REVIEW}

\title{Mean-field Modelling of Moir\'e Materials: A User's Guide with Selected Applications to Twisted Bilayer Graphene}

\author{\name{Yves H. Kwan\textsuperscript{a}, Ziwei Wang\textsuperscript{b}, Glenn Wagner\textsuperscript{c}, Nick Bultinck\textsuperscript{d}, Steven H. Simon \textsuperscript{b}, and Siddharth A. Parameswaran\textsuperscript{b}}
\affil{\textsuperscript{a}Department of Physics, University of Texas at Dallas, Richardson, Texas 75080, USA.\\\textsuperscript{b}Rudolf Peierls Centre for Theoretical Physics, Parks Road, Oxford, OX1 3PU, UK.\\
\textsuperscript{c}Institute for Theoretical Physics, ETH Zürich, 8093 Zürich, Switzerland.\\
\textsuperscript{d}Department of Physics, Ghent University, Krijgslaan 281, 9000 Gent, Belgium. }}

\maketitle

\begin{abstract}
We review the theoretical modelling of moir\'e materials, focusing on various aspects of magic-angle twisted bilayer graphene (MA-TBG) viewed through the lens of Hartree-Fock mean-field theory. We first provide an elementary introduction to the continuum modelling of moir\'e bandstructures, and explain how interactions are incorporated to study correlated states. We then discuss how to implement mean-field simulations of ground state structure and collective excitations in this setting. With this background established, we rationalize the power of mean-field approximations in MA-TBG, by discussing the idealized ``chiral-flat'' strong-coupling limit, in which ground states at  electron densities commensurate with the moir\'e superlattice are {\it exactly} captured by mean-field {\it ansätze}. We then illustrate the phenomenological shortcomings of this limit, leading us naturally into a discussion of the intermediate-coupling incommensurate Kekul\'e spiral (IKS) order and its origins in ever-present heterostrain. IKS and its placement within an expanded Hartree-Fock manifold form our first ``case study''. Our second case study involves time-dependence, and focuses on the collective modes of various broken-symmetry insulators in MA-TBG. As a third and final case study, we return to the strong-coupling picture, which can be stabilized by aligning MA-TBG to an hBN substrate. In this limit, we show how mean field theory can be adapted to the translationally non-invariant setting in order to  quantitatively study the energetics of domain walls in orbital Chern insulating states. We close with a discussion of extensions and further applications. Used either as a standalone reference or alongside the accompanying open-source code, this review should enable readers with a basic knowledge of band theory and many-body physics to systematically build and analyze detailed models of generic moir\'e systems.

\end{abstract}


\tableofcontents

\section{Introduction and Overview}\label{sec:introduction}

Two-dimensional van der Waals (vdW) materials have generated considerable attention over the last decade~\cite{Geim2013vdw,Novoselov2016vdw}. These ``engineered materials'' are composed of 2D atomic crystals~\cite{Novoselov2005atomic} that are stacked vertically and held together by interlayer vdW forces. Although the appeal of these systems is multifold, it is perhaps most vividly reflected in the combinatorial richness inherent in the choice of building blocks in each layer, including semimetals such as graphene, insulators such as hexagonal boron nitride, as well as semiconducting and metallic transition-metal dichalcogenides. An important subsidiary consideration is the ability to directly interrogate electronic, optical, magnetic, and structural properties via a range of scanning probes that afford a rare combination of real- and momentum-space resolution.

If, as is often the case,  the layers are stacked with a lattice or rotational mismatch, richer possibilities arise due to the emergence of one or more long-wavelength moir\'e pattern(s) \cite{He2021review,Carr2020review}. The interplay of electron-electron interactions with the band structure generated by such ``moiré superlattices'' has proven to be a new and fruitful setting for exploring the intersection of strong correlations, topology, and broken symmetry \cite{Andrei2020,balents2020SCcorrelationsreview,Andrei2021,mak2022semiconductor,Nuckolls2024,Adak2024review,Liu2021orbital,Lau2022reproducability}. This, coupled with the inherent tunability of 2D systems, has contributed to an explosion in experimental and theoretical activity devoted to uncovering new phenomena of these designer materials, driven by their promise to enable the long-sought objective of engineering new states of  quantum matter ``on demand'' \cite{Kennes2021simulator}.

Arguably the  most famous and best-studied example of a  moir\'e material is twisted bilayer graphene (TBG)\label{abb:TBG} in the ‘magic angle’ (MA) regime~\cite{Cao2018insulator,Cao2018SC,Yankowitz2019pressure,Lu2019orbital}. 
As early as 2009, a scanning tunneling spectroscopy study~\cite{Li2009vHs} noted that, in contrast to monolayer graphene where van Hove singularities in the dispersion lie so far from the Fermi surface as to be inaccessible to gating, they were shifted much closer to the Fermi surface in twisted bilayers, suggesting that this could be used to drive electronic instabilities to correlated states. In 2018,  experiments~\cite{Cao2018insulator,Cao2018SC} in the magic-angle regime (twist angles close to $\theta\simeq1.05^\circ$) identified correlated insulating states at filling factors $\nu=\pm2$, flanked by superconducting domes at nearby densities, with relatively large critical temperatures reaching $T_c/T_F\sim 0.1$ where $T_F$ is the Fermi temperature. The family resemblance to the phase diagram of the high-$T_c$ cuprates, and the ease of tuning superconductivity by gating, drove a series of subsequent experimental investigations that endures to the present time. Over the intervening years, these efforts uncovered a remarkable collection of correlated phenomena, which has clarified our understanding of certain aspects of the physics in this material, but also introduced new puzzles and complexities. Notable examples in TBG alone include  long-sought phases such as orbital Chern ferromagnets~\cite{Serlin2020QAH,Sharpe2019ferromagnetism,Tschirhart2021imaging,Liu2021orbital,Lu2019orbital} and (finite-field) fractional Chern insulators~\cite{Xie2021fractional,finney2025extendedfractionalcherninsulators}, both first observed in the moir\'e setting, novel types of broken-symmetry orders \cite{NuckollsTextures,Kim2023STM}, as well as new examples of longstanding experimental puzzles such as ``strange metallic'' resistivity \cite{Cao2020strange,Polshyn2019linear,Jaoui2021critical}. Importantly, several aspects of this phenomenology are challenging to access in the conventional renormalization-group-style approach of considering Fermi-surface instabilities driven by proximity to nesting or van Hove singularities, suggesting that rather different approaches may be required. Naturally, this has stimulated a large body of  theoretical work that has often drawn inspiration from other systems that fall outside the conventional paradigm,  such as quantum Hall ferromagnetism~\cite{Bultinck2020mechanism,Zhang2019hbn,Liu2019pseudo,Bultinck2020hidden,Bernevig2021TBG3,Lian2021TBG4,Chatterjee2020DMRG,Hejazi2021hybrid,Zhang2019hbn} or the theory of mixed-valence/heavy-fermion materials~\cite{SongTHF,Calugaru2023THF,HerzogArbeitman2025THF,Shi2022THF,chou2023kondo,Rai2024THF,calugaru2024thermoelectriceffectnaturalheavy,merino2024evidenceheavyfermionphysics,lau2025mixed,herzogarbeitman2024topologicalheavyfermionprinciple,Singh2024THF,Wang2025THF,haule2019mott,calderon2020interactions,Datta2023heavy,Zhou2024THF,Hu2023THF,Yu2023THF,Hu2023Kondo}.

The  richness of the moir\'e platform, first highlighted by TBG, has also catalyzed investigations into other moir\'e heterostructures, notably twisted graphene multilayers and transition metal dichalcogenide (TMD) homo- and hetero- bilayers \cite{mak2022semiconductor}. While these closely-related cousins share several traits with TBG, the introduction of extra layers and new materials also leads to complementary ingredients, such as strong spin-orbit coupling, and additional tuning parameters, such as sensitivity to an external displacement field, which can combine to produce additional effects not present in magic-angle TBG. These generalizations have significantly widened the moir\'e ecosystem, and made  research in this area a major subfield of contemporary condensed matter physics.

In this review, we give a theoretical introduction to the study of strongly-correlated physics in  moir\'e superlattices. Our discussion is anchored by concrete choices of setting and method. Namely, we choose to focus primarily on the physics of MA-TBG, since many core concepts of the moir\'e setting are rooted in an understanding of this system, which has arguably the richest links to symmetry and topology. Having made this choice, it is then quite natural to narrow our technical focus to emphasize the utility of Hartree-Fock mean-field approaches~\cite{Xie2020nature,Bultinck2020hidden,Lian2021TBG4,Bernevig2021TBG5,Cea2020insulating,Zhang2020HF,Kang2020nonabelian,Liu2021theories,Liu2021nematic,Lin2020hBN,Hejazi2021hybrid,Kwan2021domain,Zhang2021nonlinear,cea2022electrostatic,Parker2021strain,Kwan2021IKS,Xie2021weak,klebl2021longrange,gonzalez2021magnetic,Shavit2021theory,Kwan2021skyrmion,Wagner2022global,Xie2023,Blason2022Kekule,Wagner2024,Kwan2024EPC,wang2024magneticfield,adhikari2024strongly,ezzi2024hartree,Shi2025optical,sanchez2024nematic,biedermann2025twisttuned,wang_putting_2025}; these have proven, when appropriately applied and analysed, to have enormous utility in understanding several aspects of the MA-TBG normal state phase diagram, for reasons that we attempt to explain below. As such, we only briefly discuss or neglect other approaches that have often might have greater purchase on a specific subset of problems (e.g., determinantal quantum Monte Carlo at special densities \cite{Hofmann2022QMC,Zhang2021QMC,Zhang2023,huang2024evolution,daliao2021correlation,Liao2021review}, density-matrix renormalization group (DMRG) and other tensor network methods for problems with reduced degeneracy and for insulating states \cite{Soejima2020dmrg,Kang2020nonabelian,Parker2021strain,parker2021fieldtunedzerofieldfractionalchern,Faulstich2023,Wang2022IKS,Chen2021QMC,lin2022exciton}, and exact diagonalization~\cite{Xie2021TBG6,Potasz2021ED} especially with the objective of exploring fractional Chern insulators~\cite{Abouelkomsan2020particlehole,Repellin2020,wilhelm2021interplay,li2024contrasting}). 

The remainder of this review is organized as follows. In Sec.~\ref{sec:basics} we give an introduction to the continuum modelling of narrow bands, since this is the setting in which we  implement  Hartree-Fock mean-field theory. We give an introduction to Hartree-Fock in Sec.~\ref{sec:Hartree-Fock}, beginning in general terms and then specializing to the TBG context in Sec.~\ref{subsec:HF_TBG}; expert readers  may choose to simply read this final section, perhaps skimming the rest to alert themselves to any idiosyncrasies of notation. We proceed to give a brief summary of the strong-coupling picture of TBG in Sec.~\ref{sec:strong}, and thereby introduce the chiral-flat limit where Hartree-Fock becomes exact at integer filling. We then discuss three case studies that respectively illustrate the use of translational-invariant, time-dependent, and translational-breaking Hartree-Fock methods in MA-TBG. Sec.~\ref{sec:IKS} explains why the strong coupling framework is inadequate to fully describe the experimental reality in MA-TBG, and shows how the BM model and HF approach can both be extended to access the {\it incommensurate Kekul\'e spiral} (IKS) order believed to be the origin of many of the the correlated insulating states seen in experiments. Our second case study, in Sec.~\ref{sec:collective}, focuses on collective modes of the correlated states introduced previously. Sec.~\ref{sec:domain_wall}, our final case study, changes gears and discusses the mesoscopic physics of spatially inhomogeneous field configurations in a specific strong-coupling state, namely the orbital Chern insulator observed in substrate-aligned MA-TBG near filling $\nu=+3$. We close with a summary and outlook on future directions.

As a companion to this Review, we have also released an open-source numerical package to perform HF with partial symmetry breaking on MA-TBG, available at \url{https://github.com/ziweiwang-code/TBG-HF} and \url{https://doi.org/10.5281/zenodo.17701731}. The package implements the self-consistent HF method as discussed in Sec.~\ref{sec:Hartree-Fock}, and allows the user to capture the strong-coupling states discussed in Sec.~\ref{sec:strong} as well as the IKS states discussed in Sec.~\ref{sec:IKS}. As such, the reader is encouraged to refer to and experiment with the companion codes while reading the relevant sections. Note that the added functionalities required to obtain the results in Sec.~\ref{sec:collective} and Sec.~\ref{sec:domain_wall} are not currently implemented in the released code, but may be included in future versions.

\section{Moir\'e Basics: Continuum Modelling of Narrow Bands}\label{sec:basics}

In modelling moir\'e materials, we are confronted with an immediate challenge already at the single-particle level: how are we to analyze the moir\'e reconstruction? Given the utility of simple tight-binding modelling for individual 2D monolayers, one natural guess might be to simply extend this (with suitable estimates of interlayer couplings, e.g. from {\it ab initio} studies) to a large ``supercell'' at the moir\'e scale. This is quickly faced with both conceptual and practical obstacles. The conceptual obstacle is that, except for special, commensurate twist angles or lattice mismatches, a generic moir\'e problem is only {\it quasi} periodic, rather than periodic. More practically, even if one chooses a commensurate twist, at the small twist angles or lattice mismatches that are typically of the greatest interest, the moir\'e lengthscale is so large that a single unit cell typically contains $O(10^4)$ atoms, making direct tight-binding or \emph{ab initio} modelling very cumbersome. Thus, while there have been and continue to be some efforts along these directions in MA-TBG~\cite{Song2019topological,song2024collectivespin,zhu2025wavefunctiontexturestwistedbilayer,Yananose2021chirality,Lucignano2019crucial,Uchida2014moire}, alternative approaches that can sidestep these issues are very valuable.

Given this context, the ``continuum model'' approach~\cite{LopesdosSantos2007bilayer,Bistritzer2011BM} has emerged as a powerful route to analyse moir\'e systems. As we discuss below, it can be implemented readily on standalone computers, without the need for large scale numerical analysis. The key insight behind the model is that for sufficiently long moir\'e lengths, as long as one cares primarily about low-energy bands near the Fermi energy, one can take the continuum limit in each layer (e.g., so that each layer in TBG is described by Dirac fermions) and then treat the interlayer coupling as a  periodic perturbation at the moir\'e scale. Elegantly, one finds that often a ``single-harmonic'' approximation to the latter is sufficient, in essence reducing the moir\'e reconstruction to a band-folding problem that is a modest extension of one often encountered in introductory solid state courses \cite{Simon:1581455}. 

We now introduce the continuum model in the setting of MA-TBG, before discussing how it enables us to extend the model by incorporating interactions. We relegate technical details to an Appendix, where we also discuss how to adapt the modelling to incorporate strain, an important aspect to capture the phenomenology. We also briefly discuss other moir\'e systems.

\subsection{MA-TBG: General Single-Particle Considerations}

\begin{figure}[t!]
	\includegraphics[width=1\linewidth,clip=true]{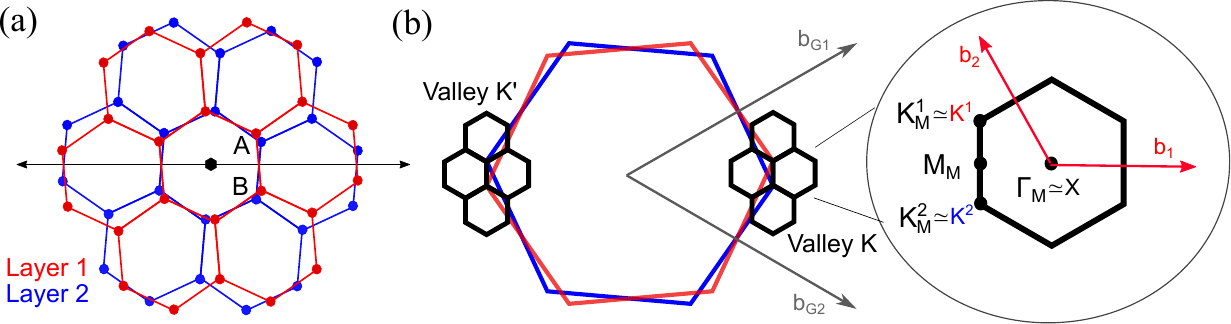}
	\caption{a) Real-space schematic of TBG with $\theta=5^\circ$ and zero interlayer shift $\bm{d}=\bm{0}$. The symmetry elements $\hat{C}_{6z}$ and $\hat{C}_{2x}$ are indicated. b) Extended zone scheme of the mBZ in each valley superposed on the rotated monolayer BZ's (red and blue). Right circle zooms into the mBZ in valley $K$, showing the high-symmetry momenta and the moir\'e reciprocal lattice vectors $\bm{b}_{1,2}$.}
	\label{fig:superlattice}
\end{figure}

We begin by outlining a procedure for specifying a twist configuration in TBG. Consider AA-stacked bilayer graphene with interlayer spacing $c_0=3.35$\,\r{A}. Without loss of generality, layers 1 and 2 are placed at $z=\pm c_0/2$, and rotated by $\pm\theta/2$ about a hexagon centre (see Fig.~\ref{fig:superlattice}a). Finally, layer 1 (the top layer) is translated by an in-plane displacement $\bm{d}$. With the appropriate choice of $\theta$ and $\bm{d}$, this is capable of representing an arbitrary stacking configuration with fixed interlayer spacing. Before discussing the electronic structure of this twisted system, three key issues need to be addressed: commensurability, symmetries, and relaxation.

The spatial motifs of the two layers interfere, causing the formation of a moir\'e pattern~\cite{amidror2009moire}. Generic twist angles are incommensurate and lead to quasiperiodicity. Strictly speaking, Bloch's theorem ceases to hold and it is therefore inappropriate to speak of energy bands in reciprocal space with a defined Brillouin zone. However, one can find specific twist angles that yield a commensurate and periodic superlattice via the following construction. Let $\bm{a}_{\text{G}1}$ and $\bm{a}_{\text{G}2}$ be the unrotated monolayer primitive lattice vectors, and $R_{\theta}$ be a rotation matrix about the stacking axis. If there exist integers $m_i,n_i$ that are not all vanishing such that \mbox{$m_1R_{\theta/2}\bm{a}_{\text{G}1}+m_2R_{\theta/2}\bm{a}_{\text{G}2}=n_1R_{-\theta/2}\bm{a}_{\text{G}1}+n_2R_{-\theta/2}\bm{a}_{\text{G}2}$}, then the combined system has discrete translation invariance~\cite{Mele2010commensuration}. Note that the statement of commensurability only depends on $\theta$ and not the interlayer shift $\bm{d}$. From an experimental standpoint, such considerations are largely irrelevant: it is impossible to determine whether a given sample has a commensurate twist, let alone deliberately fabricate a device with a precise twist. Theoretically, of course, it very helpful to be able to leverage the machinery of Bloch's theorem and think in terms of crystal momenta. Therefore some of the literature focuses on commensurate angles where a microscopic starting point can be taken, for example the large unit cell tight-binding models considered in Refs.~\cite{Kang2018Wannier,Moon2012bilayer}. A drawback of such an approach is that increasingly bulky and unwieldy Hamiltonians are required to better approximate a fixed twist angle. The problem of restricting to commensurate approximants is sidestepped in the Bistritzer-MacDonald (BM) continuum model~\cite{Bistritzer2011BM} which generates a periodic Hamiltonian for any $\theta$. Such phenomenological continuum modelling of moir\'e materials, discussed in the next subsection, is much more amenable to analytical and numerical treatment albeit at the expense of full microscopic rigor.

The finite-order symmetries of TBG depend on both $\theta$ and $\bm{d}$. For an arbitrary stacking of two graphene layers, the only guaranteed symmetry is (spinless) time-reversal symmetry (TRS)\label{abb:TRS} denoted by $\hat{\mathcal{T}}$. For $\bm{d}=\bm{0}$, the maximal subset of monolayer symmetries is preserved (Fig.~\ref{fig:superlattice}a): in addition to TRS, these include sixfold rotation about the $z$-axis ($\hat{C}_{6z}$), and twofold rotation about the $x$-axis\footnote{This is often referred to as a `mirror' symmetry $\hat{M}_y$ in the literature.} ($\hat{C}_{2x}$). The corresponding point group is $D_6$. For large twist angles, qualitative details of the band structure, such as the existence of Dirac points at charge neutrality, can depend on the symmetry group~\cite{Mele2010commensuration}. However for small twists near the magic angle regime, it is argued that such differences are invisible at experimentally resolvable energy scales, and the description in terms of $D_6$ symmetry is the most appropriate~\cite{Zou2018emergent}. This is effectively captured by the continuum model since it can shown that its spectrum is independent of $\bm{d}$~\cite{Bistritzer2011BM}. Recall that the low-energy states in monolayer graphene lie in the vicinity of the corners of the Brillouin zone, referred to as valley $K$ and $K'$~\cite{CastroNeto2009graphene}. At small twist angles, intervalley hybridization in TBG is weak because a high order of interlayer hopping is required to connect the two valleys. Therefore in addition to charge conservation $U(1)_C$, an approximate $U(1)_v$ valley conservation law emerges at low energies. As we will see, this is promoted to an exact $U(1)_v$ in the BM model, leading to a total internal symmetry $U(2)_K\times U(2)_{K'}\simeq SU(2)_K\times SU(2)_{K'}\times U(1)_C\times U(1)_v$ upon the inclusion of spin (recall that the intrinsic spin-orbit coupling is very weak in graphene~\cite{kane2005QSH,min2006intrinsic,yao2007soc,gmitra2009soc,sichau2019soc,banszerus2020soc,kurzmann2021kondo}). 

\begin{figure}
    \centering
    \includegraphics[width=0.5\linewidth]{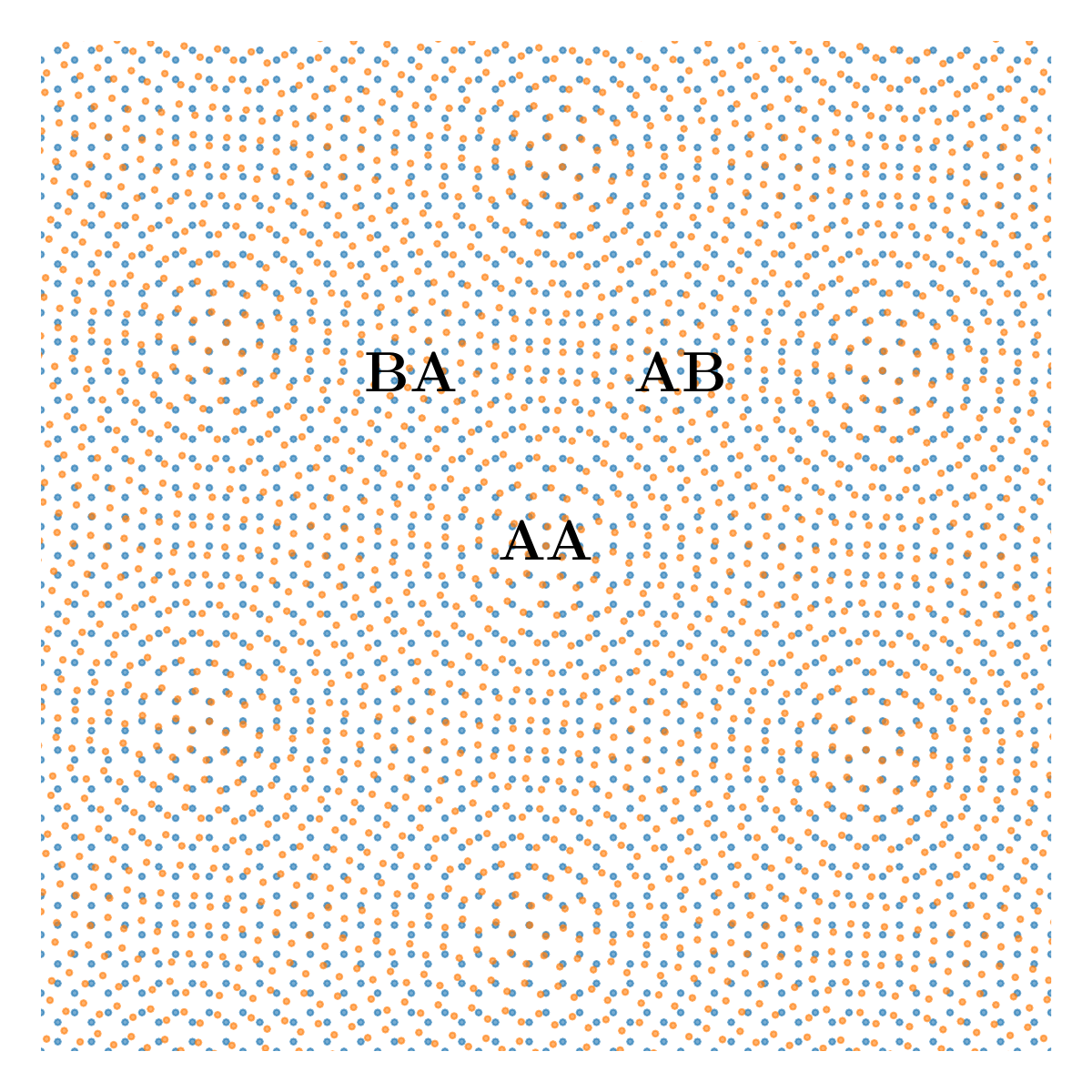}
    \caption{Rigidly twisted bilayer graphene forms regions with local AA, AB, and BA stacking configurations, named for the placement of sites in one layer relative to the other layer. Orange (blue) dots denote carbon atoms on layer 1 (2). Lattice relaxation effects cause the AB and BA regions to expand at the expense of AA regions.}
    \label{fig:stacking}
\end{figure}

The above discussion implicitly assumed a rigid rotation of layers. However, a real twisted graphene bilayer may stretch and deform to minimize the combination of elastic and interlayer potential energies. In fact, without doing an explicit moir\'e calculation, we can already qualitatively anticipate the form of such relaxation. In rigidly-twisted TBG, we can identify regions where the local stacking configuration is  AA or AB, as shown in Fig.~\ref{fig:stacking}. Recall that untwisted bilayer graphene naturally realizes the Bernal (AB-stacking) configuration, implying that it is
energetically more favourable than AA-stacking~\cite{charlier1992first}.  Therefore, the AB regions in TBG expand at the expense of AA regions, while the latter also buckle out of plane slightly to locally increase the interlayer spacing. 
Such lattice relaxation and corrugation effects can have a dramatic effect on the moir\'e band structure, and become especially pronounced for smaller twist angles. For $\theta\lesssim 0.5^\circ$, the moir\'e lattice is better thought of as large AB and BA domains separated by a triangular soliton network~\cite{SanJose2013,zhang2013valley,alden2013strain,huang2018topologically}. The out-of-plane buckling in the AA region also motivates choosing a slightly weaker interlayer tunneling amplitude $w_\textrm{AA}$ between sites on the same sublattice in the two layers relative to the amplitude $w_\textrm{AB}$ for tunneling between opposite sublattices. Their ratio $w_\textrm{AA}/w_\textrm{AB}$ controls the degree of explicit breaking of chiral sublattice symmetry\footnote{Chiral sublattice symmetry means that the single-particle Hamiltonian obeys $\{\sigma_z,H\}=0$.} in MA-TBG and plays an important role in the discussion of the exactly solvable ``chiral-flat'' limit in Section~\ref{sec:strong}.

\subsection{Bistritzer-MacDonald Continuum Model}\label{subsec:BM_model}

The Bistritzer-MacDonald model~\cite{Bistritzer2011BM} (often used to refer to the continuum model\footnote{It is worth making a point on nomenclature and history. To our knowledge, the  earliest use of a continuum model for TBG is in Lopes dos Santos {\it et al.}~\cite{LopesdosSantos2007bilayer}. However, although that work indeed predicted the suppression of the Fermi velocity, Bistritzer and MacDonald~\cite{Bistritzer2011BM} were the first to show that the bands became flat across the moiré BZ, and clearly identify the ``magic angle'' property (although this is implicit in the Fermi velocity formula of Ref.~\cite{LopesdosSantos2007bilayer}). Since the continuum model we write down is precisely that of Bistritzer and MacDonald, we will use this henceforth, in common with the prevailing convention of the field.}) is a widely used approximation to the band structure of TBG, and its central ideas apply more broadly to a wide variety of moir\'e materials in the limit of a large moir\'e unit cell (small twist angle or lattice mismatch between the layers). Specializing to TBG for concreteness, the premise is that in the low-energy limit, we can focus on momenta near one of the valleys (say valley $K$) and think of four species of fermions (two layers $\times$ two sublattices). As long as we are focused on the moir\'e bands near charge neutrality, we can use the fact that the wavefunctions are mostly built from from the monolayer states near the Dirac points, which can be well-modelled by suitably rotated versions of the monolayer Dirac Hamiltonian. Put differently, we effectively take the long-wavelength continuum limit for the intralayer problem, ignoring the microscopic lattice  beyond its role in setting the monolayer Dirac velocity (or effective mass, in the case of TMDs where each layer is a narrow-gap semiconductor), and simply ask how the interlayer coupling reconstructs the bands.
The latter takes the form of a spatially-modulating hopping amplitude between the layers, whose detailed derivation is provided in Appendix~A. In the basis of continuum plane waves $\ket{\bm{p},l\sigma}$, where $\bm{p}$ is measured with respect to the monolayer $\Gamma$-point\footnote{We do not use a layer-dependent momentum origin. Hence caution should be taken when comparing with works that measure the momentum differently in the two layers. One example is Ref.~\cite{Bistritzer2011BM}, where the momentum in each layer is measured relative to the corresponding rotated monolayer Dirac point.}, $l=1,2$ denotes the layer, and $\sigma=A,B$ indicates the microscopic sublattice, the BM Hamiltonian is\footnote{See Refs.~\cite{fang2016weakly,Carr2019continuum,Fang2019abinitio,Xie2021weak,vafek2023continuumeffective,kang2023pseudomagnetic,kang2025analytical} for extensions to the BM model that incorporate higher order terms and more severely break the approximate particle-hole symmetry.}
\begin{gather}
	\bra{\bm{p},1}H_\text{BM}\ket{\bm{p}',1} = \hbar v_F \bm{\sigma}^*_{\theta/2}\cdot(\bm{p}-\bm{K}^1)\,\delta_{\bm{p},\bm{p}'}\label{eqn:BM_start}\\
	\bra{\bm{p},2}H_\text{BM}\ket{\bm{p}',2} = \hbar v_F \bm{\sigma}^*_{-\theta/2}\cdot(\bm{p}-\bm{K}^2)\,\delta_{\bm{p},\bm{p}'}\\
	\bra{\bm{p},1}H_\text{BM}\ket{\bm{p}',2} = 
	T_1\delta_{\bm{p}-\bm{p}',\bm{0}} + T_2\delta_{\bm{p}-\bm{p}',\bm{b}_{1}+\bm{b}_2} + T_3\delta_{\bm{p}-\bm{p}',\bm{b}_{2}}\label{eqn:BM_inter}\\
	\bm{\sigma}^*_{\theta/2}=e^{-(i\theta/4)\sigma_z}(\sigma_x,\sigma_y^*)e^{(i\theta/4)\sigma_z}\label{eqn:pauli_twist}\\
	T_1 = \begin{pmatrix}w_\textrm{AA}&w_\textrm{AB}\\w_\textrm{AB}&w_\textrm{AA}\end{pmatrix},\quad T_2 = \begin{pmatrix}w_\textrm{AA}&w_\textrm{AB}e^{i\phi}\\w_\textrm{AB}e^{-i\phi}&w_\textrm{AA}\end{pmatrix},\quad	T_3 = \begin{pmatrix}w_\textrm{AA}&w_\textrm{AB}e^{-i\phi}\\w_\textrm{AB}e^{i\phi}&w_\textrm{AA}\end{pmatrix}\label{eqn:BM_end},
\end{gather}
where $\phi=\frac{2\pi}{3}$, $v_F$ is the monolayer Dirac velocity\footnote{We use $v_F=8.8\times 10^5\,\text{ms}^{-1}$ in this work. Note that the choice of $v_F$ is not completely trivial, since it is known that the effective velocity in graphene is scale-dependent~\cite{gonzalez1994nonfermi,elias2011dirac,stauber2017interacting,lee2025revealingelectron}.}, and $\bm{K}^l$ is the rotated Dirac wavevector of valley $K$ in layer $l$. The sublattice degree of freedom has been absorbed into the $2\times 2$ matrix structure. Due to the layer rotation, the intralayer kinetic terms pick up a `Pauli twist', which represents a small correction in the small-angle limit. If neglected, this results in a particle-hole symmetry $\mathcal{P}$ (PHS). The expressions for valley $K'$ can be deduced using TRS, which takes the explicit form $\hat{\mathcal{T}}=\tau_x\mathcal{K}$ where $\tau_x$ is a Pauli matrix in valley space. In the Appendix, we also discuss how the effects of uniaxial heterostrain can be modelled. We also list the actions of the discrete symmetries of the BM Hamiltonian in Appendix~B.

The interlayer term is parameterized by tunneling amplitudes $w_\textrm{AA}$ and $w_\textrm{AB}$, which as mentioned earlier roughly correspond to same- and opposite-sublattice interlayer tunneling. We have also introduced the basis moir\'e reciprocal lattice vectors (RLVs) $\bm{b}_1,\bm{b}_2$ that connect different moir\'e Brillouin zones (mBZ) in the extended zone scheme
\begin{equation}
		 \begin{gathered}\label{eqn:moire_b1}
		\bm{b}_{1}=(R_{\theta/2}-R_{-\theta/2})(\bm{b}_{\text{G}2}-\bm{b}_{\text{G}1})=\sqrt{3}k_{\theta}(1,0)\\
		\bm{b}_{2}=(R_{\theta/2}-R_{-\theta/2})\bm{b}_{\text{G}2}=\sqrt{3}k_{\theta}(-\frac{1}{2},\frac{\sqrt{3}}{2}),
		\end{gathered}
		\end{equation}
where $k_\theta=2k_D\sin\theta/2$ is the moir\'e wavevector, $\bm{b}_{\text{G}1}$ and $\bm{b}_{\text{G}2}$ are the untwisted monolayer basis RLVs, and $k_D=\frac{4\pi}{3\sqrt{3}a_{\text{CC}}}$ is the monolayer Dirac wavevector in terms of the graphene C$-$C bond length $a_\text{CC}\simeq 1.42\,$\r{A} (see Fig.~\ref{fig:superlattice}b). Inspection of the moir\'e RLVs reveals that the mBZ is the monolayer BZ rotated by $90^\circ$ and shrunk by the moir\'e scale factor $2\sin(\theta/2)$. Therefore, the real-space superlattice is similarly rotated and expanded by $1/(2\sin(\theta/2))$. A good rule of thumb is that the moir\'e lattice constant is $a_\text{M}\simeq 14\,$nm at the magic angle $\theta\simeq 1.05^\circ$. The real-space moir\'e lattice vectors associated with $\bm{b}_1,\bm{b}_2$ are $\bm{a}_1,\bm{a}_2$.

With the mBZ specified, we can now define the filling factor $\nu$, which is 0 at charge neutrality, and increments by 1 whenever we add one electron per moir\'e unit cell.

\subsubsection{Solving the Continuum Model}\label{subsubsec:solve_continuum}

As written above, the BM Hamiltonian has been expressed as a matrix in the plane-wave basis, in which a plane-wave state with momentum $\bm{p}$, layer $l$ and sublattice $\sigma$ is denoted as $\ket{\bm{p}, l\sigma}$. (This is a Dirac-electron analog of the familiar textbook problem of `folding' and then reconstructing a free electron dispersion in the presence of a weak periodic potential, with the added wrinkle that the periodic perturbation is a sublattice-dependent interlayer tunneling term and hence acts non-trivially in the layer-sublattice space.) Note that only momenta $\bm{p},\bm{p}'$ that differ from each other by some moir\'e RLV can hybridize within the BM model. To numerically solve the continuum model, it is necessary to impose an ultraviolet cutoff. 
The choice of cutoff should be consistent with the symmetry of the Hamiltonian.
One simple choice is a circular cutoff\footnote{This is the choice of cutoff that is implemented in the companion code.}, i.e.~only retaining plane wave states with
\begin{align}
    |\bm{p} - \tau \bm{K}^1| &< \Lambda \quad \text{for layer 1, and} \\
    |\bm{p} - \tau \bm{K}^2| &< \Lambda \quad \text{for layer 2},
\end{align}
where the cutoff wavevector $\Lambda$ satisfies $k_\theta \ll\Lambda <k_D$. $\tau=\pm$ is a valley index corresponding to $K$ or $K'$, respectively, and its appearance in the constraints above correctly filters the plane waves into the corresponding valleys. 

After diagonalizing the Hamiltonian in the plane wave basis, we obtain the moir\'e band structure $\epsilon^\mathrm{SP}_{\tau a}(\bm{p})$ and moir\'e Bloch wavefunctions
\begin{equation}
    \label{eq:plane_wave_expansion_p}
    \ket{\psi_{\tau a}(\bm{p})}=\sum_{\bm{G},l,\sigma}u_{\tau a;l\sigma}(\bm{p},\bm{G})\ket{\bm{p}+\bm{G},l\sigma},
\end{equation}
where $\bm{G}$ runs over the moir\'e reciprocal lattice. $u_{\tau a;l\sigma}(\bm{p},\bm{G})$ are known as the Bloch coefficients. Note that $u_{\tau a;l\sigma}(\bm{p},\bm{G}) = 0$ if $\bm{p}+\bm{G}$ falls outside the cutoff. $a$ indexes the moir\'e bands in each valley.

We now make some important remarks regarding the momentum structure of the moir\'e bands and associated conventions:
\begin{enumerate}
\item We note that the moir\'e bands carry a conserved valley index $\tau$. This is sensible at the small energy scales of interest here, since for small twist angles all the states in any single band that lie near the Fermi energy in TBG originate near only one of underlying monolayer valleys. The key distinction with a moir\'e-less graphene monolayer is that in the latter, such a valley label cannot be applied across the microscopic BZ; it is the smallness of the moir\'e BZ that enables {\it all} the states in a given band to be assigned a consistent valley label. As a strict matter, of course, valley $U(1)$ is only an approximate symmetry and can never be exact at the atomic scale, but it is a well-defined emergent symmetry at the moir\'e scale\footnote{We note that Umklapp processes reduce the valley $U(1)$ symmetry to a $\mathbb{Z}_3$ symmetry \cite{Bultinck2020hidden,Wu2019interacting,Aleiner2007spontaneous}.}. 

\item While $\bm{p}$ represented a plane wave momentum in all preceding discussion, it now also plays the role of a moir\'e crystal momentum in Eq.~\ref{eq:plane_wave_expansion_p}. Given the separation of scales between atomic and moir\'e length scales, it is useful to redefine the moir\'e crystal momentum so that it is measured relative to the moir\'e $\Gamma_M$-point of a suitably-defined mBZ. Since the rotated Dirac momenta of the monolayers are invariant under $\hat{C}_{3z}$ symmetry, which is preserved in (non-strained) TBG, it is natural to associate the mBZ corners $K_M$ and $K_M'$ with these momenta. The convention that we use for the 1st mBZ in valley $K$ is depicted in Fig.~\ref{fig:superlattice}b, with the situation in valley $K'$ obtained by applying time-reversal. This means that $\Gamma_M$ of the 1st mBZ lies just to the right of the monolayer valley-$K$ Dirac momenta for our geometry conventions. This motivates a new definition where the moir\'e momentum, now labelled $\bm{k}$, is zero at $\Gamma_M$, and the 1st mBZ corresponds to $\bm{G}=0$.  To implement this reframing, we define a new notation $\ket{\bm{k},\bm{G},\tau l\sigma}$ for plane wave states
\begin{equation}
    \ket{\bm{k},\bm{G},\tau l\sigma}\equiv \ket{\tau\bm{X}+\bm{k}+\bm{G},l\sigma},
\end{equation}
where $\bm{X}$ denotes the plane wave momentum associated with $\Gamma_M$ of the 1st mBZ in valley $K$. We now recast the moir\'e Bloch wavefunctions as
\begin{equation}
    \label{eq:plane_wave_expansion_k}
    \ket{\psi_{\tau a}(\bm{k})}=\sum_{\bm{G},l,\sigma}u_{\tau a;l\sigma}(\bm{k},\bm{G})\ket{\bm{k},\bm{G},\tau l\sigma},
\end{equation}
with corresponding kinetic energy $\epsilon^\text{SP}_{\tau a}(\bm{k})$. The BM Hamiltonian can be expressed in second quantization as 
\begin{equation}\label{eq:HBM_second_quantized}
    \hat H_\textrm{BM}=\sum_{\bm{k}\tau sa}\epsilon^\mathrm{SP}_{\tau a}(\bm{k})\hat{c}^\dagger_{\tau a s}(\bm{k})\hat{c}_{\tau a s}(\bm{k}),
\end{equation}
where $\hat{c}^\dagger_{\tau a s}(\bm{k})$ is the creation operator for a moir\'e Bloch state with moir\'e momentum $\bm{k}$ in valley $\tau$, spin $s$ (relative to some prescribed spin quantization axis), and BM band $a$\footnote{For a given $\bm{k}$, the dimension of the index $a$ is equal to four times the number of moir\'e RLVs $\bm{G}$ kept within the momentum cutoff per valley and layer. Note that for our circular cutoff, the number of plane waves, and hence the number of `bands', depends on $\bm{k}$. In fact, there is no symmetric cutoff that enables all $\bm{k}$ to have the same number of bands. This is not an issue since any discontinuities in the bands only occurs for kinetic energies far away from the low-energy states near charge neutrality.}.
\item We adopt the \emph{periodic gauge} corresponding to the convention $\ket{\psi_{\tau a}(\bm{k}+\bm{G})}\equiv \ket{\psi_{\tau a}(\bm{k})}$, which simplifies the book-keeping in numerical calculations\footnote{This simplification is very convenient when considering many-body physics. The Fock space can be generated by operators $\hat{c}^\dagger_{\tau a s}(\bm{k})$ acting on the Fock vacuum, for a set of $\bm{k}$ lying within some choice of the first mBZ (see e.g.~Eq.~\ref{eq:n1n2mtm}). In the periodic gauge, we can freely identify $\hat{c}^\dagger_{\tau a s}(\bm{k})=\hat{c}^\dagger_{\tau a s}(\bm{k}+\bm{G})$ in formal manipulations. If we did not use the periodic gauge, then $\hat{c}^\dagger_{\tau a s}(\bm{k})$ is not necessarily equal to $\hat{c}^\dagger_{\tau a s}(\bm{k}+\bm{G})$ because they could differ by a phase. This would lead to an additional layer of book-keeping when doing many-body calculations.}. This implies the relation $u_{\tau a;l\sigma}(\bm{k}+\bm{G}',\bm{G})=u_{\tau a;l\sigma}(\bm{k},\bm{G}+\bm{G}')$.
\item For the purposes of numerical calculations, we consider a periodic system with $N_1\times N_2$ moir\'e unit cells along the $\bm{a}_1$ and $ \bm{a}_2$ directions respectively. This discretizes the moir\'e momenta according to
\begin{equation}\label{eq:n1n2mtm}
    \bm{k}=\frac{n_1 \bm{b}_1}{N_1}+\frac{n_2 \bm{b}_2}{N_2},
\end{equation}
where $n_1=0,\ldots,N_1-1$ and $n_2=0,\ldots,N_2-1$.
\item Given Bloch coefficients $u_{Ka;l\sigma}(\bm{k},\bm{G})$ for valley $K$, it is possible to use time-reversal (if it is a good symmetry, e.g.~in the absence of magnetic fields) to choose the Bloch coefficients for valley $K'$ according to
\begin{equation}
    u_{K'a;l\sigma}(\bm{k},\bm{G})=u^*_{Ka;l\sigma}(-\bm{k},-\bm{G}).
\end{equation}
The kinetic energies obey $\epsilon_{K'a}^\textrm{SP}(\bm{k})=\epsilon_{Ka}^\textrm{SP}(-\bm{k})$.
\end{enumerate}

\begin{figure}[t!]
	\includegraphics[width=1\linewidth,clip=true]{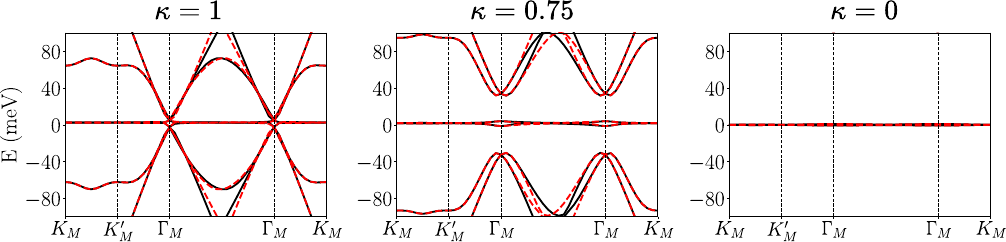}
	\caption{Band structure of the BM model along a cut in the mBZ for different values of the chiral ratio $\kappa=w_{\text{AA}}/w_{\text{AB}}$. The twist angle is held constant at the magic angle appropriate for the chiral limit $\kappa=0$ ($\theta\simeq 1.06^\circ$). Black (red) lines denote valley $K$ ($K'$).}
	\label{fig:BM_bandstructure}
\end{figure}

Figure~\ref{fig:BM_bandstructure} plots the band structure of the BM model, focusing on the low-energy bands closest to neutrality. For the `isotropic' limit $w_\text{AA}=w_\text{AB}$ originally considered by Bistritzer and MacDonald~\cite{Bistritzer2011BM}, it was found that the Dirac points were left intact and remained at the charge neutrality point (CNP)\footnote{$\hat{C}_{2z}\hat{\mathcal{T}}$ quantizes the Berry phase of an arbitrary loop in momentum space to $0$ or $\pi$, depending on whether it encloses an even or odd number of Dirac points. This is a $\mathbb{Z}_2$ topological invariant which prevents the Dirac points from gapping out individually~\cite{Goerbig2017Dirac}. In fact, a stronger condition holds as long as the central pair of bands remain energetically isolated. For a two-band model there exists a $\mathbb{Z}$ winding number, which takes the same value for the two Dirac points in a single valley of TBG. Therefore they cannot annihilate without interference from remote bands~\cite{Po2018mott,Goerbig2017Dirac,deGail2011protected,Ahn2019failure} or breaking of $U(1)_v$. In addition, the Dirac points are pinned at the mBZ corners due to $\hat{C}_{3z}$ symmetry, since the degenerate states form a doublet under the $D_3$ symmetry group relevant for a single valley.}. Reduction of the twist angle suppressed the renormalized Dirac velocity $v_F^*$ and flattened the central eight bands (including spin and valley) to $\lesssim 1\,\text{meV}$. At the magic angle\footnote{Note that generally there is no unique definition for the magic angle. For example one could have defined it as the angle for which the central bands have the minimal bandwidth. In any case, its precise value depends on parameters, and what is important is that the band structure is such that interactions play a dominant role.} of $\theta\simeq 1.05^\circ$, $v_F^*$ vanished, but the central band manifold was not energetically isolated from neighbouring remote bands, in contrast to experiments which strongly suggested single-particle band gaps of the order of tens of meV~\cite{Cao2018insulator} at filling $\nu=\pm4$. This conundrum is resolved by accounting for lattice relaxation effects discussed previously, which leads to $w_\text{AA}<w_\text{AB}$ and cleanly isolates the central bands from higher-energy states. The real-space weight of the central band wavefunctions is concentrated around the AA regions.

As noted previously, a  key dimensionless parameter is the chiral ratio $\kappa=w_{\text{AA}}/w_{\text{AB}}$. Its precise value in experiments is debatable, but is expected to lie in the range $0.5<\kappa<0.8$~\cite{Koshino2018Wannier,Carr2019continuum,Nam2017relaxation,Ledwith2021TB}. Interactions have been argued to renormalize this down further~\cite{Vafek2020RG}. In Ref.~\cite{Tarnopolsky2019chiral}, this was taken to the extreme by setting $\kappa=0$. In this so-called chiral limit, the model acquires a particle-hole symmetric spectrum due to the chiral symmetry $\{\hat{H}_\text{BM},\sigma_z\}=0$. This can be thought of as an idealized version of realistic TBG. In particular, at the magic angle, the central bands become degenerate, exactly flat, and energetically isolated across the entire mBZ at the magic angle. This limiting \emph{chiral-flat} case will be considered in more detail with electron-electron interactions in Sec.~\ref{sec:strong}.

\subsection{Adding Interactions}\label{subsec:interactions}

The usual way to add interactions in a  correlated electronic system is to construct an effective Hubbard model on a tight-binding lattice of the Wannier orbitals of the bands proximate to the Fermi energy. In the MA-TBG setting, at the very least the latter should include the eight bands (two bands per spin and valley) of the BM model. However, the topology of the energy bands makes it challenging to implement this while preserving various exact and emergent symmetries that are important to understanding the problem. In particular, while the spin is a conventional $SU(2)$ symmetry owing to the smallness of the spin-orbit coupling, the  $\hat{C}_{2z}\hat{\mathcal{T}}$ and $\hat{D}_3$ symmetries of the single valley problem lead to a topological obstruction: it can be shown (e.g.~by analyzing the representations of the bands at high-symmetry points) that it is impossible to construct exponentially-localized Wannier orbitals within a single valley while preserving these symmetries \cite{Song2019topological,Po2019faithful,Kang2018Wannier,Koshino2018Wannier,Liu2019pseudo,yuan2018model,Zou2018emergent,Ahn2019failure,vafek2021lattice}. Thus, any symmetry-preserving tight binding model requires additional bands\footnote{And, given that promoting the approximate particle-hole symmetry to an exact symmetry leads to ``strong topology'' \cite{Song2019topological,Song2021stable}, must also break PHS.} beyond the central octet. However, one can sidestep this problem by working directly in momentum space, as we now summarize.

Recall from Section~\ref{subsubsec:solve_continuum} that Bloch eigenstates of the single-particle BM Hamiltonian are given by $\ket{\psi_{\tau a}(\bm{k})}$,
where $\tau=\pm1$ indicates valley $K$ or $K'$, and $a$ is a BM band label. The associated creation operator is $\hat{c}^\dagger_{\tau a}(\bm{k})$. 
In the periodic gauge, the moir\'e cell-periodic functions $\ket{u_{\tau a}(\bm{k})}=e^{-i\bm{k}\cdot\hat{\bm{r}}}\ket{\psi_{\tau a}(\bm{k})}$ are \emph{not} periodic in its momentum argument. In particular, we have $\ket{u_{\tau,a}(\bm{k}+\bm{G})}=e^{-i\bm{G}\cdot\hat{\bm{r}}}\ket{u_{\tau,a}(\bm{k})}$, where $\bm{G}$ is a moir\'e RLV. 
We define the continuum model form factor
\begin{align}
\lambda_{\tau,ab}(\bm{k},\bm{q})&=\braket{u_{\tau,a}(\bm{k})|u_{\tau,b}(\bm{k}+\bm{q})}=\sum_{\bm{G},l,\sigma}u^*_{\tau a;l\sigma}(\bm{k},\bm{G})u_{\tau b;l\sigma}(\bm{k}+\bm{q},\bm{G})\\
&=\lambda^*_{\tau,ba}(\bm{k}+\bm{q},-\bm{q}),\label{eqn:form_factors}\end{align}
where $\bm{k}$ is a mBZ momentum, but $\bm{q}$ can be any (intravalley) momentum transfer. This is an important quantity since it encodes the Bloch structure of the wavefunctions and determines the flavour-diagonal density operators
\begin{equation}
    \hat{\rho}_{\tau s}(\bm{q})=\sum_{\mystackrel{\bm{k}\in \mathrm{mBZ}}{a b}}\hat{c}^\dagger_{\tau a s}(\bm{k})\lambda_{\tau,ab}(\bm{k},\bm{q})\hat{c}_{\tau b s}(\bm{k}+\bm{q}),
\end{equation}
where we have reintroduced the spin label $s$. Note that in the periodic gauge, $\lambda_{\tau,ab}(\bm{k},\bm{q})$ is periodic (in the mBZ) in its first momentum argument, but not its second momentum argument.

In terms of the form factors, the flavor-conserving part of the long-range density-density Coulomb Hamiltonian is
\begin{align}
\begin{split}\label{eq:HCoul}
    \hat{H}_{\text{int}}&=\frac{1}{2A}\sum_{\tau\tau' ss'}\sum_{\bm{q}\in \mathrm{all}}V(\bm{q}):\hat{\rho}_{\tau s}(\bm{q})\hat{\rho}^\dagger_{\tau' s'}(\bm{q}):_{\rho_0}\\
    &=\frac{1}{2A}\sum_{\mystackrel{ss'\tau\tau'}{abcd}}\sum_{\bm{q} \in \mathrm{all}}\sum_{\bm{k}^\alpha,\bm{k}^\beta\in\text{mBZ}}V(\bm{q})\lambda_{\tau,ab}(\bm{k}^\alpha,\bm{q})\lambda^*_{\tau',dc}(\bm{k}^\beta,\bm{q})\\
    &\quad\quad\quad\times :\hat{c}^\dagger_{\tau as}(\bm{k}^\alpha)\hat{c}^\dagger_{\tau' cs'}(\bm{k}^\beta+\bm{q})\hat{c}_{\tau' ds'}(\bm{k}^\beta)\hat{c}_{\tau bs}(\bm{k}^\alpha+\bm{q}):_{\rho^0},
\end{split}
\end{align}
where $A$ is the total area of the system\footnote{Recall from Section~\ref{subsubsec:solve_continuum} that this discretizes the momenta.}. We emphasize that, in the above expression, the summations over $\bm{k}^\alpha$ and $\bm{k}^\beta$ are restricted to the mBZ, whereas $\bm{q}$ is summed over {\it all} intravalley momentum transfers (i.e.~not restricted to only the mBZ momenta). Expressing $\hat{H}_\text{int}$ in the BM band basis is reasonable since it is often desirable to truncate to a subset of low-energy bands (often just the central bands) rather than keep all degrees of freedom up to the plane wave cutoff used to diagonalize $H_{\text{BM}}$. This is known as \emph{band projection}.
The $:\ldots:_{\rho^0}$ notation specifies the \emph{interaction scheme}, which has to do with the `normal-ordering' of the interaction. We defer a detailed discussion of interaction schemes and band projection to Section~\ref{subsubsec:subtraction_schemes} and \ref{subsubsec:band_projection}.

Typically, the interaction potential\footnote{In the companion code, we impose a circular ultraviolet cutoff on the momentum transfer $\bm{q}$ in $V(\bm{q})$ for computational convenience. This means that the form factors do not need to be computed for $\bm{q}$ lying beyond the cutoff. The contributions from large $|\bm{q}|$ are suppressed both by the decay of $V(\bm{q})$ and the form factors.} $V(q)$ is chosen as the dual or single gate-screened form\footnote{We have assumed here a purely 2D interaction which neglects the layer structure of the wavefunctions. This is a good approximation for TBG, but may fail for heterostructures with many layers~\cite{Kolar2023}.} given by  $\frac{e^2}{2\epsilon_0\epsilon_r q}\tanh{qd_{\text{sc}}}$ or  $\frac{e^2\left(1-e^{-2qd_{\text{sc}}}\right)}{2\epsilon_0\epsilon_r q}$ respectively, with screening distance $d_{\text{sc}}\sim 5-40\,\text{nm}$ and relative permittivity $\epsilon_r\sim 4 - 30$\footnote{This wide range of $\epsilon_r$ encompasses the extremes that have been considered in the literature, and reflects the current lack of consensus regarding the appropriate value (which may depend on the type of calculation being performed). For TBG, we recommend using $\epsilon_r\sim 10$, and scanning a range of interaction strengths to evaluate the robustness of the phenomena under consideration to this modelling uncertainty. }~\cite{Bultinck2020hidden}. The screening distance $d_{\text{sc}}$ is related to the physical gate distance $d_g$ via $d_{\text{sc}}=d_g\sqrt{\epsilon_\perp/\epsilon_\parallel}$, where $\epsilon_\perp$ ($\epsilon_\parallel$) is the dielectric constant of hBN perpendicular (parallel) to the atomic plane \cite{Soejima2020dmrg}. The value of $\epsilon_r$ not only encompasses the contribution of the hBN dielectric, but also captures screening and renormalization effects due to the remote bands. The latter can be accounted for more quantitatively by down-folding methods such as the constrained random phase approximation \cite{Goodwin2019,Vanhala2020,Pizarro2019}. 

\subsubsection{Interaction Schemes}\label{subsubsec:subtraction_schemes}
For the purposes of discussing the interaction scheme\footnote{In the literature, this has also been referred to as subtraction scheme or normal-ordering prescription.}, we consider keeping all degrees of freedom within the plane wave cutoff used to diagonalize $H_\text{BM}$ (i.e.~we do not consider band projection or truncation for the moment). In Eq.~\ref{eq:HCoul}, the quartic term is accompanied by a `normal-ordering' symbol $:\ldots:_{\rho^0}$. It is tempting to simply pick the conventional normal-ordering so that all annihilation operators appear to the right of all creation operators, i.e.~$\hat{H}_\text{int}\sim c^\dagger c^\dagger c c$. As we will see below, this corresponds to choosing the reference state $\rho^0$ to be the Fock vacuum. However, this runs into issues to do with double-counting interactions. 

To illustrate this, we consider the simpler example of modelling the hole-doping regime of a (non-moir\'e) semiconductor whose single valence band maximum has kinetic energy $E(\bm{p})=-\frac{p^2}{2m^*}$, where $m^*$ is the effective mass. We consider the conduction band to be sufficiently far separated in energy so that it is never occupied. If $c^\dagger(\bm{p})$ is the electron creation operator for a valence state, then it is tempting to normal-order the interacting Hamiltonian as
\begin{equation}\label{eq:H_normal_order_wrong}
    \hat{H}\stackrel{?}{=}-\sum_{\bm{p}}'\frac{p^2}{2m^*}c^\dagger(\bm{p})c(\bm{p})+\frac{1}{2A}\sum_{\bm{p},\bm{p}',\bm{q}}'V(\bm{q})c^\dagger(\bm{p}+\bm{q}) c^\dagger(\bm{p}'-\bm{q}) c(\bm{p}') c(\bm{p}),
\end{equation}
where the primes indicate that the summations are restricted so that the arguments of all fermion operators lie within the circular cutoff $\Lambda$, and we have ignored any non-trivial Bloch structure/form factors. $A$ is the total area of the system. For the band insulator $\ket{\nu=0}$ which corresponds to fully occupying the valence band, the effective electronic band structure including Fock self-energy corrections\footnote{For this example, the Hartree correction is momentum-independent.} is
\begin{equation}
    E_\text{eff}(\bm{p})=-\frac{p^2}{2m^*}+\frac{1}{A}\sum_{|\bm{p}+\bm{q}|\leq\Lambda}V(\bm{q}).
\end{equation}
The second term is an interaction-induced correction to the effective dispersion at charge neutrality, which can e.g.~renormalize the effective mass of a single hole. However, typically $m^*$ would have been extracted from \emph{ab initio} calculations or experimental input, and thus already implicitly accounts for such effects. The problem then is that Eq.~\ref{eq:H_normal_order_wrong} overcounts these contributions. To remedy this, we add an extra one-body term to the Hamiltonian that precisely cancels (subtracts off) the spurious correction
\begin{align}
    \hat{H}&=-\sum_{\bm{p}}'\frac{p^2}{2m^*}c^\dagger(\bm{p})c(\bm{p})+\frac{1}{2A}\sum_{\bm{p},\bm{p}',\bm{q}}'V(\bm{q})c^\dagger(\bm{p}+\bm{q}) c^\dagger(\bm{p}'-\bm{q}) c(\bm{p}') c(\bm{p})\\
    &-\frac{1}{A}\sum_{\bm{p}}'\sum_{|\bm{p}+\bm{q}|\leq\Lambda}V(\bm{q})c^\dagger(\bm{p})c(\bm{p})\label{eq:subtraction_term}\\
    &=-\sum_{\bm{p}}'\frac{p^2}{2m^*}c^\dagger(\bm{p})c(\bm{p})+\frac{1}{2A}\sum_{\bm{p},\bm{p}',\bm{q}}'V(\bm{q})c(\bm{p})c(\bm{p}') c^\dagger(\bm{p}'-\bm{q}) c^\dagger(\bm{p}+\bm{q})+\ldots
\end{align}
where $\ldots$ includes unimportant constants and a momentum-independent chemical potential shift. In this particular case, the final line shows that the interaction has simply been normal-ordered with respect to $\ket{\nu=0}$ instead of the Fock vacuum. This example shows that the interaction scheme is not specific to Hartree-Fock calculations, and is rather an issue of how to specify the many-body Hamiltonian, even for non-moir\'e systems.

Generalizing to the TBG setting, we define an interaction scheme by specifying a reference density matrix $\rho^0_{\tau s;ab}(\bm{k})=\langle \hat{c}_{\tau bs}^\dagger(\bm{k})\hat{c}_{\tau as}(\bm{k})\rangle_0$. The interaction term Eq.~\ref{eq:HCoul} becomes
\begin{equation}\label{eq:Hint_corrected}
    \hat{H}_{\text{int}}=\hat{H}_{\text{int, n.o.}}+\hat{H}_\text{sub}
\end{equation}
\begin{align}
\begin{split}\label{eq:Hint_n.o.}
    \hat{H}_{\text{int, n.o.}}
    &=\frac{1}{2A}\sum_{\mystackrel{ss'\tau\tau'}{abcd}}\sum_{\bm{q} \in \mathrm{all}}\sum_{\bm{k}^\alpha,\bm{k}^\beta \in \mathrm{mBZ}}V(\bm{q})\lambda_{\tau,ab}(\bm{k}^\alpha,\bm{q})\lambda^*_{\tau',dc}(\bm{k}^\beta,\bm{q})\\
    &\quad\quad\quad\times \hat{c}^\dagger_{\tau as}(\bm{k}^\alpha)\hat{c}^\dagger_{\tau' cs'}(\bm{k}^\beta+\bm{q})\hat{c}_{\tau' ds'}(\bm{k}^\beta)\hat{c}_{\tau bs}(\bm{k}^\alpha+\bm{q})
\end{split}
\end{align}
\begin{align}
\begin{split}\label{eq:Hsub}
    \hat{H}_{\text{sub}}
    &=-\frac{1}{A}\sum_{\mystackrel{\bm{k}^\alpha \in \mathrm{mBZ}}{ s\tau ab}}\left[\sum_{\bm{G}}\sum_{\mystackrel{\bm{k}^\beta \in \mathrm{mBZ}}{s'\tau' cd}}V(\bm{G})\lambda^*_{\tau',dc}(\bm{k}^\beta,\bm{G})\rho^0_{\tau s,dc}(\bm{k}^\beta)\right]\lambda_{\tau,ab}(\bm{k}^\alpha,\bm{q}) \hat{c}^\dagger_{\tau as}(\bm{k}^\alpha)\hat{c}_{\tau bs}(\bm{k}^\alpha)\\
    &+\frac{1}{A}\sum_{\mystackrel{\bm{k}^\alpha \in \mathrm{mBZ}}{s\tau ab}}\sum_{\mystackrel{\bm{q}\in\text{all}}{cd}}V(\bm{q})\lambda^*_{\tau,bc}(\bm{k}^\alpha,\bm{q})\rho^0_{\tau s,dc}(\bm{k}^\alpha+\bm{q})\lambda_{\tau,ad}(\bm{k}^\alpha,\bm{q})\hat{c}^\dagger_{\tau as}(\bm{k}^\alpha)\hat{c}_{\tau bs}(\bm{k}^\alpha).
\end{split}
\end{align}
We have split $\hat{H}_\text{int}$ into a piece $\hat{H}_\text{int, n.o.}$ corresponding to the interaction normal-ordered with respect to the Fock vacuum, and a one-body `subtraction' correction $\hat{H}_\text{sub}$, which corresponds the term in Eq.~\ref{eq:subtraction_term} with the proper form factors and other extra degrees of freedom of TBG restored. More precisely, $\hat{H}_\text{sub}$ is given by the negative of the Hartree and Fock mean-field potentials (see Section~\ref{sec:Hartree-Fock} for an introduction to Hartree-Fock) of the state $\rho^0$ for the vacuum-normal-ordered interaction $\hat{H}_\text{int, n.o.}$. This implies that $\hat{H}_\text{BM}$ would be the mean-field Hamiltonian in the state $\rho^0$ --- the interpretation is that interactions are `measured relative' to $\rho^0$. 

How is the reference state $\rho^0$ determined? To our knowledge, there is no general prescription for $\rho^0$, and its resolution is an open question. The choice should preserve the physical symmetries of the system, as well as any desired emergent symmetries (such as PHS). In TBG, since these $\rho^0$-dependent corrections take the interaction scale~$\sim 30\,$meV, the effective dispersion can be significantly broadened compared to the bare BM bands. Theoretical modelling of interaction physics in TBG is complicated by the lack of a unique prescription for $\rho^0$, which can affect not only quantitative details but also qualitative trends. Such subtleties plague other graphene-based moir\'e materials, but are less problematic for some other systems\footnote{One example where the choice of $\rho^0$ is natural and agreed upon in the literature is semiconducting moir\'e TMDs, which exhibit an eV scale band gap at charge neutrality. In this case $\rho^0$ corresponds to the neutrality state obtained by fully filling the valence bands.}. Below, we list some commonly used interaction schemes for TBG by specifying the reference density matrix in the BM band basis.
\begin{itemize}
    \item \underline{Full average scheme:} This corresponds to $\rho^0_{\tau s;ab}=\frac{1}{2}\delta_{ab}$ for all bands. This is typically utilized in many-body calculations that work in the plane wave basis and retain all BM bands (i.e.~no band projection). 
    \item \underline{Central average scheme:} This corresponds to $\rho^0_{\tau s;ab}=\frac{1}{2}\delta_{ab}$ for the central bands and $\rho^0_{\tau s;ab}=\delta_{ab}$ for the remote valence bands. The remote conduction bands are unoccupied in the reference state. This scheme is often called just `average scheme' in TBG, and has also been referred to as the `infinite temperature' scheme. This is the most commonly adopted option, and has the added bonus of enabling the interacting chiral-flat strong-coupling limit described in Section~\ref{sec:strong}.
    \item \underline{Graphene scheme:} $\rho^0$ corresponds to occupying all valence bands of decoupled graphene layers (i.e.~interlayer tunneling is switched off) at charge neutrality.
    \item \underline{Charge neutrality scheme:} $\rho^0_{\tau s;ab}=\delta_{ab}$ for all valence bands, and $\rho^0_{\tau s;ab}=0$ for all conduction bands.
\end{itemize}
Several works, including Refs.~\cite{Kwan2021skyrmion,Kwan2021IKS,Wagner2022global,Hejazi2021hybrid,parker2021fieldtunedzerofieldfractionalchern,Faulstich2023,kwan2025MFCI3,Yu2025MFCI4,yu2024versus}, have examined the consequences of utilizing different interaction schemes in TBG and other moir\'e systems.

\subsubsection{Band Projection}
\label{subsubsec:band_projection}

For many-body numerical calculations, it is usually intractable to explicitly keep all degrees of freedom. In any case, a large fraction of the single-particle states have kinetic energies far from the Fermi level so that they are not expected to participate non-trivially in low-energy physics. To account for this, we perform \emph{band projection} to reduce the effective degrees of freedom~\cite{parker2021fieldtunedzerofieldfractionalchern,kwan2025MFCI3}. This involves first identifying a subset of \emph{active} (act.) single-particle states, with the remaining \emph{frozen} single-particle states being divided into frozen valence (f.v.) and frozen conduction (f.c.) ones. The idea is that we restrict to the many-body Hilbert space spanned by the wavefunctions
\begin{equation}
    \ket{\Psi}=\mathcal{O}(\hat{c}^\dagger_{a\in\text{act.}})\prod_{v\in\text{f.v.}}\hat{c}^\dagger_{v}\ket{\text{vac}},
\end{equation}
where $\mathcal{O}(\hat{c}^\dagger_{a\in\text{act.}})$ is an arbitrary function of the active states, and $\ket{\text{vac}}$ is the Fock vacuum. This means that all f.v. (f.c.) states are enforced to be  occupied (unoccupied), while the active degree of freedom are left unconstrained. The effective Hamiltonian for the active degrees of freedom will experience additional one-body terms arising from the static Hartree and Fock potentials of the filled bands\footnote{Note that these potentials are not due to virtual (particle-hole) excitations between active and frozen states. The latter instead represent many-body processes that would stem from relaxing the strict frozen constraint.}.

In the TBG context, the BM bands are divided into active and frozen bands. The one-body correction $\hat{H}_\text{f.v.}$ to the active bands generated by the frozen valence bands can be expressed as
\begin{align}
\begin{split}\label{eq:Hfv}
    \hat{H}_{\text{f.v.}}
    &=\frac{1}{A}\sum_{\mystackrel{\bm{k}^\alpha \in \mathrm{mBZ}}{s\tau ,ab\in\text{act.}}}\left[\sum_{\bm{G}}\sum_{\mystackrel{\bm{k}^\beta \in \mathrm{mBZ}}{ s'\tau',v\in\text{f.v.}}}V(\bm{G})\lambda^*_{\tau',vv}(\bm{k}^\beta,\bm{G})\right]\lambda_{\tau,ab}(\bm{k}^\alpha,\bm{q}) \hat{c}^\dagger_{\tau as}(\bm{k}^\alpha)\hat{c}_{\tau bs}(\bm{k}^\alpha)\\
    &-\frac{1}{A}\sum_{\mystackrel{\bm{k}^\alpha \in \mathrm{mBZ}}{s\tau ,ab\in\text{act.}}}\sum_{\mystackrel{\bm{q}\in\text{all}}{v\in\text{f.v.}}}V(\bm{q})\lambda^*_{\tau,bv}(\bm{k}^\alpha,\bm{q})\lambda_{\tau,av}(\bm{k}^\alpha,\bm{q})\hat{c}^\dagger_{\tau as}(\bm{k}^\alpha)\hat{c}_{\tau bs}(\bm{k}^\alpha).
\end{split}
\end{align}
The band-projected Hamiltonian $\hat{H}_\text{proj.}$ can then be expressed as 
\begin{equation}\label{eq:Hproj}
    \hat{H}_\text{proj.}=\hat{H}_{\text{BM}}\bigg|_\text{act.}+\hat{H}_\text{int, n.o.}\bigg|_\text{act.}+\hat{H}_\text{sub}\bigg|_\text{act.}+\hat{H}_\text{f.v.},
\end{equation}
where $\hat{H}_\text{BM}$ is given in Eq.~\ref{eq:HBM_second_quantized}, $\hat{H}_\text{int, n.o.}$ is given in Eq.~\ref{eq:Hint_n.o.}, and $\hat{H}_\text{sub}$ is given in Eq.~\ref{eq:Hsub}. The notation $\hat{H}_\text{int, n.o.}\bigg|_\text{act.}$ means that we simply restrict the creation/annihilation operators in $\hat{H}_\text{int, n.o.}$ (as written in Eq.~\ref{eq:Hint_n.o.}, so without performing any additional anticommutations) to those belonging to active bands only --- terms that contain any frozen band operators are removed. Note that $\hat{H}_\text{f.v.}$ takes the same form as $\hat{H}_\text{sub}\bigg|_\text{act.}$ except that $\rho^0_{\tau s,ab}(\bm{k})$ is replaced by $-\delta_{a,b}\delta_{a\in\text{f.v.}}$. For certain interaction schemes, this leads to a large cancellation in the contributions from frozen states. For example, if we take the active bands to be the central bands, then the remote bands (which are all frozen) do not contribute at all to $\hat{H}_\text{proj.}$ for the central average scheme and the charge neutrality scheme.

\subsubsection{Additional Perturbations and Other Approaches}
The interaction Hamiltonian $\hat{H}_\text{int}$ can be augmented with additional subleading contributions beyond the long-range Coulomb part. For instance, intervalley Coulomb scattering at momentum transfers $\bm{q}\sim \bm{K}$ is power-law suppressed at small angles~\cite{Bultinck2020mechanism}. However, this term is important for helping to break the $SU(2)_K\times SU(2)_{K'}$ symmetry to the physical $SU(2)_S$ and resolving some of the spin degeneracies of the correlated states. It typically leads to a ferromagnetic `Hund's coupling' that prefers to align the spins in the two valleys. Coupling to graphene-scale phonons provides a comparable antiferromagnetic contribution, such that the overall sign of the Hund's coupling is uncertain~\cite{Chatterjee2020skyrmionic}, although experiments suggest the overall Hund's coupling is antiferromagnetic \cite{Morissette2023}. Onsite Hubbard interactions on the graphene scale can also be incorporated.  
Our presentation of the interacting continuum model in momentum and band space is convenient because it is relatively tractable, and sidesteps Wannier obstructions arising from the fragile topology\footnote{This becomes strong topology in the presence of PHS \cite{Song2019topological,Song2021stable}.} of the central bands. It also highlights the resemblance of TBG to a multi-component quantum Hall system in certain limits. This connection will be elaborated on in Section~\ref{sec:strong}.

To close this section, we also briefly comment on other perspectives in the literature. In fact, a real-space Hubbard-like approach based on `fidget-spinner' Wannier orbitals can be pursued, though one has to contend with extended interactions and non-trivial representations of some symmetry operations \cite{Zou2018emergent,Kang2018Wannier,Koshino2018Wannier,Seo2019ferromagnetic,Kang2019strong,xu2018kekule}.  A very different `heavy-fermion' approach resolves the topological obstruction by borrowing band representations from the remote bands, and emphasizes the emergence of local moment physics arising from localized (on the moir\'e scale) effective $f$-modes, and dispersive and topologically-anomalous $c$-fermions \cite{SongTHF,Wang2025THF,Calugaru2023THF,HerzogArbeitman2025THF,Shi2022THF,Rai2024THF,calugaru2024thermoelectriceffectnaturalheavy,merino2024evidenceheavyfermionphysics,herzogarbeitman2024topologicalheavyfermionprinciple,Singh2024THF}. This has the appeal of giving a physical way to reconcile the apparent dichotomy between the emergence of strong local correlations and ``moment formation''  with the topologically-enforced itineracy of the underlying free fermion bands. Operationally, one often still uses Hartree-Fock to analyze the role of interactions within this description, but certain parametric dependencies can be more ``physical'' in this approach \cite{HerzogArbeitman2025kekule,HerzogArbeitman2025THF}. Furthermore, the heavy-fermion  representation is also a convenient starting point for implementing other methods, such as dynamical mean-field theory (DMFT), which can yield detailed quantitative information on local correlations \cite{Rai2024THF}.   A more recent proposed approach that closely resembles the heavy fermion theory is to identify `quasi-local moments', that while not exponentially localized only have a small power-law weight, and work within the basis where these are manifest \cite{LedwithNLM}.  Yet another proposal is to introduce an ancilla fermion which hybridizes with the physical electron to give rise to Mott and pseudogap physics \cite{zhao2025ancillatheorytwistedbilayer}. Finally, several works have investigated correlated physics starting from a full microscopic tight-binding model \cite{klebl2021longrange,sanchez2024nematic,gonzalez2021magnetic}, although as discussed earlier this entails a vastly greater computational overhead.

\subsection{Beyond TBG: Other Moir\'e Materials}

We briefly discuss how the continuum model framework presented above for TBG can be generalized to other moir\'e materials, and highlight a few representative examples. The key steps are to identify the relevant continuum degrees of freedom describing each layer of the material stack, and the symmetries of the heterostructure. The latter are important to constrain the form of the moir\'e potentials and interlayer moir\'e hopping terms. A simple generalization of TBG is alternating-twist multilayer graphene, which consists of $N$ layers of graphene that are rotated from the untwisted configuration by angles $\theta/2,-\theta/2,\theta/2,-\theta/2,\ldots$ respectively \cite{Khalaf2019hierarchy,Carr2020sandwiched,Mora2019trilayer}. Despite the presence of multiple twisted layers, the continuum model can still be described by a single moir\'e period. This family of systems shares similarities with TBG, including $\hat{C}_{2z}$ symmetry and superconducting domes \cite{Park2021TTG,Hao2021electric,park2021magicanglemultilayergraphenerobust,Zhou2025,Cao2021Pauli}, but has the added feature that the band structure and phenomenology can be tuned by applying an external displacement field.   

The untwisted components that come together to build a moir\'e heterostructure do not have to be monolayers. In twisted monolayer-bilayer graphene, the (say) top layer is monolayer graphene, but the bottom two layers form Bernal-stacked bilayer graphene. In the continuum model, the $4\times 4$ sub-block corresponding to the Bernal stack consists of non-moir\'e hoppings that reproduce the low-energy band structure for Bernal bilayer graphene, while the moir\'e periodicity is reflected in the interlayer tunneling between the monolayer graphene and the upper layer of the Bernal bilayer \cite{Ledwith2022family,Liu2019quantum}.  

The low-energy physics of hexagonal semiconducting moir\'e TMDs often also lies at the $K$ and $K'$ valleys. For moir\'e bilayers, we distinguish between homobilayers where the layers are identical, and heterobilayers where the layers are different compounds\footnote{Given that each monolayer does not possess $\hat{C}_{2z}$ symmetry, one has to distinguish between AA- and AB-stacking.}. In the former, the low-energy description of each monolayer depends on whether we are interested in electron- or hole-doping. For hole-doping, strong Ising spin-orbit coupling splits the spin-valley degrees of freedom, such that only the spinful time-reversal partners $K\uparrow$ and $K'\downarrow$ appear at low energies~\cite{xiao2012coupled}. Within a given flavour sector, each monolayer is typically described by a single-component effective mass parabola, and the interlayer physics is introduced via intralayer moir\'e potential and interlayer moir\'e tunneling terms \cite{Wu2019}. TMD homobilayers have been experimentally shown to exhibit physics such as superconductivity \cite{Guo2025superconductivity,xia2024unconventionalsuperconductivitytwistedbilayer,xu2025signaturesreentrant}, and the integer and fractional quantum anomalous Hall effects at zero magnetic field~\cite{cai2023signatures,zeng2023integer,Park2023,Xu2023,park2024ferromagnetism,xu2024interplay}. Heterobilayers are distinct in that a moir\'e pattern is present even without twisting due to the lattice mismatch, and the low-energy physics is confined to one of the layers owing to the band offsets between the two constituents. In this case, the simplest continuum model within each flavour consists of a single effective mass parabola subject to a moir\'e-periodic potential~\cite{Wu2018hubbard,Zhang2020moire,Xie2022valley}. 

There exist moir\'e materials where the low-energy physics is not centered at the $K$-points, but other high-symmetry points such as the $\Gamma$-point \cite{Angeli2021Gamma,Classen2022ultra} or $M$-points \cite{Calugaru2025Mpoint,ingham2025moiremvalleybilayersquasionedimensional}, or even less-symmetric momentum regions~\cite{Wu2024possible,wang2022one}. These settings can exhibit phenomenology not present in $K$-valley materials, such as quasi-1D behaviours. In some of these cases, the absence of a topological obstruction means that it may be preferable to Wannierize the problem and study an effective tight-binding description, rather than resort to a continuum description. Furthermore, the nature of correlations in these systems may limit the utility of mean-field descriptions, though they still provide useful insight and are often a good first pass at the problem.

Finally, we comment on supermoir\'e platforms where an even longer lengthscale emerges on top of the moir\'e lengthscale. Consider twisting three identical monolayers with interlayer twists $\theta_{12}$ and $\theta_{23}$. Each adjacent pair of monolayers ($\{1,2\}$ and $\{2,3\}$) forms its own moir\'e pattern. Alternating-twist trilayer corresponds to $\theta_{12}=-\theta_{23}$ where the moir\'e Brillouin zones of two pairs align, such that the system can be described by a single moir\'e period. On the other hand, generic $\theta_{12},\theta_{23}$ lead to two moir\'e patterns that interfere and form a moir\'e-of-moir\'e pattern~\cite{nakatsuji2023multiscale,Devakul2023magic,yang_multi-moire_2024,popov_magic_2023,mao2023supermoireffective,Datta2024HTG,Guerci2024mosaic}. One special limit is the equi-angle helical trilayer, where $\theta_{12}=\theta_{23}$ and the supermoir\'e lengthscale $a_{\text{MM}}\sim a_{\text{M}}/(2\sin(\theta/2))$ is enhanced by a further factor of the moir\'e scale factor. Lattice relaxation calculations show that the supermoir\'e relaxes into large domains where the system locally recovers approximate moir\'e translation symmetry \cite{Devakul2023magic}. Hence, one approach to understand such systems is to first investigate the local physics within a domain, which can be described with a moir\'e continuum model like the previous examples \cite{xia_topological_2025,Kwan2024strong,kwan2025fractionalmosaic}, and then analyze how the tiling of domains affects mesoscopic physics.

\section{Hartree-Fock Mean-Field Theory}\label{sec:Hartree-Fock}
We now turn to a discussion of the Hartree-Fock (HF) mean-field approach. Although this is a standard technique taught in most courses on many-body theory, we present a brief self-contained introduction to time-independent HF, before we highlight subtleties of the modelling in the moir\'e setting by discussing the implementation of these techniques in TBG. We also discuss time-dependent HF.

\subsection{Hartree-Fock Basics}\label{subsec:basics}

In this section, we introduce the basic setup of self-consistent Hartree-Fock mean field theory. Readers familiar with HF may wish to skip or skim this section, and jump straight to the details of the implementation in TBG (see Sec.~\ref{subsec:HF_TBG}).

Consider a Hamiltonian with kinetic energy $\hat{T}$ and two-particle interaction $\hat{V}$,
\begin{equation}\label{eq:TV}
    \hat{H}=\hat{T}+\hat{V}=\sum_{ab}T_{ab}\hat{c}^\dagger_a\hat{c}^{\phantom{\dagger}}_b+\frac{1}{2}\sum_{abcd}V_{ab,cd}\hat{c}^\dagger_a\hat{c}^\dagger_b\hat{c}^{\phantom{\dagger}}_d\hat{c}^{\phantom{\dagger}}_c, 
\end{equation}
where $a,b,...$ denote single-particle orbitals, and the two-electron interaction integrals are $V_{ab,cd} := \braket{ab|\hat{V}|cd}$. These interaction matrix elements satisfy $V_{ab,cd}=V_{ba,dc}=V^*_{cd,ab}$ but are not anti-symmetrized for fermion exchange.  Hartree-Fock is a variational approximation that optimizes the ground state energy within the manifold of Slater determinants: states that can be written as $\ket{\Psi} = \prod_{1 \leq i \leq N_e} \hat{d}^\dagger_i \ket{0}$ for $N_e$ particles and some electron creation operators $\hat{d}^\dagger_i$ that satisfy $\{\hat{d}_i, \hat{d}^\dagger_j\} = \delta_{ij}$. (These are sometimes also referred to as {\it Gaussian} states, since they satisfy Wick's theorem: the expectation value of $2n$-fermion operators factor into sums of permutations of all the different pairings into products of $n$ two-fermion operator expectation values.)

A Slater determinant is fully specified by its one-body density matrix\footnote{We caution that sometimes this is defined instead as $P_{ab}=\langle \hat{c}_a^\dagger \hat{c}_b\rangle$, such as in the companion codes.} $P_{ab}=\langle \hat{c}_b^\dagger \hat{c}^{\phantom{\dagger}}_a\rangle$, which is Hermitian ($P^\dagger = P$), idempotent ($P^2 = P$) and whose trace is equal to the number of particles ($\tr(P) = N_e$). The energy expectation value of a Slater determinant can be evaluated using Wick's theorem
\begin{equation}
    E[P]=\bra\Psi\hat{H}\ket\Psi=\sum_{ab}T_{ab}P_{ba}+\frac{1}{2}\sum_{abcd}(V_{ac,bd}-V_{ac,db})P_{ba}P_{dc}.
\end{equation}
We define the Hartree-Fock Hamiltonian\footnote{One way to motivate this definition is to consider a ``partial Wick contraction" given by $\hat{c}^\dagger_a\hat{c}^\dagger_b\hat{c}^{\phantom{\dagger}}_d\hat{c}^{\phantom{\dagger}}_c \rightarrow \braket{\hat{c}^\dagger_a\hat{c}^{\phantom \dagger}_c}\hat{c}^\dagger_b\hat{c}^{\phantom\dagger}_d + \braket{\hat{c}^\dagger_b\hat{c}^{\phantom \dagger}_d}\hat{c}^\dagger_a\hat{c}^{\phantom\dagger}_c - \braket{\hat{c}^\dagger_a\hat{c}^{\phantom \dagger}_d}\hat{c}^\dagger_b\hat{c}^{\phantom\dagger}_c - \braket{\hat{c}^\dagger_b\hat{c}^{\phantom \dagger}_c}\hat{c}^\dagger_a\hat{c}^{\phantom\dagger}_d$. This is the mathematical way of expressing the notion that every electron experiences the mean-field generated by the other electrons.}
\begin{equation}\label{eq:H_HF}
    [H^{\text{HF}}[P]]_{ab}=T_{ab}+\sum_{cd}(V_{ac,bd}-V_{ac,db})P_{dc},
\end{equation}
allowing us to rewrite  the energy functional  as
\begin{equation}
    E[P] = \frac{1}{2}\tr\left(\left(T + H^{\text{HF}}[P]\right)P\right).
\end{equation}
It will be useful to also define the Hartree and Fock contributions to $H^\text{HF}[P]$ separately
\begin{gather}
[H^\text{H}[P]]_{ab}=\sum_{cd}V_{ac,bd}P_{dc}\\
[H^\text{F}[P]]_{ab}=-\sum_{cd}V_{ac,db}P_{dc}.
\end{gather}
In terms of these, the energy functional can be written as\footnote{We emphasize the presence of the factor of $\frac{1}{2}$ in front of the Hartree and Fock terms. Recall that these terms can be thought of as describing the mean field generated by the particles in the system due to their Coulomb interactions. However, the total energy only counts the interaction between each pair of particles once, and so the $\frac{1}{2}$ is needed to prevent double-counting.}
\begin{equation}
    E[P]=\text{tr}(TP)+\frac{1}{2}\text{tr}\left((H^\text{H}[P]+H^\text{F}[P])P\right).
\end{equation}

It can be shown that, for $P$ to correspond to a local \textit{minimum} of $E[P]$ within the variational manifold of $N_e$-particle Slater determinants, a necessary condition is that $P$  be given by occupying the $N_e$ {lowest-energy} eigenstates of $H^{\text{HF}}[P]$ (this is known as the ``aufbau principle'' analogous to atomic physics)~\cite{lions_solutions_1987}. This self-consistency condition is the basis for the iterative Hartree-Fock optimization procedure that we outline next.

\subsection{Solving the Self-Consistent Mean-Field Equations}

\subsubsection{Outline of the Iterative Procedure}
In order to find a density matrix $P$ that satisfies the self-consistency condition, the basic procedure is as follows. 

In the {\bf initialization step}, we choose some initial trial state $P_0$ (we will discuss this choice in Sec.~\ref{subsubsec:initialization}). Denoting the number of electrons as $N_e$ and the number of orbitals as $N$, the $k$-th {\bf iteration step} ($k \ge 0$) consists of the following sub-tasks:
\begin{enumerate}
    \item Construct the mean-field Hamiltonian from the current state $P_k$, i.e. $H_{k} \equiv H^{\text{HF}}[P_{k}]$.
    \item Diagonalize $H_{k}$, i.e.~find eigenvalues $\epsilon_i$ and eigenvectors $\psi_{i,a}$ that satisfy \mbox{$\sum_b[H_{k}]_{ab}\psi_{i,b} = \epsilon_i\psi_{i,a}$} for $1 \leq i \leq N$, where we assume that the eigenvalues are ordered such that $\epsilon_i \leq \epsilon_{i+1}$.
    \item Construct the new state $P_{k+1}$ from the $N_e$ lowest-energy eigenstates of $H_{k}$, i.e. $[P_{k+1}]_{ab} = \sum_{1 \leq i \leq N_e} \psi^{\phantom*}_{i,a}\psi^*_{i,b}$.
    \item If converged (we discuss convergence criteria in Sec.~\ref{subsubsec:conv_criteria}), terminate and return the final state $P_f = P_{k+1}$. Otherwise, proceed to the next iteration.
\end{enumerate}
However, this straightforward iterative procedure is known to sometimes oscillate between two or more states, and fail to converge. In order to prevent such situations, we slightly modify the iteration, as we now describe.

\subsubsection{Optimal Damping Algorithm}
The optimal damping algorithm (ODA)~\cite{cances_can_2000} provides a simple means to ensure (local) convergence of the HF self-consistency procedure. The central idea behind the ODA is that, after constructing $P_{k+1}$ in step (3) of iteration $k$, instead of directly using it as the input step of iteration $k+1$, we first find the \textit{optimal} (i.e.~lowest energy) linear combination of $P_{k+1}$ and $P_k$, and use this as the input for iteration $k+1$. We take
\begin{equation}
    P_\textrm{new}(\lambda)=(1-\lambda)P_k+\lambda P_{k+1},
\end{equation}
where we treat $0\leq\lambda\leq1$ as a variational parameter and compute the energy
\begin{equation}
    E[P_\textrm{new}(\lambda)]=E[P_k+\lambda(P_{k+1}-P_k)]=E[P_k]+\frac{s}{2}\lambda+\frac{1}{2}c\lambda^2,
\end{equation}
where
    \begin{align}
        s &= 2\tr\left(H_k(P_{k+1} - {P}_k)\right)\\
        c &= \tr\left((H_{k+1} - {H}_k)(P_{k+1} - {P}_k)\right).
    \end{align}
Since the HF energy is a quadratic function of the density matrix $P$, $E[P_\textrm{new}(\lambda)]$ is a quadratic function of $\lambda$. Minimizing the energy by setting $\partial E[P_\textrm{new}(\lambda)]/\partial\lambda=0$ leads to $\lambda=-s/2c$. However, since we fix $\lambda$ to be in the domain $0\leq\lambda\leq1$, if $c \leq -s/2$, we should set $\lambda = 1$. Since the optimization problem has a simple analytical solution, it can be performed efficiently. To summarize, the ODA modifies the basic HF iterative step as follows:
\begin{enumerate}
    \item Construct the mean-field Hamiltonian from the current state, i.e. $\tilde{H}_{k} \equiv H^{\text{HF}}[\tilde{P}_{k}]$.
    \item Diagonalize $\tilde{H}_{k}$, i.e.~find eigenvalues $\epsilon_i$ and eigenvectors $\psi_{i,a}$ that satisfy \mbox{$\sum_b[\tilde{H}_{k}]_{ab}\psi_{i,b} = \epsilon_i\psi_{i,a}$} for $1 \leq i \leq N$, where we assume that the eigenvalues are ordered such that $\epsilon_i \leq \epsilon_{i+1}$.
    \item Construct $[P_{k+1}]_{ab} = \sum_{1 \leq i \leq N_e} \psi_{i,a}\psi^*_{i,b}$.
    \item If converged, terminate and return the final state $P_f := P_{k+1}$.
    \item Compute $H_{k+1} = H^{\text{HF}}[P_{k+1}]$
    \item Compute 
    \begin{align}
        s &= 2\tr\left(H_k(P_{k+1} - \tilde{P}_k)\right)\\
        c &= \tr\left((H_{k+1} - \tilde{H}_k)(P_{k+1} - \tilde{P}_k)\right)
    \end{align}
    \item If $c \leq -s/2$, let $\lambda = 1$. Otherwise, let $\lambda = -s/2c$. Set $\tilde{P}_{k+1} = (1 - \lambda)\tilde{P}_k + \lambda P_{k+1}$ and go to the next iteration.
\end{enumerate}
We mention in passing that ODA is only one of the options to accelerate convergence of the HF optimization. For example, another method is EDIIS (Energy-Dependent Direct Inversion in the Iterative Subspace), which uses the results of successive HF iterations to build a picture of the local ``energy landscape'', and leverages this to construct an improved input to future iterations as a linear combination of previous solutions~\cite{Kudin2002}.

\subsubsection{Initialization}\label{subsubsec:initialization}

The HF iteration procedure is seeded by an initial density matrix $P_0$. The choice of $P_0$ is informed by the purpose of the HF calculation and prior knowledge of the physics of the system. A key aspect is whether we wish to enable symmetries of $\hat{H}$ to be broken. If $P_0$ respects a particular symmetry of $\hat{H}$, then that symmetry will be preserved in the HF Hamiltonian $H^\text{HF}[P_0]$ and hence future iterations. In Sec.~\ref{subsec:HF_TBG}, we will discuss how to handle symmetries in the context of TBG. For now, we consider \emph{unrestricted} HF calculations with no restriction on the symmetry of the variational state (apart from particle-number conservation). 

The most common application of HF is to study the mean-field phase diagram. While the main goal is to obtain the variational ground state, understanding the competing states (which may be local but not global energy minima) is also useful. In most scenarios, the HF practitioner does not have prior knowledge of the ground state or relevant low-energy phases. Performing multiple calculations starting with \emph{random} $P_0$ increases the likelihood of identifying the important phases, in case some were not known beforehand. The global mean-field ground state is then obtained by minimizing the converged total energy $E[P]$ over the set of HF calculations. A random trial state  (that is Hermitian, idempotent, and describes $N_e$ electrons) can be conveniently parametrized as  $P_0 = UDU^\dagger$, where $D$ is a diagonal matrix with $D_{ab} = \delta_{ab}$ for $a,b \leq N_e$ and 0 otherwise, and $U$ is a Haar-random $SU(N)$ matrix. It should be checked that enough\footnote{There is no general rule for determining how many seeds are enough. If the phase diagram is being computed across a range of parameters, then our recommendation is to check that the ground state energies across the phase diagram do not change when slightly increasing the number of random seeds (combined with any specially-designed initial states as described below).} random seeds are considered to fully capture the set of low-energy phases.

To efficiently evaluate the phase diagram, it is also useful to supplement the above unbiased random seeds with specially-designed initial states $P_0$ targeted at specific phases. This is because some (local) minima can have small basins of attraction that are only accessed by a small proportion of random states\footnote{Symmetries can lead to issues accessing relevant low-energy states in the HF procedure. For example, for Hamiltonians with a conserved flavor index, fully flavour-polarized phases can be difficult to reach from starting points with small flavour polarization. In this case, it is prudent to ensure that the set of initial states $P_0$ span the breadth of possible symmetry sectors.}. If the objective is to analyze a specific mean-field phase, then specially-designed initial states are appropriate.

\subsubsection{Convergence Criteria}\label{subsubsec:conv_criteria}
A key point of a self-constent HF calculation is to decide when it has converged, so that the iterative loop terminates. 
In the simplest implementation, we simply impose some convergence threshold $\epsilon \ll 1$, such that the procedure is deemed to have converged if
\begin{equation}
    \left\lVert P_{k+1} - \tilde{P}_k \right\rVert  < \epsilon,
\end{equation}
where $\left\lVert \cdot \right\rVert$ denotes the Frobenius norm of a matrix. 

\subsection{HF Implementation in MA-TBG: Generalities}\label{subsec:HF_TBG}
In this subsection, we discuss how  the general framework of HF discussed in the preceding subsections can be applied to MA-TBG.

\subsubsection{Projected Hamiltonian with Interaction Scheme}

We have introduced the interaction term (Eq.~\ref{eq:Hint_corrected}) in the many-body Hamiltonian, paying attention to subtleties associated with the interaction scheme. 
We have also discussed band projection, leading to the projected Hamiltonian in Eq.~\ref{eq:Hproj}. In the following, we project to the central bands of TBG, i.e.~8 total bands including spins and valleys, though the formalism straightforwardly generalizes to cases where a larger number of active bands are retained. In the general formulation of Eq.~\ref{eq:TV}, we assign the `kinetic energy' $\hat{T}$ and `two-particle interaction' $\hat{V}$ according to\footnote{There is an alternative parameterization that exists if $\hat{H}_\text{f.v.}=0$. Some works in the literature instead describe the decomposition as $\hat{T}=\hat{H}_\text{BM}$ and $\hat{V}=\hat{H}_\text{int, n.o.}\bigg|_\text{act.}$. In the HF equations, $\hat{V}$ is then mean-field decoupled using $\delta P=P-\rho^0$ instead of $P$, where $P$ is the one-body density matrix in the active subspace. The resulting contributions to the HF Hamiltonian that are proportional to the reference state $\rho^0$ are equivalent to $\hat{H}_\text{sub}\bigg|_\text{act}$. This perspective is implemented in the companion code. Correspondingly, the total energy is re-defined as $E[P]=\frac{1}{2}\text{Tr}(TP)+\frac{1}{2}\text{Tr}\left((P - \rho^0)H^{\mathrm{HF}}[P - \rho^0]\right)$, which is equivalent to the definition in the main text up to an unimportant overall constant.}
\begin{gather}
    \hat{T}=\hat{H}_{\text{BM}}\bigg|_\text{act.}+\hat{H}_\text{sub}\bigg|_\text{act.}+\hat{H}_\text{f.v.}\\
    \hat{V}=\hat{H}_\text{int, n.o.}\bigg|_\text{act.}.
\end{gather}
Note that while $\hat{T}$ is a one-body operator, it contains terms that depend on the interaction potential $V(q)$ due to $\hat{H}_\text{sub}\bigg|_\text{act.}$ (which is the projected version of Eq.~\ref{eq:Hsub}) and $\hat{H}_\text{f.v.}$ (Eq.~\ref{eq:Hfv}). The two-body term $\hat{V}$ is expressed purely in the form $\propto \hat{c}^\dagger\hat{c}^\dagger\hat{c}\hat{c}$.

\subsubsection{Mean-field Equations: Translationally-invariant Case}
The specific HF equations that must be solved for MA-TBG depend on which symmetries are enforced to be preserved, and which symmetries are allowed to be broken. This describes the `restricted' HF variational manifold. Evidently, the choice of symmetries to impose depends on the physics of interest. For example, when discussing the IKS state in Sec.~\ref{sec:IKS}, we will be interested in considering states that break valley $U(1)_v$ but at a finite wavevector. For domain wall configurations in Sec.~\ref{sec:domain_wall}, we only enforce translational symmetry parallel to the domain wall, while translation symmetry is broken in the other direction. We will defer discussion of these more involved situations to the specific case studies, and here simply discuss the moir\'e translation-invariant problem.

In the following, we allow intervalley coherence (breaking of $U(1)_v$ symmetry), but impose $S_z$ conservation. The latter means we restrict to spin-collinear states along the (arbitrarily chosen) $z$-axis in spin space.
In terms of the one-body density matrix, we have
\begin{equation}
P_{\tau'bs'\bm{k}^\beta, \tau a s \bm{k}^\alpha} \equiv \braket{\hat{c}^\dagger_{\tau a s}(\bm{k}^\alpha)\hat{c}_{\tau'bs'}(\bm{k}^\beta)} =0\text{\,\,\,\,if\,\,\,\,} s\neq s'\text{\,\,or\,\,}\bm{k}^\alpha\neq\bm{k}^\beta,
\end{equation}
where $\tau = \pm 1$ for valley $K$ and $K'$ respectively. This block-diagonal structure allows us to parameterize the one-body density matrix as\footnote{We caution that a different indexing convention for the projector is used in the companion code. In particular, we use $P_{\tau a,\tau' b}(\bm{k},s)\equiv \braket{\hat{{c}}^\dagger_{\tau a s}(\bm{k})\hat{{c}}_{\tau'bs}(\bm{k})}$ there.}
\begin{equation}
    {P}_{\tau'b,\tau a}(\bm{k},s) \equiv \braket{\hat{{c}}^\dagger_{\tau a s}(\bm{k})\hat{{c}}_{\tau'bs}(\bm{k})}.
\end{equation}
Due to the block-diagonal structure of one-particle density matrix and the symmetry of the Hamiltonian, the HF Hamiltonian is also diagonal in $\bm{k}$ and $s$, i.e.
\begin{equation}
    {H}^{\text{HF}}_{\tau a s \bm{k}^\alpha, \tau'bs'\bm{k}^\beta} = \delta_{ss'}\delta_{\bm{k}^\alpha\bm{k}^\beta}{H}^{\text{HF}}_{\tau a, \tau' b}(\bm{k}^\alpha, s)
\end{equation}
for some ${H}^{\text{HF}}_{\tau a, \tau' s}(\bm{k}^\alpha, s)$. The interacting Hartree and Fock contributions can be expressed as
\begin{align}
\begin{split}
\label{eq:Hartree_and_Fock_term}
{H}^\text{H}_{\tau a, \tau' b}(\bm{k}, s) &= 
\frac{\delta_{\tau \tau'}}{A}\sum_{s'\tau''cd}\sum_{\bm{G}}\sum_{\bm{k}'\in\text{mBZ}}V(\bm{G}){\lambda}_{\tau,ab}(\bm{k},\bm{G}){\lambda}^*_{\tau'',dc}(\bm{k}',\bm{G}){P}_{\tau''d,\tau'' c}(\bm{k}',s') \\
    {H}^{\text{F}}_{\tau a , \tau' b}(\bm{k},s)
    &=-\frac{1}{A}\sum_{cd}\sum_{\bm{q}\in\text{all}}V(\bm{q}){\lambda}_{\tau,ad}(\bm{k},\bm{q}){\lambda}^*_{\tau',bc}(\bm{k},\bm{q}){P}_{\tau d, \tau' c}(\bm{k}+\bm{q},s).
\end{split}
\end{align}
The HF Hamiltonian is then
\begin{equation}\label{eq:HHF_T_H_F}
    \hat{H}^\text{HF}=\hat{T}+\hat{H}^\text{H}+\hat{H}^\text{F}.
\end{equation}

\subsubsection{Initialization in the Presence of Symmetry}\label{subsubsec:TBG_initialization}
In Sec.~\ref{subsubsec:initialization}, we explained how to construct initial density matrices for the HF iteration procedure for a generic model. We now discuss the changes in the current setting where $\bm{k}$ and $s$ are conserved. We can still use specially designed initial states as long as they preserve momentum and spin. To initialize random states, we use a modified procedure where we randomly distribute (the precise random distribution is not important) the $N_e$ particles across the different $(\bm{k}, s)$ sectors. For each sector, we apply the method in Sec.~\ref{subsubsec:initialization} to generate a random one-body density matrix $P_{\tau a,\tau'b}(\bm{k},s)$ of the specified particle number. 

\subsubsection{Symmetry-preserving Diagonalization and the Aufbau Principle}
Since the HF Hamiltonian can be decomposed into symmetry sectors labeled by $(\bm{k}, s)$, we can simply diagonalize the Hamiltonian of each sector, i.e.~$\sum_{\beta}{H}^{\text{HF}}_{\alpha,\beta}(\bm{k},s)\psi_{i,\beta}(\bm{k},s) = \epsilon_i(\bm{k},s)\psi_{i,\alpha}(\bm{k},s)$ for $1 \leq i \leq 4$ (assuming we are projecting only into the central bands), where $\alpha,\beta$ are composite indices for band and valley. This represents a significant speed-up in the computational time compared to the case where no symmetries are enforced. For a calculation that includes $N_k$ momentum points, if one does not preserve any symmetry, the time complexity (per iteration) would be $O(N_k^3)$, from the matrix diagonalization. In contrast, the time complexity is $O(N^2_k)$ for the symmetry-preserving procedure, as the construction of $H^{\text{HF}}[P]$ takes $O(N^2_k)$ time. After diagonalizing the HF Hamiltonian matrices, we occupy the $N_e$ lowest-energy states out of \textit{all} eigenstates\footnote{One can also specify a target symmetry sector that does not necessarily include the ground state. This is useful for assessing the competition between different phases. For example, one can enforce vanishing spin polarization by occupying the lowest $N_e/2$ energy states in both spin sectors at each iteration.}.

\subsection{Interpreting and Analysing Hartree-Fock Results}

\subsubsection{Observables in Hartree-Fock}
\label{sec:observables}
At the end of the HF iteration procedure, the one-body density matrix $P$ can be examined to understand its properties. Consider a one-body observable $\hat{\mathcal{O}}=\sum_{\bm{k},s\tau \tau'ab}\mathcal{O}_{\tau a ,\tau'b}(\bm{k},s)\hat{c}^\dagger_{\tau, a}(\bm{k},s)\hat{c}_{\tau', b}(\bm{k},s)$ that preserves momentum and spin. The expectation value of this operator in the HF state is
\begin{equation}
    \langle \mathcal{O}\rangle_P=\frac{1}{N_1N_2}\mathrm{Tr}(P\mathcal{O})=\frac{1}{N_1N_2}\sum_{\mystackrel{\bm{k}\in \mathrm{mBZ}}{\tau\tau' sab}}P_{\tau'b, \tau a }(\bm{k},s) \mathcal{O}_{ \tau a ,\tau'b}(\bm{k},s),
\end{equation}
where the normalization is chosen so $\langle \mathcal{O}\rangle_P$ is defined per moir\'e unit cell. 
We can use $\mathcal{O}_{\tau a ,\tau'b}(\bm{k},s)=[\tau_z]_{\tau\tau'}\delta_{ab}$ to define the valley polarization $\textrm{VP}=|\langle \tau_z\rangle_P|$. For example, $\textrm{VP}=1$ if all the electrons fill one moir\'e band in valley $K$. Similarly we define the spin polarization 
$\textrm{SP}=|\langle s_z\rangle_P|$. 
The amount of intervalley coherence can be quantified by
\begin{equation}
    \textrm{IVC}=\frac{1}{N_1N_2}\
\sum_{\mystackrel{\bm{k}\in \mathrm{mBZ}}{sab}}|P_{K'b,Ka}(\bm{k},s)|^2,
\end{equation}
which probes the valley off-diagonal part of the HF projector. 

More generally, given a symmetry of the original Hamiltonian represented by $\hat{U}$, we can characterize the degree of symmetry-breaking in the HF state via 
\begin{equation}\label{eq:SU}
    S_{\hat{U}}=\frac{1}{N_1N_2}\lVert P-P^{\hat{U}}\rVert^2.
\end{equation}
This quantifies the difference between the density matrix $P$, and $P^{\hat{U}}$ obtained after applying the symmetry operation to $P$. We need to distinguish the two cases depending on whether $\hat{U}$ is unitary or anti-unitary. If $\hat{U}$ corresponds to an anti-unitary symmetry, then the projector will also be complex conjugated under the action of the symmetry. If we define the symmetry operations as 
\begin{gather}\label{eq:sym_op}
    \hat{O}\hat{c}^\dagger_{\alpha s}(\bm{k})\hat{O}^{-1} = \sum_\beta U^{\hat{O}}_{\beta\alpha}(\bm{k},s) \hat{c}^\dagger_{\beta s}(\hat{O}\bm{k}), \\
    \hat{A}\hat{c}^\dagger_{\alpha s}(\bm{k})\hat{A}^{-1} = \sum_\beta U^{\hat{A}}_{\beta\alpha}(\bm{k},s) \hat{c}^\dagger_{\beta s}(\hat{A}\bm{k})\label{eq:sym_opA}
\end{gather}
for unitary $\hat{O}$ and anti-unitary $\hat{A}$, then the corresponding transformations on the density matrix are\footnote{The rotation is defined in an ``active" sense.}
\begin{gather}
    P^{\hat{O}}_{\alpha\alpha'}(\hat{O}\bm{k},s)=\sum_{\beta\beta'}U^{\hat{O}}_{\alpha\beta}(\bm{k},s)\left[U^{\hat{O}}_{\alpha'\beta'}(\bm{k},s)\right]^*P_{\beta\beta'}(\bm{k},s),\\
    P^{\hat{A}}_{\alpha\alpha'}(\hat{A}\bm{k},s)=\sum_{\beta\beta'}U^{\hat{A}}_{\alpha\beta}(\bm{k},s)\left[U^{\hat{A}}_{\alpha'\beta'}(\bm{k},s)\right]^*P^*_{\beta\beta'}(\bm{k},s),
\end{gather}
where $\alpha,\beta$ are combined band and valley indices, and we have assumed that the transformation does not act on spin, as will be the case for the symmetries we are interested in ($\hat C_{2z}$, spinless $\hat{\mathcal{T}}$, $\hat C_{3z},\dots$).
If the state $P$ preserves the symmetry, it will transform into itself under the symmetry operation so that $S_{\hat{U}}=0$, where $S_{\hat{U}}$ is defined in Eq.~\ref{eq:SU}. This quantity will increase as the degree of symmetry-breaking increases. We can evaluate this quantity for e.g.~$\hat U=\hat C_{2z}\hat{\mathcal{T}}$, which is an example of an antiunitary symmetry. Explicitly, this takes the form
\begin{equation}
    U^{\hat C_{2z}\hat{\mathcal{T}}}_{\tau' b,\tau a}(\bm{k},s)=\delta_{\tau,\tau'}\sum_{\bm{G}, l, \sigma}u_{\tau a;l\sigma}(\bm{k},\bm{G})^*u_{\tau b;l\bar\sigma}(\bm{k},\bm{G})^*,
\end{equation}
where $\bar{\sigma}$ takes the opposite value to $\sigma$. See Appendix~C for a derivation of this equation. 
Note that $\hat C_{2z}\hat{\mathcal{T}}$ preserves the momentum $\bm{k}$, spin $s$, and valley $\tau$, but has a non-trivial action on the microscopic sublattice. On the other hand, spinless time-reversal $\hat{\mathcal{T}}$ takes $\bm{k}\rightarrow -\bm{k}$ and flips the valley, but has a trivial action on the microscopic sublattice:
\begin{equation}
    U^{\hat{\mathcal{T}}}_{\tau' b,\tau a}(\bm{k},s)=\delta_{\bar\tau,\tau'}\sum_{\bm{G}, l, \sigma}u_{\tau a;l\sigma}(\bm{k},\bm{G})^*u_{\bar\tau b;l\sigma}(-\bm{k},-\bm{G})^*.
\end{equation}
Various mean-field gaps can be directly extracted from the HF eigenvalues\footnote{A physical interpretation of the HF eigenvalues can be given in terms of Koopman's theorem \cite{KOOPMANS1934104}. This states that when removing a particle from an occupied HF orbital or adding a particle to an unoccupied HF orbital, the change in total energy is given by the corresponding HF eigenvalue, assuming that the other particles in the system do not alter their wavefunctions. We caution though that this should be only considered a crude heuristic for understanding the excitations of the HF state, and can qualitatively break down in certain situations. For example, doping charges into a ferromagnetic Chern band can lead to skyrmion textures where the many-body wavefunction is substantially reconstructed, as demonstrated in e.g.~Ref.~\cite{Kwan2021skyrmion}. Also, the HF band gaps introduced in Eq.~\ref{eq:bandgaps} are often overestimates of the true neutral gap of the system, partly because they neglect the attraction between the created electron and hole.} $\epsilon_i(\bm{k},s)$, which form the mean-field band structure. The direct and indirect HF band gaps are
\begin{gather}\label{eq:bandgaps}
    \Delta^\text{HF}_{\text{direct}}=\min_{\bm{k} \in \mathrm{mBZ}}\bigg[
    \min_{s,i}\epsilon_i(\bm{k},s)\bigg|_\text{unocc.} - \max_{s,i}\epsilon_i(\bm{k},s)\bigg|_\text{occ.}
    \bigg]\\
    \Delta^\text{HF}_{\text{indirect}}=\min_{\mystackrel{\bm{k} \in \mathrm{mBZ}}{s,i}}\epsilon_i(\bm{k},s)\bigg|_\text{unocc.} - \max_{\mystackrel{\bm{k} \in \mathrm{mBZ}}{s,i}}\epsilon_i(\bm{k},s)\bigg|_\text{occ.}.
\end{gather}
Note that these are only meaningful if the occupation number in every $\bm{k}$ sector is equal within each spin sector, otherwise the state is gapless.
HF interpolation (see Sec.~\ref{sec:HF_interpolation}) can be used to refine these gaps. We caution however, that the HF method tends to overestimate band gaps, such that $\Delta^\mathrm{HF}$ is often significantly larger than the experimentally extracted gaps. $\Delta^\mathrm{HF}$ may also be sensitive to the number of active bands included---central-band only calculations often lead to overestimates of the band gap compared ones that include more active bands.

The Chern number and Berry curvature of a gapped HF state can be obtained using the numerical method outlined in Ref.~\cite{Fukui2005Chern}.

\subsubsection{Hartree-Fock Interpolation}
\label{sec:HF_interpolation}
The self-consistency iterations are the most computationally expensive part of the HF calculations. For some settings, up to several thousand iterations may be required for convergence, thus limiting the system sizes that can be accessed\footnote{Using the companion code, a single iteration for a $24 \times 24$ system and only including the central bands takes about $0.4$ seconds. For much larger system sizes, the time-complexity is only quadratic in the system size, but a more memory-efficient implementation may be necessary, at the expense of time-efficiency.}. On the other hand, obtaining high-resolution band structures and Fermi surfaces may be crucial, e.g.~to study superconducting instabilities~\cite{Wagner2024} of a doped correlated insulator. One solution is to use two different momentum meshes, a coarse mesh and a fine mesh, an approach termed \emph{Hartree-Fock interpolation}~\cite{TSTG2}. The coarse mesh is used for the HF self-consistency procedure. After convergence, the Hartree and Fock terms in Eq.~\ref{eq:Hartree_and_Fock_term} are then re-evaluated on the fine mesh with the density matrix on the coarse mesh. More specifically, we construct
\begin{align}
\begin{split}
\label{eq:Hartree_and_Fock_term_interpolated}
{H}^\text{H}_{\tau a, \tau' b}(\bm{k}, s) &= 
\frac{\delta_{\tau \tau'}}{A}\sum_{s'\tau''cd}\sum_{\bm{G}}\sum_{\bm{k}'\in\text{c.m.}}V(\bm{G}){\lambda}_{\tau,ab}(\bm{k},\bm{G}){\lambda}^*_{\tau'',dc}(\bm{k}',\bm{G}){P}_{\tau''d,\tau'' c}(\bm{k}',s') \\
    {H}^{\text{F}}_{\tau a , \tau' b}(\bm{k},s)
    &=-\frac{1}{A}\sum_{cd}\sum_{\bm{G}}\sum_{\bm{k}'\in\text{c.m.}}V(\bm{k}'-\bm{k}+\bm{G}){\lambda}_{\tau,ad}(\bm{k},\bm{k}'-\bm{k}+\bm{G}) \\ & \quad \quad \times {\lambda}^*_{\tau',bc}(\bm{k},\bm{k}'-\bm{k}+\bm{G}){P}_{\tau d, \tau' c}(\bm{k}',s).
\end{split}
\end{align}
where c.m.~denotes the coarse mesh in the mBZ, $A$ is the area of the system of the coarse mesh, while $\bm{k}$ is allowed to take any value on the fine mesh. The HF Hamiltonian on the fine mesh is then diagonalized to obtain the mean-field band structure. Since this calculation only needs to be performed once after the final iteration, it does not add significantly to the computational cost.    

\subsubsection{Shell-filling Effects}

One type of finite-size effect that can be significant in HF calculations is related to ``shell-filling''. For example, even if the ground state in the thermodynamic limit should be a spin-unpolarized state,  a HF calculation with an odd number of electrons  necessarily yields a spin imbalance of at least one. A finite-size HF solution may thus break certain symmetries that would in fact be preserved in the thermodynamic limit. While in the above example, one can mitigate the issue simply by choosing an even number of electrons, in general one cannot determine the appropriate numbers of electrons to include solely based on symmetry considerations. To illustrate this point, consider a single-band problem with a symmetry, say time-reversal symmetry $\hat{\mathcal{T}}$, that maps $\bm{k} \rightarrow -\bm{k}$. For a state to preserve $\hat{\mathcal{T}}$ and translation symmetries, every pair of states that are related by $\mathcal{T}$ must be either both occupied or both empty,  except the states with time-reversal-invariant-momenta (TRIM), i.e.~$\bm{k} \equiv -\bm{k}$ up to a reciprocal lattice vector\footnote{Denoting the primitive reciprocal lattice vectors as $\bm{b}_{1,2}$, in general there are 4 nonequivalent TRIMs given by $\bm{0}, \bm{b}_1/2, \bm{b}_2/2, \bm{b}_1/2 + \bm{b}_2/2$. Depending on the choice of the momentum grid, they may not be all present.}, which can be either occupied or unoccupied. Therefore, a $\hat{\mathcal{T}}$-symmetric state needs to have an even (odd) number of electrons if an even (odd) number of TRIM orbitals are occupied. 

More generally, we call a set of states related by symmetry a ``shell", just like how an electronic shell consists of states related by rotation symmetry in atomic physics. For the many-body state to preserve a symmetry, each shell must be fully occupied or fully empty. For the first example of spin polarization, every shell consists of two states, so that one should choose an even number of electrons to avoid shell-filling effects. For the second example of $\hat{\mathcal{T}}$-symmetry, each shell can have either one (TRIM) or two (non-TRIM) states, making the parity of a symmetric state dependent on the energetics. In general, this issue is absent if one finds the ground state to be a correlated insulator, but poses a problem when the ground state has a Fermi surface (compensated metals or generic non-integer filling factors). One should therefore be cognizant of the fact that small symmetry breaking effects for metallic states obtained from HF can arise from shell-filling effects, which can be confirmed, if necessary, by inspecting the Fermi surfaces. We also note that, irrespective of the shell filling effect, gapless states with Fermi surfaces will exhibit finite gaps (between the highest occupied HF energy and the lowest unoccupied HF energy) due to the discrete momentum mesh.

\subsection{Time-dependent Hartree-Fock and the Random-Phase Approximation}\label{subsec:TDHF}

The Hartree-Fock self-consistency equation has a natural generalization to incorporate time evolution. In general, the Schrödinger-picture time-evolution of a density matrix $\hat{\rho}$ under a Hamiltonian $\hat{H}$ is given by $\hat{\rho} = e^{-i\hat{H}t}\hat{\rho} e^{i\hat{H}t}$. Equivalently, we can write
\begin{equation}
    i\dot{\hat{\rho}} = [\hat{H}, \hat{\rho}].
\end{equation}

Drawing on the ideas underlying the Hartree-Fock treatment where individual particles see an effective one-body Hamiltonian generated by the many-body state, we can write down the following time-evolution equation under the\textit{ mean-field approximation} for the \textit{single-particle} density matrix of a \textit{many-body} interacting system,
\begin{equation}\label{TDHF}
i\dot{P} = [H^{\text{HF}}[P],P],
\end{equation}
where $H^{\text{HF}}[P]$ is defined in Eq.~\ref{eq:H_HF}. This ``Time-Dependent Hartree-Fock (TDHF) equation'' was  first written down by Dirac in 1930~\cite{Dirac}. Importantly,  the TDHF time evolution conserves energy:
\begin{eqnarray}
\dot{E} & = & \frac{\mathrm{d}}{\mathrm{d}t}\tr\left(P\left[T + \frac{1}{2} (H^{\text{H}}[P] + H^{\text{F}}[P])\right] \right) \\
 & = & \tr\left( \dot{P} \left[ T+\frac{1}{2} (H^{\text{H}}[P] + H^{\text{F}}[P]) \right] \right) + \frac{1}{2}\tr\left(P( H^{\text{H}}[\dot{P}] +  H^{\text{F}}[\dot{P}] )\right) \\
  &  = &\tr\left( \dot{P} \left[ T+\frac{1}{2} (H^{\text{H}}[P] + H^{\text{F}}[P]) \right] \right) + \frac{1}{2}\tr\left((H^{\text{H}}[P] + H^{\text{F}}[P]  )\dot{P}\right) \\
   & = & \tr\left(\dot{P}H^{\text{HF}}[P] \right) \\
   & = & 0 \, ,
\end{eqnarray}
where the last line follows directly from $\dot{P}=-i[H^{\text{HF}}[P],P]$. This energy conservation is no coincidence, but follows from the fact that the TDHF equation can be interpreted as a time-dependent variational equation, as we explain in the following section. We will closely follow the exposition of Ref.~\cite{Kramer}.

\subsubsection{Time-dependent Variational Principle with Slater Determinants}
Given a Slater determinant $|\psi_0\rangle$ (for example the one which solves the Hartree-Fock self-consistency equation for the problem of interest), one can parametrize all Slater determinants \emph{which have non-zero overlap with $|\psi_0\rangle$} as
\begin{equation}
|\psi(z)\rangle = \exp\left(\sum_{m,i}z_{im}\hat{c}^\dagger_m \hat{c}_i\right)|\psi_0\rangle = \exp\left(\hat{c}^\dagger Z^T \hat{c}\right)|\psi_0\rangle. \label{Thouless1}
\end{equation}
Here we have introduced the index notation $i,j$ for single-particle states that are occupied in $|\psi_0\rangle$, and $m,n$ denote states that are empty in $|\psi_0\rangle$. Eq.~\eqref{Thouless1} is called the \emph{Thouless parametrization} of Slater determinants~\cite{ThoulessStability}.

We now promote the coefficients $z_{im}(t)$ to general time-dependent functions, with the time evolution determined by the Time-Dependent Variational Principle (TDVP). In general, consider the Lagrangian 
\begin{equation}
    L = \frac{i}{2}\frac{\braket{\psi|\dot{\psi}}-\braket{\dot{\psi}|\psi}}{\braket{\psi|\psi}} - \frac{\braket{\psi|H|\psi}}{\braket{\psi|\psi}},
\end{equation}and the time-evolution of $\ket{\psi}$ from the variational optimization of the Lagrangian. If the variational manifold of $\ket{\psi}$ is the entire Hilbert space, the above procedure reproduces the exact Schrödinger time-evolution. More generally, in the TDVP, we can restrict the variational manifold of $\ket{\psi}$ to a more tractable sub-manifold of the Hilbert space. For our present purposes, we use $|\psi(z(t))\rangle$ as an ansatz to approximate the exact time-evolved state $e^{-iHt}|\psi(z(0))\rangle$. The optimal approximation can be found by constructing the Lagrangian
\begin{equation}\label{eq:Lagrangian}
L = \frac{i}{2}\langle\psi(z)|\psi(z)\rangle^{-1}\sum_{m,i} \left[\dot{z}_{im} \langle \psi(z)|\partial_{im}\psi(z)\rangle - \dot{\bar{z}}_{im} \langle \partial_{im} \psi(z)|\psi(z)\rangle \right] - \frac{\langle \psi(z)|H|\psi(z)\rangle}{\langle\psi(z)|\psi(z)\rangle}\,,
\end{equation}
where $\partial_{im}=\frac{\partial}{\partial z_{im}}$ (and similarly for $\bar{\partial}_{im}$). Formally treating $z$ and $\bar{z}$ as independent variables, we can rewrite this as
\begin{equation}
L = \frac{i}{2}\sum_{m,i}\left(\dot{z}_{im}\partial_{im} - \dot{\bar{z}}_{im}\bar{\partial}_{im} \right)\ln N(\bar{z},z) - \mathcal{H}(\bar{z},z)\, ,\label{Lag1}
\end{equation}
with
\begin{eqnarray}
N(\bar{z},z) & = & \langle \psi(z)|\psi(z)\rangle = \det\left(\mathds{1} + ZZ^\dagger \right), \\
\mathcal{H}(\bar{z},z) & = & \frac{ \langle \psi(z)|H|\psi(z)\rangle}{\langle \psi(z)|\psi(z)\rangle}. 
\end{eqnarray}
The functions $z_{im}(t)$ that realize the optimal approximation for the time-evolved state are obtained by solving the equations of motion obtained from the Lagrangian in Eq.~\eqref{Lag1}~\cite{Kramer}. These equations of motion are
\begin{eqnarray}
\dot{z}_{im} & = & i\{\mathcal{H},z_{im}\}, \label{eom1}\\
\dot{\bar{z}}_{im} & = & i\{\mathcal{H},\bar{z}_{im}\}. \label{eom2}
\end{eqnarray}
Here we have introduced the Poisson bracket
\begin{eqnarray}
    \{f,g\} & := & \sum_{im,jn} C^{-1}_{im,jn}\left(\partial_{im}f\, \bar{\partial}_{jn}g -  \bar{\partial}_{jn} f \, \partial_{im} g \right), \\
    C_{im,jn} & = & \bar{\partial}_{im}\partial_{jn}\ln N(\bar{z},z) = (\mathds{1}+Z^\dagger Z)^{-1}_{mn} (\mathds{1}+Z Z^\dagger)^{-1}_{ji}.
\end{eqnarray}
The equations of motion~\eqref{eom1}-\eqref{eom2} show that the TDVP indeed produces an energy-conserving dynamics. The single-particle density matrix of $|\psi(z)\rangle$ is given by
\begin{equation}\label{defP}
P = \left(\begin{matrix} (\mathds{1}+ZZ^\dagger)^{-1} & (\mathds{1}+ZZ^\dagger)^{-1}Z \\
Z^\dagger (\mathds{1}+ZZ^\dagger)^{-1} & Z^\dagger (\mathds{1}+ZZ^\dagger)^{-1}Z \end{matrix}\right)\,,
\end{equation}
where $\mathds{1}$ is the $N_e \times N_e$ identity matrix. The first $N_e$ indices of the rows and columns of $P$ run over the occupied orbitals of the reference state $\ket{\psi_0}$, and the rest of the indices run over the empty orbitals. Via a straightforward (but tedious) calculation one finds that Eqs.~\eqref{eom1}-\eqref{eom2} imply that the above single-particle density matrix evolves according to the TDHF equation in Eq.~\eqref{TDHF}. In other words, time-evolution under the Lagrangian given in Eq.~\ref{eq:Lagrangian} is equivalent to the TDHF introduced in Eq.~\ref{TDHF}.

The TDVP dynamics is non-linear. However, if we are only interested in Gaussian fluctuations around the mean-field saddle point (i.e. keeping only the quadratic terms in the Lagrangian~\eqref{Lag1}), we obtain linear equations of motion. The linearized equations of motion take on a particularly simple form, as we now show. First, using Eq.~\eqref{eq:TV} we find that to second order in $z,\bar{z}$ the TDVP Hamiltonian can be written as
\begin{equation}
\mathcal{H}(\bar{z},z) = E_0 + \frac{1}{2} (\bar{z} \;z) \left(\begin{matrix} A & B \\ B^* & A^* \end{matrix} \right) \left(\begin{matrix}z \\ \bar{z} \end{matrix}\right) + \mathcal{O}(z^3) \,,\label{Hz}
\end{equation}
where $E_0 = \langle \psi_0|H|\psi_0\rangle$, and we have defined the matrices $A$ and $B$ as
\begin{eqnarray}
A_{im,jn} & = & (E_m - E_i) \delta_{ij}\delta_{mn} + V_{mj,in} - V_{mj,ni} \label{defA1}, \\
B_{im,jn} & = & V_{mn,ij} - V_{mn,ji}. \label{defB1}
\end{eqnarray}
Note that $A$ is Hermitian, and $B$ is symmetric. 

From Eq.~\eqref{Hz}, we find that the linearized equations of motion are
\begin{equation}
\left(\begin{matrix} \dot{z} \\ \dot{\bar{z}}\end{matrix}\right) =- i\left(\begin{matrix} A & B \\ -B^* & -A^*\end{matrix}\right)\left(\begin{matrix} z \\ \bar{z} \end{matrix} \right). \label{linTDVP}
\end{equation}
Stationary solutions of the linearized TDVP equation are obtained by diagonalizing the matrix on the right-hand side this equation. At this point it is important to remember that Eq.~\eqref{linTDVP} was obtained by treating $z$ and $\bar{z}$ as independent variables. From the structure of the matrix in Eq.~\eqref{linTDVP} we see that if $(v_1,v_2)$ is an eigenvector with eigenvalue $\omega$, then $(\bar{v}_2,\bar{v}_1)$ is an eigenvector with eigenvalue $-\omega$. We can now again impose that $z$ and $\bar{z}$ are related by complex conjugation by combining the solutions $(v_1,v_2)$ and $(\bar{v}_2,\bar{v}_1)$:
\begin{equation}
\left(\begin{matrix} z(t) \\ \bar{z}(t) \end{matrix}\right) = \left(\begin{matrix}v_1 \\ v_2 \end{matrix} \right)e^{-i\omega t} + \left(\begin{matrix}\bar{v}_2 \\ \bar{v}_1 \end{matrix} \right)e^{i\omega t}.
\end{equation}
So to reiterate, our solution strategy was to first solve the linearized dynamics in a larger space where $z$ and $\bar{z}$ are independent variables, and then project back to the physical space where $z$ and $\bar{z}$ are related by complex conjugation. 

As a final comment, let us mention that the matrix in Eq.~\eqref{linTDVP} is non-Hermitian. Nevertheless, we have assumed that $\omega$ is real. This is indeed guaranteed to be true when $|\psi_0\rangle$ is a (local) minimum in the variational energy landscape, i.e. when~the matrix in quadratic term Eq.~\eqref{Hz} is positive definite~\cite{ThoulessStability}. Computing the eigenvalues of the Hermitian quadratic matrix in Eq.~\ref{Hz} is a useful way to assess the local stability of the HF state $\ket{\psi_0}$ in the manifold of Slater determinants. A negative eigenvalue signals that $\ket{\psi_0}$ must be locally unstable to a different HF solution.

\subsubsection{Linearized TDHF and the Random-Phase Approximation}
There is a close connection between TDHF and the Random-Phase Approximation\footnote{We caution that our usage of the term RPA is distinct from the standard `Bohm-Pines' treatment of the homogeneous electron gas~\cite{bohm1953collective,giuliani2008quantum,wolf2024quasiboson}. In the latter, only bubble diagrams are retained, and the propagators do not include Hartree or Fock self-energy corrections.} (RPA), as first pointed out in Ref.~\cite{Goldstone}. To explain this connection, we define the particle-hole operators
\begin{equation}
\hat{O}_{ab} = \hat{c}^\dagger_a \hat{c}_b - \langle  \hat{c}^\dagger_a \hat{c}_b \rangle \,,
\end{equation}
where $\langle \cdot \rangle$ is the ground state expectation value in the system under consideration. Next, we consider the zero-temperature particle-hole Green's function
\begin{equation}
\left[G_2(\omega)\right]_{ab,cd} = -i \int_{-\infty}^{+\infty}\mathrm{d}t\, e^{i\omega t} e^{-\epsilon |t|}\langle \hat{T} \hat{O}_{ab}(t) \hat{O}_{cd}\rangle\,,
\end{equation}
with $\hat{T}$ the bosonic time-ordering operator, and $\epsilon>0$ a small regulator. The particle-hole Green's function can be obtained as a solution to the Bethe-Salpeter equation~\cite{Salpeter}, which is easily derived using Feynman diagram techniques. The exact Bethe-Salpeter equation is of course intractable for a general interacting system, so one has to resort to approximate solutions. 

One such approximation is provided by RPA, where the interaction kernel in the Bethe-Salpeter equation is taken to be a combination of a particle-hole scattering event, and a particle-hole annihilation and creation event --- both due to the density-density interaction. In the terminology of diagram-enthusiasts: the approximate Bethe-Salpeter equation is obtained by summing the geometric series generated by both \emph{bubble} and \emph{ladder} diagrams. Crucially, the RPA approximation is a \emph{conserving approximation} in the sense of Kadanoff and Baym~\cite{Baym}, meaning that it preserves all symmetries and hence respects the Ward identities.

Poles in $G_2(\omega)$ give access to the spectrum of particle-hole excitations. It turns out that the equation which determines the locations of the poles in the RPA approximation is exactly the same as the linearized TDHF equation~\cite{ThoulessGF}. This equivalence between RPA and TDHF makes it clear that (1) RPA is indeed a conserving approximation, as the TDVP formalism explicitly preserves all symmetries, and (2) solutions to the linearized TDHF equation provide an approximation for the energies of particle-hole excitations. The latter include neutral collective modes, which can be interpreted as particle-hole bound states in the TDHF/RPA framework. These collective modes include the Goldstone modes when $|\psi_0\rangle$ breaks a continuous symmetry, and these are guaranteed to be gapless due to the conserving nature of the approximation.

\subsubsection{TDHF Equations for MA-TBG}
We now apply the general formalism of the previous subsections to MA-TBG. First we define the particle-hole operators with momentum $\bm{q}$ as
\begin{equation}
\sum_{\bm{k} \in \mathrm{mBZ}}\sum_{\mu,\nu} \varphi^m_{\bm{q},\mu\nu}(\bm{k}) \hat{d}^\dagger_{\mu}(\bm{k}+\bm{q}) \hat{d}_{\nu}(\bm{k})\,,
\end{equation}
where $\mu,\nu$ label the Hartree-Fock orbitals, occupied or unoccupied, with creation operator defined as $\hat{d}^\dagger_\mu(\bm{k}) = \sum_{\tau a s}\psi_\mu^{\tau a s}(\bm{k})\hat{{c}}^\dagger_{\tau a s}(\bm{k})$, and $m$ labels the excitation mode. The particle-hole wavefunctions satisfy
\begin{equation}
\varphi^m_{\bm{q},\mu\nu}(\bm{k}) = 0 \;\;\text{ if }\;\; n_\mu(\bm{k}+\bm{q}) = n_\nu(\bm{k}).
\end{equation}
Here we have used $n_\mu(\bm{k}) \equiv \braket{\hat{d}^\dagger_\mu \hat{d}_\mu}_{\psi_0}\in \{0,1\}$ to denote the occupation of the single-particle states in the Slater determinant $|\psi_0\rangle$.

If we denote the form factors of the HF orbitals as $\Lambda_{\mu\nu}(\bm{k}, \bm{q}) = \sum_{\tau s a b}(\psi^{\tau a s}_\mu(\bm{k}))^*\psi^{\tau b s}_\nu(\bm{k}+\bm{q}) \lambda_{\tau,ab}(\bm{k},\bm{q})$, where $\lambda_{\tau, ab}$ are the single-particle form factors defined in Eq.~\ref{eqn:form_factors}, the linearized TDHF equation~\eqref{linTDVP} for MA-TBG can then be written as following generalized eigenvalue equation
\begin{align}\label{collmode}
    & \left(n_\mu(\bm{k}+\bm{q}\right) -  n_\nu(\bm{k}))\varphi^m_{\bm{q},\mu\nu}(\bm{k})\eta_{\bm{q},m}\omega_{\bm{q},m}   =   |E_\mu(\bm{k}+\bm{q})- E_\nu(\bm{k})|\varphi^m_{\bm{q},\mu\nu}(\bm{k}) \\ 
     & + \frac{1}{A}\sum_{\bm{G}}V(\bm{q}+\bm{G})\sum_{\bm{k}'\in \mathrm{mBZ}}\tr\left(\varphi^m_{\bm{q}}(\bm{k}')\Lambda(\bm{k}',\bm{q}+\bm{G}) \right)\left[ \Lambda^\dagger(\bm{k},\bm{q}+\bm{G})\right]_{\mu\nu} \nonumber \\
     & - \frac{1}{A}\sum_{\bm{q}'\in \mathrm{all}}V(\bm{q}')  \left[\Lambda(\bm{k}+\bm{q},\bm{q}')\varphi^m_{\bm{q}}(\bm{k}+\bm{q}')\Lambda^\dagger(\bm{k},\bm{q}') \right]_{\mu\nu}\,, \nonumber
\end{align}
where $\eta_{\bm{q},m}\in\{-1,+1\}$ is defined such that $\omega_{\bm{q},m}\geq 0$. The factor $n_\mu(\bm{k}+\bm{q}) -  n_\nu(\bm{k})$ on the left-hand side of this equation comes from the minus sign in front of the lower two blocks in Eq.~\eqref{linTDVP}. The first term on the right-hand side contains the absolute value of the difference in mean-field band energies. We recognize the remaining two terms containing the interaction potential $V(\bm{q})$ as finite-momentum generalizations of the Hartree and Fock potentials.

\section{Why Hartree-Fock? Exact Results in the Chiral-Flat Strong Coupling Limit}\label{sec:strong}

Given that TBG hosts narrow bands and strong Coulomb interactions, the reader may question whether Hartree-Fock theory (Sec.~\ref{sec:Hartree-Fock}), which is restricted to uncorrelated Slater determinants, has any hope of capturing the many-body physics. On the other hand, many groups have studied TBG and related materials using the HF method \cite{Kang2021cascades,Zhang2020HF,Bultinck2020hidden,Liu2021nematic,Xie2020nature,Xie2021weak,Cea2020insulating,zhang2022correlaed,Lian2021TBG4,Liu2021theories,Kwan2021IKS,Faulstich2023,Wagner2022global,Cea2019electronic,Guinea2018electrostatic,Xie2023,Hejazi2021hybrid,Kwan2024EPC,Kwan2021skyrmion,Kwan2021domain,WangCTI,Wang2024Tri,HerzogArbeitman2025kekule,TSTG2,Christos2022TTG,Breio2023chern}. A hint towards a resolution lies in experiments that observe the QAH at $\nu=+3$~\cite{Serlin2020QAH,Sharpe2019ferromagnetism} and a proliferation of multiple Chern insulating phases at small magnetic field $B$~\cite{Tomarken2019compressibility,Wong2020cascade,Park2021flavour,Xie2019spectroscopic,Nuckolls2020Chern,Pierce2021unconventional,Xie2021fractional,Saito2021hofstadter,Yu2021hofstadter,Lu2019orbital,Das2021symmetry,Stepanov2021competing,Wu2021Chern,Choi2021correlation,Grover2022mosaic,Uri2020mapping,Das2021reentrant}, pointing to the relevance of band topology and flavor symmetry-breaking. Such ingredients are also inherent to the phenomenon of quantum Hall ferromagnetism (QHFM), which arises in flavor-degenerate Landau levels. There, mean-field theory has proven remarkably useful for understanding symmetry-broken quantum Hall states at integer filling factors~\cite{Yang1994quantum,Girvin1999QHE,Nomura2006graphene,Eisenstein2004}. In this section, we highlight the similarities between QHFM, and TBG in the `strong-coupling' regime, and show how insights from the former can shed light on the physics of the latter. This connection is facilitated by working in the Chern basis of TBG, which is introduced in Sec.~\ref{subsec:Chern_basis}. In Sec.~\ref{subsec:generalized}, we show how under certain assumptions, the interacting Bistritzer-MacDonald model yields exact Slater determinant ground states, which are interpreted as generalized ferromagnets. While much of the phenomenology of TBG departs from this idealized  limit, it has a very important {\it conceptual} role underpinning the use of mean-field studies: it provides a limit where the {\it exact} ground states are  Slater determinants. Thus, working sufficiently close to this limit, we might expect that ground states  continue to be well-approximated by Slater determinants --- an intuition borne out both by exact diagonalization and DMRG studies, and by making predictions that are consistent with experiments. In Sec.~\ref{subsec:strong_complications}, we discuss complications to this strong-coupling picture, but defer a treatment of \emph{intermediate-coupling} to Sec.~\ref{sec:IKS}.
We only highlight some of the main facets of the strong-coupling perspective; readers may consult Ref.~\cite{strongcoupling_review} for a more detailed and technical review.

\subsection{Chern Basis}\label{subsec:Chern_basis}

Near the magic angle $\theta\simeq 1.05^\circ$ where the moir\'e lattice constant $a_M\simeq 14\,\text{nm}$, the typical Coulomb scale can be estimated as $U\simeq \frac{e^2}{4\pi\epsilon_0\epsilon_r a_M}\simeq 20\,\text{meV}$ with $\epsilon_r\simeq 5$ capturing the dielectric screening from the hBN environment. Since this exceeds the kinetic bandwidth $W\lesssim 10\,\text{meV}$, the BM band basis may not be the most natural one for interaction physics. Indeed, as we highlighted in the introduction (Sec.~\ref{sec:introduction}), there is ample experimental evidence for flavour polarization and band reconstruction, such that the concepts of non-interacting valence and conduction bands may not be appropriate for understanding interaction effects. This affords us the freedom to depart from the kinetic basis.  

In the following, we focus exclusively on the Hamiltonian projected onto the central bands. An alternative single-particle basis that more explicitly reveals the underlying topological character of the central bands is motivated from considering the role of the substrate in the observation of the QAH at $\nu=+3$ in Ref.~\cite{Serlin2020QAH}. There, alignment to the hBN substrate effectively generates a sublattice splitting term $\Delta\sigma_z$, which imposes a relative bias between the microscopic $A$ and $B$ sublattices\footnote{More involved theoretical treatments find that the coupling of TBG to the hBN can be substantially more complicated than the simplified treatment mentioned here, and the physics can be highly sensitive to the precise relative angle with the substrate~\cite{Cea2020hBN}. This may be relevant for the `Chern mosaic' regime of TBG where the system breaks up into a mesoscopic array of domains with different local Chern number \cite{Grover2022mosaic}. In Sec.~\ref{sec:domain_wall}, we discuss how to address domain walls induced by spatial variations in the local sublattice splitting within Hartree-Fock.}. If we rediagonalize the continuum model with this term, the Dirac points at the $K_M,K'_M$ corners become gapped, and the central valence and conduction bands are Chern bands with $C=\pm1$~\cite{Zhang2019hbn,Bultinck2020mechanism}.

For small $\Delta$ though, the valence and conduction bands are individually not well-behaved near the $K_M,K'_M$ corners where the gap is small. Hence, the non-interacting kinetic basis for small $\Delta$ will not suffice. However, there is experimentally a proliferation of Chern insulating phases at finite magnetic field $B$ even in the absence of explicit hBN alignment. This hints that even when $\Delta=0$, the sublattice degree of freedom plays a non-trivial role and may be driven to be `ferromagnetic' by strong interactions.

The above considerations prompt the introduction of the \emph{Chern basis}, obtained by directly diagonalizing the sublattice operator $\sigma_z$ in the subspace of central bands at $\Delta=0$. To do this, within each valley sector $\tau$, we construct the $2\times2$ matrix 
\begin{equation}
    [\Gamma_\tau(\bm{k})]_{ab}=\sum_{\bm{G},\sigma,l}\sigma [u_{\tau a;l\sigma}(\bm{k},\bm{G})]^*u_{\tau b;l\sigma}(\bm{k},\bm{G}),
\end{equation}
with eigensolutions labelled by $\tilde{\sigma}=\pm1$. We choose $\tilde{\sigma}=+1$ to correspond to the eigenvalue close to $+1$, and $\tilde{\sigma}=-1$ to correspond to the eigenvalue close to $-1$. Denoting the eigenvectors as $w_{\tau\tilde{\sigma},a}(\bm{k})$, the Bloch functions in the Chern basis are uniquely\footnote{Up to an overall $\bm{k}$-dependent gauge phase. We refer readers to Refs.~\cite{Bultinck2020hidden,Bernevig2021TBG3,Song2021stable} for details of the gauge-fixing procedure. We will assume that such a sensible choice of gauge has been fixed in the following discussion.} defined as
\begin{equation}\label{eq:Chern_basis_Bloch}
    \ket{\phi_{\tau\tilde{\sigma}}(\bm{k})}=\sum_{a}w_{\tau\tilde{\sigma},a}(\bm{k})\ket{\psi_{\tau a}(\bm{k})}.
\end{equation}
Hence, the Chern band with $\tilde{\sigma}=+1$ ($-1$) is mostly polarized on microscopic sublattice $A$ ($B$). Despite the fact that the microscopic sublattice polarization is not perfect for generic parameters, we will often refer to these two Chern bands as `sublattice' bands labelled by $A$ and $B$. If needed, the use of the index $\sigma$ or $\tilde{\sigma}$ will clarify whether we are referring to the microscopic sublattice, or the Chern basis. 
Crucially, the Bloch states of Eq.~\ref{eq:Chern_basis_Bloch} have non-trivial Chern numbers $C=\tilde{\sigma}\tau$~\cite{Bultinck2020hidden,Liu2019pseudo,Lian2021TBG4}. Hence, including the spin degree of freedom, we can group these Chern bands into `Chern quartets' with $C=\pm1$. The $C=+1$ quartet is spanned by $\ket{K\uparrow A}, \ket{K\downarrow A},\ket{K'\uparrow B},\ket{K'\downarrow B}$, while the $C=-1$ quartet is spanned by $\ket{K\uparrow B}, \ket{K\downarrow B},\ket{K'\uparrow A},\ket{K'\downarrow A}$.

\subsection{Generalized Ferromagnets: Exact Slater Determinant Ground States}\label{subsec:generalized}

\begin{figure}
    \centering
    \includegraphics[width=\linewidth]{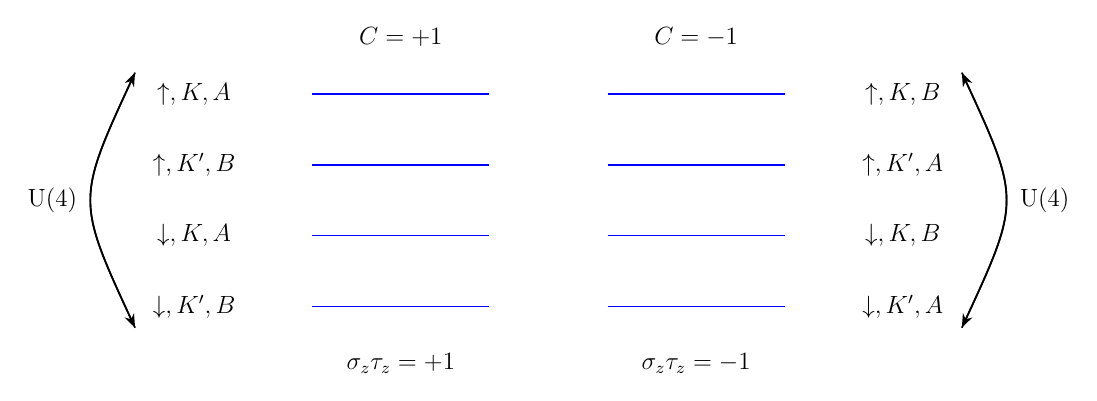}
    \caption{The Hamiltonian $\hat{H}_{U_\text{S}}$ satisfies a $U(4)_{C=1}\times U(4)_{C=-1}$ symmetry corresponding to arbitrary rotations between bands in the same Chern sector.}
    \label{fig:Strong_coupling_schematic}
\end{figure}

The utility of the Chern basis for addressing many-body physics is most apparent when the chiral ratio $\kappa=w_\text{AA}/w_\text{AB}$ is set to zero. In this so-called \emph{chiral limit}, the BM model gains a chiral symmetry $\{\sigma_z,H_\text{BM}\}=0$, and the Chern basis introduced above becomes completely sublattice polarized. Furthermore, at the magic angle, the central bands are exactly flat and degenerate through the entire mBZ~\cite{Tarnopolsky2019chiral}. In this \emph{chiral-flat} limit, the Chern basis itself becomes a valid kinetic basis for the central subspace. As we have frozen out the remote bands, the projected Hamiltonian consists only of the interaction term, which is reminiscent of the Landau level problem. The link to QHFM is sharpened by recognizing that $\hat{C}_{2z}\hat{\mathcal{T}}$ and particle-hole symmetry $\mathcal{P}$, in concert with chiral symmetry, constrain the form factors to be diagonal in the Chern basis and only depend on the Chern number. In particular, we can parameterize~\cite{Bultinck2020hidden}
\begin{equation}\label{eq:symmetric_form_factor}
    \Lambda^{\text{S}}_{\alpha\beta}(\bm{k},\bm{q})=\bra{\phi_\alpha(\bm{k})}e^{-i\bm{q}\cdot\hat{\bm{r}}}\ket{\phi_\beta(\bm{k}+\bm{q})}=F^{\text{S}}(\bm{k},\bm{q})e^{i\phi^{\text{S}}(\bm{k},\bm{q})C_\alpha}\delta_{\alpha\beta},
\end{equation}
where $\alpha,\beta$ are composite indices for the sublattice $\tilde{\sigma}$, spin $s$, and valley $\tau$. The use of the uppercase $\Lambda$ denotes that the form factor is expressed in the Chern basis. The superscript `S' is short-hand for `symmetric', and reflects the large $U(4)_{C=1}\times U(4)_{C=-1}$ symmetry of the density operator\footnote{$\bm{d}^\dagger(\bm{k})$ is an eight-component vector that collects the Chern basis creation operators including all flavours, and matrix-vector product over $\alpha,\beta$ is implied.} $\hat{\rho}(\bm{q})=\sum_{\bm{k}}\hat{\bm{d}}^\dagger(\bm{k})\Lambda^{\text{S}}(\bm{k},\bm{q})\hat{\bm{d}}(\bm{k}+\bm{q})$ under independent Chern-number preserving $U(4)$ rotations. We schematically illustrate these symmetry operations in Fig.~\ref{fig:Strong_coupling_schematic}.

In the `central average' subtraction scheme\footnote{As a reminder, the reference state in the central average scheme consists of fully occupied (unoccupied) remote valence (conduction) bands. The central bands are taken to have occupation $1/2$ in the reference density matrix.} defined in Sec.~\ref{subsubsec:subtraction_schemes}, the projected Hamiltonian in the chiral-flat limit can be expressed as 
\begin{gather}\label{eq:H_US}
    \hat{H}_{U_\text{S}}=\frac{1}{2A}\sum_{\bm{q}}V(\bm{q})\left[\delta{\hat{\rho}}(\bm{q})\right]^\dagger\delta\hat{\rho}(\bm{q})\\
    \delta{\hat{\rho}}(\bm{q})=\hat{\rho}(\bm{q})-\bar{\rho}(\bm{q}), \quad 
    \bar{\rho}(\bm{q})=\frac{1}{2}\sum_{\bm{k},\bm{G}}\delta_{\bm{G},\bm{q} }\tr\Lambda^\text{S}(\bm{k},\bm{G}),
\end{gather}
Remarkably, as we briefly review below, the strong-coupling Hamiltonian $\hat{H}_{U_\text{S}}$ yields {\it exact} Slater determinant ground states in the chiral-flat limit at $\nu=0$ for any repulsive interaction $V(\bm{q})$. These involve uniformly (in $\bm{k}$-space) polarizing any four orthogonal directions within the Chern quartets. One example would be polarizing into all the $A$ bands, which would yield a valley Hall state with net vanishing Chern number. Acting with independent $U(4)$ rotations within the Chern quartets generates the continuous manifold of $C=0$ states. We can similarly obtain ground states with total Chern number $|C|=2,4$ by unequally occupying the two Chern quartets\footnote{If the Chern sector $C$ has $n_C$ occupied bands, then its configuration can be described by a matrix spinor living in the Grassmannian projective space~\cite{Atteia2021zoo,Kwan2021skyrmion}
\begin{equation}
    \frac{SU(4)}{SU(n_C)\times SU(4-n_C)\times U(1)}.
\end{equation}}. These Slater determinants can all be referred to as \emph{generalized ferromagnets}~\cite{Bultinck2020hidden,Lian2021TBG4}, and highlight the resemblance to QHFM. In the latter setting of Landau levels, a typically $SU(2)$ index\footnote{This could be the spin, valley, or layer degree of freedom.} ferromagnetizes due to Coulomb exchange. The analogy in TBG  is that the Chern quartets comprise a \emph{pair} of $SU(4)$-degenerate Landau levels with \emph{opposite} effective magnetic fields in order to match the opposite Chern numbers.

Hybridization between bands with opposite Chern numbers in TBG is disfavored in the strong-coupling limit: the mismatch of Chern numbers leads to topologically-enforced vortices in the order parameter that are energetically costly~\cite{Bultinck2020mechanism}. (However, inter-Chern hybridization can be induced away from the strong-coupling limit. As discussed in Refs.~\cite{KwanCTI,WangCTI,zou2025valleyordermoiretopological}, kinetic dispersion or interaction anisotropies can nevertheless stabilize an inter-Chern coherent state which has been dubbed the \emph{Chern-texture insulator.})

To show that Eq.~\ref{eq:H_US} hosts Slater determinant ground states, we note that it is manifestly positive semi-definite so that any zero-energy state is necessarily a ground state. Consider the action of $\delta\hat{\rho}(\bm{q})$ on any generalized ferromagnet $\ket{\Psi}$ at $\nu=0$. This vanishes for $\bm{q}\neq\bm{G}$ because we have filled Chern bands and the density operator is diagonal in this basis. For $\bm{q}=\bm{G}$, $\delta\hat{\rho}(\bm{q})$ annihilates $\ket{\Psi}$ if it has total $C=0$ owing to the compensating $\bar{\rho}(\bm{q})$ term. If $\ket{\Psi}$ has non-zero Chern number, then we could have terms like $\sim \sum_{\bm{k}}F^\text{S}(\bm{k},\bm{G})\sin\phi^\text{S}(\bm{k},\bm{G})\ket{\Psi}$. However TRS $\hat{\mathcal{T}}=\tau_x \mathcal{K}$, where $\mathcal{K}$ is complex conjugation, constrains $F^\text{S}(-\bm{k},-\bm{q})=F^\text{S}(\bm{k},\bm{q})$ and $\phi^\text{S}(-\bm{k},-\bm{q})=\phi^\text{S}(\bm{k},\bm{q})$. Using the general property $\Lambda^S(\bm{k},\bm{q})=[\Lambda^S(\bm{k}+\bm{q},-\bm{q})]^\dagger$, we obtain $F^\text{S}(-\bm{k},\bm{G})=F^\text{S}(\bm{k},\bm{G})$ and $ \phi^\text{S}(-\bm{k},\bm{G})=-\phi^\text{S}(\bm{k},\bm{G})$. Therefore the summation over $\bm{k}$ vanishes, meaning that we have an exact zero-energy ground state. In fact, the above construction for exact generalized ferromagnetic ground states can be extended to non-zero integer fillings if one imposes an additional `flat-metric condition' on the form factors~\cite{Lian2021TBG4}\footnote{The flat metric condition requires that $\lambda_{\alpha\beta}(\bm{k},\bm{G})=\xi(\bm{G})\delta_{\alpha\beta}$. This is trivially true for $\bm{G}=0$, and is approximately satisfied for large $\bm{G}$ due to the decaying Bloch function overlaps. Hence the largest violations are for the smallest shell of non-zero RLVs. We note that the strong-coupling states are still eigenstates of the many-body Hamiltonian when the flat metric condition is not satisfied, but one cannot prove that they are {\it ground} states.}.
In this way, we have analytic solutions for strong-coupling insulators at every integer $\nu$, which can take various Chern numbers $C=4-|\nu|,2-|\nu|,\ldots,|\nu|-4$. Note that the states at even (odd) integer filling necessarily have even (odd) Chern number. Each combination of filling and Chern number comprises a degenerate manifold of states related by rotations within each Chern quartet. Similar strong-coupling analyses has been applied to other moir\'e graphene platforms~\cite{Kwan2024strong,strongcoupling_review,Ledwith2021TB,lee2019theory,calugaru2021TSTG1,Christos2022TTG}.

While the the chiral-flat limit does not describe realistic TBG, its utility lies in the existence of a hierarchy of scales \footnote{Explicit expressions for these scales can be found in Refs.~\cite{Bultinck2020hidden,Lian2021TBG4,Kwan2021skyrmion}.} at strong coupling which permits corrections to be treated perturbatively within the manifold of $U(4)\times U(4)$ generalized ferromagnets~\cite{Bultinck2020hidden,Lian2021TBG4}. The energy scale of the full symmetry group is the `symmetric' Coulomb scale $U_\text{S}\sim20\,$meV. For realistic chiral ratio $\kappa\simeq 0.5-0.8$, there are new terms in the Hamiltonian that break the $U(4)\times U(4)$ symmetry. The main effect of the dispersion is to enable single-particle tunneling between the Chern sectors with energy scale\footnote{Note that more general interaction schemes lead to a finite inter-Chern tunneling even in the chiral limit with a vanishing kinetic bandwidth. The reason is that `flat' in chiral-flat is defined as the absence of additional one-body terms with respect to the strong-coupling form in Eq.~\ref{eq:H_US}. Interaction schemes other than the average scheme will generate such additional terms when rewritten in this way. Furthermore, as touched upon in Sec.~\ref{subsec:strong_complications}, the effective $t_\text{S}$ is enhanced at non-zero integer fillings owing to the Hartree contribution from the extra occupied electron or hole bands.
} $t_\text{S}\sim 5\,\text{meV}$. 
Deviation from the chiral limit also renders the Chern basis states no longer perfectly sublattice-polarized, and produces $U(4)$-violating corrections to the symmetric form factor in Eq.~\ref{eq:symmetric_form_factor}. In the density-density interaction, this leads to corrections with energy scale $U_\text{A}\sim5\,$meV. 

Focusing for simplicity on the spinless\footnote{This means that we drop the spin index $s$ completely.} Hamiltonian at neutrality such that the parent symmetry group is $U(2)\times U(2)$, the effects of the above perturbations can be captured via anisotropies with positive strengths $J$ and $\lambda$ (i.e.~energy scales $E_J=A_\text{UC}J$ and $E_\lambda=A_\text{UC}\lambda$ with $A_\text{UC}$ the area of the moir\'e unit cell). It turns out that the generalized ferromagnets that get energetically selected by these terms have total $C=0$. Defining an intra-Chern Pauli triplet $\bm{\eta}$ and an inter-Chern Pauli triplet $\bm{\gamma}$
\begin{equation}\label{eq:Chern_triplet}
  \bm{\eta}=(\sigma_x\tau_x,\sigma_x\tau_y,\tau_z),\quad \bm{\gamma}=(\sigma_x,\sigma_y\tau_z,\sigma_z\tau_z),
\end{equation}
we can parameterize the strong-coupling ferromagnets residing in the $C=0$ sector with two Chern-filtered pseudospins  
\begin{equation}\label{eq:npm}
    \bm{n}_\pm(\bm{k})\equiv  \bra{\Psi}\hat{\bm{d}}^\dagger(\bm{k}) \bm{\eta} \frac{1\pm\sigma_z\tau_z}{2}  \hat{\bm{d}}(\bm{k})\ket{\Psi}
\end{equation}
which are independent of $\bm{k}$. $\bm{n}_C(\bm{k})$ parameterizes the direction that $\ket{\Psi}$ polarizes in within Chern sector $C$. Inter-Chern tunneling $t_\text{S}$ favors maximizing the number of flavors that are singly-occupied, since this enables virtual tunneling leading to a superexchange $J\sim \frac{t_\text{S}^2}{U_\text{S}}$. Within our space of spinless states with total $C=0$, this manifests as an anti-ferromagnetic (AFM) coupling between the two pseudospins. As discussed above, a finite chiral ratio $\kappa$ causes the density operator to develop less symmetric parts. For instance the $KA$ Chern band now has a small component in the  microsopic $B$ sublattice, such that there is a finite exchange gain between the $KA$ and $KB$ Chern bands. It can be shown that the effect on the pseudospins is a coupling $\lambda$ that is AFM in-plane and FM out-of-plane. Both break the chiral-flat $U(2)\times U(2)$ symmetry to distinct $U(2)$ subgroups~\cite{Bultinck2020hidden}. The resulting easy-plane ground state with $\bm{n}_+=-\bm{n}_-$ (and hence valley-$U(1)_v$ degeneracy) is known as the KIVC~\cite{Bultinck2020hidden}, so called because it preserves a modified Kramers TRS $\hat{\mathcal{T}}'=\tau_y\hat{\mathcal{K}}$, and possesses intervalley coherence (IVC). An equivalent description of the spinless KIVC is via its density matrix $P_{\text{KIVC}}=\frac{1}{2}(1+Q_{\text{KIVC}})$, where
\begin{equation}\label{eq:QKIVCspinless}
    Q_{\text{KIVC}}=\sigma_y(\tau_x\cos\phi_\text{IVC}+\tau_y\sin\phi_\text{IVC})
\end{equation}
is parametrized through the IVC angle $\phi_\text{IVC}$.

A similar analysis, with spin reintroduced, can be performed at the other integer fillings to deduce the specific ferromagnets selected by the anisotropies~\cite{Lian2021TBG4,Bernevig2021TBG5}. 
The anisotropies reduce the symmetry group from $U(4)\times U(4)$ to $SU(2)_K\times SU(2)_{K'}\times U(1)_v\times U(1)_C$, where the $SU(2)$ pieces correspond to independent spin rotations in the two valley sectors.
We outline a simple recipe for recovering the final results for any integer filling $\nu$. First, as many occupied bands as possible are grouped into spin-polarized versions of the spinless KIVC bands described above. This maximizes the exchange gain due to $\lambda$ and superexchange due to $J$. For even $\nu$, the construction ends here, yielding a maximally intervalley-coherent state with $C=0$. For odd $\nu$, the leftover occupied band is chosen to be valley-polarized, leading to a total $|C|=1$~\cite{Lian2021TBG4}. Tab.~\ref{tab:strongcoupling_summary} summarizes the result at each filling assuming spin-collinear density matrices (though recall that the residual $SU(2)_K\times SU(2)_{K'}$ flavor symmetry can be used to generate degenerate solutions).

\begin{table}[h!]
\centering
\begin{tabular}{c|c|c|c|c}
\hline
$|\nu|$ & State & Spin pol. & Valley pol. & $|C|$ \\ 
\hline
\hline
$0$ & Spin-singlet KIVC  & 0 & 0 & 0 \\ 
\hline
$1$ & Partially intervalley-coherent  & 1 & 1 & 1 \\ 
\hline
$2$ & Spin-polarized KIVC& 2 & 0 & 0 \\ 
\hline
$3$ & Sublattice-polarized QAH & 1 & 1 & 1 \\ 
\hline
\end{tabular}
\vspace{0.2cm}
\caption{Summary of strong coupling states stabilized away from the chiral-flat limit at different integer fillings $\nu$, with the respective spin polarization, valley polarization and Chern number $C$.}\label{tab:strongcoupling_summary}
\end{table}

The energetic ordering between closely-competing states implied above can be be altered by external perturbations. Consider the sublattice bias term $\Delta\sigma_z$ arising from hBN alignment.  A sufficiently strong $\Delta$ can unwind the IVC of some strong-coupling states in order to induce sublattice polarization. For instance at $\nu=-2$, the KIVC state, which hybridizes Chern bands with opposite valley and sublattice, gives way to the $U(1)_v$-symmetric valley Hall (VH) state that polarizes into (say) $KB\uparrow$ and $\bar{K}B\uparrow$ Chern bands. The orbital part of a perpendicular magnetic field can also couple to the strong-coupling states differently depending on $C$. A magnetic field is useful in splitting the Chern insulators in filling space according to the Streda formula \cite{Streda1982}. In particular, the density $n$ of a state with Chern number $C$ obeys $\frac{\partial n}{\partial B}=\frac{eC}{h}$. The resulting Chern numbers extracted from finite field measurements are largely consistent with the even/odd dichotomy of the strong-coupling framework \cite{Lian2021TBG4}.

In summary, the strong-coupling picture yields an elegant perspective on TBG: namely, that there is a controllable limit (chiral-flat) in which the ground states are exact Slater determinants, and hence lie within the variational subspace of the many-body Hilbert space  accessible via Hartree-Fock mean field theory. Furthermore, perturbing the chiral-flat limit to depart from the artificial enhanced $U(4)\times U(4)$ symmetry and  restore the physical symmetries  selects  specific broken-symmetry states from within this strong-coupling Slater determinant (``generalized ferromagnet'') manifold. These analytical predictions are indeed borne out by numerical Hartree-Fock studies of the interacting BM model~\cite{Kwan2021IKS,Bultinck2020hidden,Xie2020nature,Cea2020insulating,Zhang2020HF,Parker2021strain,Lian2021TBG4}. Additionally, various numerical techniques that complement Hartree-Fock, such as DMRG \cite{Parker2021strain}, quantum Monte Carlo \cite{Zhang2021QMC,Hofmann2022QMC}, and exact diagonalization \cite{Potasz2021ED,Xie2021TBG6}, obtain ground states that are broadly similar to generalized ferromagnets, thereby validating the strong coupling treatment of the interacting BM model. 

\subsection{Complications to the Strong-Coupling Framework}\label{subsec:strong_complications}

The strong-coupling limit of the BM model provides a range of insights into its phase diagram, which is corroborated by numerical studies of the interacting BM model. However, the agreement worsens away from charge neutrality, and the discrepancy becomes increasingly severe at large fillings. For example, there are several alternative non-strong-coupling states at $|\nu|=3$~\cite{Xie2023,Kang2020nonabelian,Zhang2021nonlinear}, such as various density wave orders, and their competition can depend on details of the modelling. Generally, a larger value of the chiral ratio favours these competing states~\cite{Xie2023,Kang2020nonabelian}.

One significant origin of the breakdown of strong-coupling theory lies in the so-called `Hartree' corrections to the interacting band structure\footnote{Note that Fock corrections are also present, which tend to counteract the Hartree contribution somewhat.}. Typically in crystalline materials, small amounts of doping can be theoretically accounted for by just shifting the chemical potential. It is not obvious that this is valid in TBG, since entire moir\'e bands worth of electrons can be doped into or out of the system due to the small mBZ. Furthermore, we have argued above for strong-coupling orders which reshuffle the flavor occupations with large exchange splittings. This is exacerbated by the real-space localization of flat-band charge density near the AA-stacking regions of TBG, which leads to electrostatic distortions of the effective dispersion. In momentum space, this distortion arises from the non-trivial $\bm{k}$-dependent structure in the Bloch functions. For example, the single-particle wavefunctions near $K_{\text{M}},K'_\text{M}$ are sharply peaked at the AA-stacking regions, while the wavefunctions near $\Gamma_{\text{M}}$ are more diffuse in real-space. The result is that for (say) electron-doping away from charge neutrality, there is a relative downward renormalization of the quasiparticle energy around $\Gamma_{\text{M}}$. Since this is controlled by the interaction strength, which is typically larger than the bare kinetic scale, the interacting bands become substantially deformed. There is a pronounced asymmetry (relative to the Fermi level) of this renormalization:  the conduction bands further away from charge neutrality develop a pronounced minimum at $\Gamma_{\text{M}}$ (the so-called `Hartree dip') and enhanced bandwidth, while the valence bands remain comparatively flat~\cite{Pierce2021unconventional,Guinea2018electrostatic,Rademaker2019charge,Cea2019electronic,Goodwin2020hartree,Kang2021cascades}. At large enough integer fillings, this Hartree dip can even fully close the exchange splitting induced by flavor ferromagnetism\footnote{This occurs for the $|\nu|=3$ QAH state in the average subtraction scheme.}.

This band renormalization has been argued to explain some of the experimental phenomenology. One feature that is shared by almost all experiments is the asymmetry of Landau fans in transport. In the presence of correlated insulators at finite integer fillings, the Shubnikov-de Haas oscillations only develop for densities away from charge neutrality~\cite{Lu2019orbital,Cao2018insulator,Cao2018SC,Yankowitz2019pressure,Sharpe2019ferromagnetism,Serlin2020QAH,Park2021flavour,Zondiner2020cascade,Uri2020mapping,Saito2020independent,Saito2021pomeranchuk,Stepanov2020untying,Wu2021Chern,Saito2021hofstadter}. This can be rationalized by the considerably lighter effective masses for this direction of doping. On the other hand, the other direction involves doping into heavy quasiparticles, whose effective masses may preclude visible oscillations, or may even involve instabilities of the Fermi liquid. These Hartree effects have also been proposed to be deleterious to more exotic fractionalized phases. Several theoretical studies have investigated the possibility of fractional Chern insulators when fractionally filling the topological conduction band of the QAH at $\nu=+3$~\cite{Repellin2020,Abouelkomsan2020particlehole,wilhelm2021interplay,Ledwith2020FCI}. The quantum geometry and non-interacting flatness of the moir\'e bands should provide a favorable arena for such phases. However, the enhanced bandwidth due to band renormalization suppresses their formation~\cite{parker2021fieldtunedzerofieldfractionalchern}, such that the assistance of finite perpendicular magnetic fields is required to stabilize a fractional Chern insulator in experiments~\cite{Xie2021fractional,finney2025extendedfractionalcherninsulators}.

We now briefly comment on some other terms in the Hamiltonian that influence the strong-coupling perturbation analysis even for a \emph{pristine} twisted bilayer, i.e.~in the absence of imperfections like heterostrain\footnote{The effect of heterostrain will be extensively discussed in Sec.~\ref{sec:IKS}.} and disorder, and external fields. As mentioned in Sec.~\ref{subsec:interactions}, there are Hund's couplings that break the $SU(2)_K\times SU(2)_{K'}$ symmetry, which can be treated perturbatively within the lowest-energy strong-coupling states. The most interesting case is at $\nu=-2$ since this is near the fillings where superconductivity is most prominent. Taking the VH state as an example, a ferromagnetic (anti-ferromagnetic) Hund's coupling would lead to a spin-ferromagnetic (spin-valley-locked) state. This distinction is crucial for constraining the properties of the proximate superconducting domes, if the correlated insulator is indeed the parent state. Recent experiments have addressed this using electron spin resonance \cite{Morissette2023} and twist-decoupled heterostructures~\cite{Hoke2024decoupled}. These studies suggest that the Hund's coupling is anti-ferromagnetic. (We return to the effective Hund's coupling at some length when discussing the spin structure of the IKS state below.)

Beyond settling the spin degeneracies, additional terms may even invert the hierarchy of pseudospin ordering in the strong-coupling insulators. This is the case when accounting for the coupling to graphene zone-corner optical $K$-phonons, which can scatter electrons between the valleys~\footnote{Angle-resolved photoemission spectroscopy yields signatures that have been linked to a strong coupling to these $K$-phonons~\cite{chen2024strong}.}. Focusing again on $\nu=-2$, it has been theoretically proposed that this electron-phonon coupling can lower the energy of a different strong-coupling state below that of the KIVC~\cite{Blason2022Kekule,Kwan2024EPC,Shi2025optical}. This new so-called TIVC state\footnote{Note that Ref.~\cite{Kwan2024EPC} showed that there are actually two closely related phases, the TIVC-QSH and the IVC-QAH, which are degenerate at Hartree-Fock level. The former exhibits a quantized spin Hall effect and spinful TRS, while the latter has $|C|=2$. Both have a non-zero Kekul\'e charge density.} also possesses intervalley coherence, but its symmetries permit a finite Kekul\'e charge distortion at wavevector $q=2K_D$ on the graphene scale, which crucially is absent in the other candidate strong-coupling orders\footnote{Valley-diagonal orders like the VH clearly lack a Kekul\'e distortion due to the $U(1)_v$-symmetry. The KIVC has intervalley coherence, but its $\hat{\mathcal{T}}'$ symmetry means that only a Kekul\'e \emph{current} density shows up \cite{Calugaru2021KD,Hong2021KD}.}. As a result, the $K$-phonons can directly couple to TIVC order, leading to a lattice distortion energy that can potentially overcome the perturbative exchange scale $\lambda$ \cite{Kwan2024EPC}. Since the two occupied Chern bands in the TIVC reside in opposite spins, the superexchange scale $J$ does not play a role.

\section{Case Study I: Kekul\'e Spiral Order}\label{sec:IKS}

In this section, we revisit the applicability of the strong-coupling theory in Sec.~\ref{sec:strong} to realistic TBG devices. We discuss why the strong-coupling picture, while theoretically appealing, fails to fully capture the experimental phenomenology. We then go on to discuss the resolution to this puzzle, which turns out to be twofold: to add a seemingly small amount of heterostrain (which has been experimentally characterized in many samples) and to consider a wider class of variational states than the QH ferromagnet manifold. In this first case study, we motivate and discuss these modifications, and how the resulting modifications can be studied using the HF machinery built up in the preceding sections.

\subsection{Motivation: Challenges to Strong Coupling}
The strong coupling picture of TBG is microscopically concrete and conceptually appealing, and has been successful in understanding many experimental aspects. It supplies a wealth of candidate correlated insulators at every integer filling, unified by the common theme of generalized flavor ferromagnetism. This can explain at least some of the observed correlated insulators, and provides a foundation for the cascade of transitions and resets at low temperatures. The strongest evidence in favor of strong-coupling theory is the topology of various correlated states. The early experimental observation of a QAH resistance in hBN-aligned TBG at $\nu=+3$~\cite{Serlin2020QAH,Sharpe2019ferromagnetism} is naturally framed in terms of a spin, valley, and sublattice-polarized Chern insulator. Further experimental support is given by the observed stabilization of  QAH insulators\footnote{Strictly speaking, these are not anomalous since they only emerge in a finite magnetic field. However the small fields required (significantly smaller than $\sim20$\,T corresponding to one flux per unit cell) suggest that these Chern states remain competitive at zero field.} with Chern numbers $C = \pm 3, \pm 2, \pm 1$ at $\nu = \pm 1, \pm 2, \pm 3$  on applying a small out-of-plane magnetic field as low as 0.1~T, even in the absence of substrate alignment \cite{Nuckolls2020Chern,Wu2021Chern,Das2021symmetry,Stepanov2021competing,Saito2021hofstadter}. Evidence for spin skyrmion behaviour \cite{Yu2023spin} and (finite-field) fractional Chern insulators \cite{Xie2021fractional,finney2025extendedfractionalcherninsulators} also highlight the utility of the Chern basis in understanding the physics.

However, there are also  inconsistencies between the strong-coupling picture and several aspects of the experimental phenomenology. This is not surprising given the heterogeneous collection of experimental results on this material---no single theory could possibly account for the often contrasting behaviours across different samples. The most salient example of sample-to-sample variations is the nature of the ground state at charge neutrality, which is predicted to be a robust insulating KIVC within the strong coupling perspective~\cite{Bultinck2020hidden,Lian2021TBG4}. While some groups indeed detect a clear gap~\cite{Lu2019orbital,Sharpe2019ferromagnetism,Serlin2020QAH,Stepanov2020untying,Wu2021Chern,Pierce2021unconventional,Polshyn2019linear,Stepanov2021competing}, most observe semimetallic or metallic behaviour~\cite{Cao2018insulator,Cao2018SC,Yankowitz2019pressure,Park2021flavour,Cao2021nematicity,Liu2021tuning,Zondiner2020cascade,Uri2020mapping,Saito2020independent,Das2021symmetry,Saito2021pomeranchuk,Rozen2021pomeranchuk}, a conflict which cannot be fully resolved by considering alignment with the hBN substrate. Similarly, strong-coupling predicts symmetry-broken topological insulators at $|\nu|=1$, but the majority of experiments see at most a small resistivity anomaly on top of otherwise metallic behaviour.

One possible resolution to this discrepancy is that the strong-coupling theory fails to capture the actual low-energy physics of the interacting BM model. This is unlikely due to the broad agreement between perturbative analyses and numerical calculations of the Hamiltonian at integer fillings and $T=0$. Indeed, they match best near charge neutrality, but this is precisely where there are the most serious disagreements with experiment. Another possibility is that the starting Hamiltonian does not accurately reflect the situation relevant to realistic TBG devices, and suitable amendments are required to recover consistency between theory and experiment. It is this path that we follow in the rest of this section.

Before continuing, we remind the reader that our discussion is focused on the low-temperature normal state phase diagram, especially at integer fillings. While the scope may seem rather restrictive, the nature of the commensurate states is expected to provide insights into other regions of the phase diagram. The cascade transitions and Landau fans suggest that the integer orders control the properties of the Fermi surface and its quasiparticles at non-integer doping. Furthermore, the proximity of correlated insulators and superconducting domes hints at their possible relationship. A better handle on the correlated normal states would provide guidelines as to the mechanisms that aid, or perhaps hinder, superconductivity in TBG. It is also useful to consolidate our understanding of the connections between theory, which has much to say about correlated insulators, and experiments, without the complications of thermal fluctuations and other complex phenomena.

\subsubsection{Heterostrain}
The deviation from the ideal BM Hamiltonian that we consider here is uniaxial heterostrain, where the two layers are stretched along a common axis but with opposite signs\footnote{The orthogonal direction is also stretched/compressed according to the Poisson ratio $\nu_\text{P}\simeq 0.16$ \cite{Cao2014poisson}.}. In the setting of van der Waals homobilayers, it is important to distinguish this from homostrain, where strain is applied identically to both layers. Homostrain, to first order, does not lead to a distortion of the moir\'e lattice, and has a substantially smaller impact on the electronic structure~\cite{Huder2018strain}. We refer the reader to Ref.~\cite{zhang2023generalstrain} for discussions of how more general types of strain affect the non-interacting band structure, and henceforth use the term strain to exclusively mean uniaxial heterostrain. We emphasize that these strains are distinct from the moir\'e-periodic lattice relaxations that generally occur within a moir\'e unit cell.

Scanning tunneling microscopy/spectroscopy (STM/STS) probes can directly access the triangular moir\'e superlattice in single-gated samples by imaging the charge density peaks at the AA-stacking regions. By comparing the moir\'e lattice constants along different directions, these studies routinely measure strains of strength $\epsilon=0.1-0.7\%$ \cite{Kerelsky2019maximized,Choi2019correlations,Xie2019spectroscopic}. Evidence of strain is also seen in magnetotransport experiments \cite{Wang2023magnetotransport,finney2022unusual}. While the magnitude of the strain appears small, its ubiquity in TBG devices calls for careful examination of the potential effects on the band structure and correlated states. 

As described in Appendix~A, strain can be incorporated into the non-interacting BM model and enters in two ways~\cite{Bi2019strain}. Firstly, the moir\'e lattice vectors are deformed slightly, and the mBZ becomes a stretched hexagon\footnote{For larger heterostrains, the mBZ can change in more substantial ways~\cite{escudero2024designingstrain}.}. Secondly, the stretching of the graphene bonds leads to an effective vector potential that takes opposite signs in the two layers and valleys\footnote{This to be contrasted to an in-plane magnetic field, which enters as a vector potential that takes opposite signs in the two layers but the same sign in the two valleys, thereby breaking time-reversal symmetry.}. 
Strain preserves $\hat{C}_{2z}$ but breaks $\hat{C}_{3z}$ and $\hat{C}_{2x}$. Hence the Dirac points which are protected by $\hat{C}_{2z}\hat{\mathcal{T}}$ remain intact, but are unpinned from the $K_\text{M}$ and $K'_\text{M}$ points and migrate towards the mBZ center. The Dirac points also separate in energy leading to compensated Fermi pockets at charge neutrality, and the overall bandwidth of the central bands increases considerably. While the vector potential shift for realistic strains is small at the level of the graphene BZ, its scale is effectively magnified by the moir\'e effect. The enhancement of the non-interacting bandwidth is the first hint that strain may challenge the validity of the strong-coupling perspective.

The first theoretical study of correlations in the strained interacting BM model was undertaken in Ref.~\cite{Parker2021strain}, which performed a combined HF and DMRG study at even integer fillings. At charge neutrality and in the absence of strain, the ground state was the KIVC state with a large gap within HF. The KIVC order parameter and charge gap were quickly suppressed in the presence of a finite strain. For sufficiently strong strains that are still well within the experimentally relevant regime, the symmetry-breaking disappeared completely, and the ground state became a symmetric semimetal with compensated Fermi pockets. These findings were invoked to explain the puzzles surrounding transport experiments---since most samples are strained, their transport characteristics are then semimetallic owing to the strain-induced destruction of the strong-coupling KIVC insulator. However, these same calculations observed destruction of the KIVC order and a closing of the charge gap at $|\nu|=2$ for similar threshold strain values, in contradiction with the experimental consensus that the correlated insulator at this filling is the most robust. 

For the strain-based hypothesis to hold, either the relative positions of the phase boundaries at $|\nu|=0$ and $2$ were not faithfully captured, or an alternative gapped symmetry-broken phase emerges at $|\nu|=2$ that is stable up to a much larger value of strain. The latter possibility is not surprising given that departures from the flat-band/QHFM limit are non-negligible, and that the competition between itineracy and localization characteristic of Hubbard physics may be relevant to TBG. 

\subsection{IKS States: Motivation and Overview}
Motivated by the above discussion, our next step is to investigate the ground states of the strained BM model. Unsurprisingly, given the focus of this review,  our tool of choice for doing so will be HF mean-field theory. The rationale behind this choice is as follows. Recall that the strong-coupling picture introduced in the preceding section builds on the fact that in the chiral-flat limit, the {\it exact} ground state is a Slater determinant. Due to the enormous $U(4)\times U(4)$ symmetry of this limit, there are many degenerate choices, but on perturbing away by adding nonzero $w_\textrm{AA}$ and reinstating the single-particle dispersion, specific Slater determinants are selected at each integer filling. Various aspects of the phenomenology of states within the QH ferromagnetic manifold disagree with experiment, suggesting that we need to look further. Rather than moving directly to tackle the full many-body problem, for which techniques are limited, it is reasonable to first investigate a wider range of Slater determinants, i.e. enlarge our considerations to a broader class of mean-field states. In other words, the central premise we adopt is that as we add perturbations such as strain to move away from the strong coupling picture, the ground state leaves the manifold of strong-coupling ferromagnets, but the Hartree-Fock approximation that it is faithfully described by a Slater determinant holds true.

What sorts of states should we consider? An initial guess is to consider orders that break moir\'e translational invariance. Such states are arguably natural from two perspectives. The first is one from the quantum Hall effect:  various perturbations  away from the idealized holomorphic structure of the lowest Landau level can favour spatially inhomogeneous ground states, albeit for partial filling of a {\it single} Landau level. These can often be captured in a Hartree-Fock treatment \cite{Koulakov1996cdw,Fogler1996ground}, which for certain choices of interaction becomes exact in the limit of large Landau level index $n$ \cite{Moessner1996Landau}, in consonance with the experimental observation of stripe and bubble phases in higher Landau levels \cite{Fogler2002bubble}. Given that the chiral-flat limit has a close connection to the lowest Landau level, we might expect that a similar situation might arise on moving away from this limit (although in this case we are working at integer filling, so the analogy is not immediate). The second comes from considering extended Hubbard models: owing to the topological obstructions of the flat bands, any Hubbard-type description naturally has extended interactions, which are known to stabilize stripe-ordered states \cite{Zaanen1989charged,Inui1991Hubbard,Poilblanc1989charged,Giamarchi1990Hubbard,Emery1999stripe,KivelsonEmeryTranquada,RevModPhys.87.457,RevModPhys.75.1201,MACHIDA1989,Machida1990,DevereauxStripes,KEFLiquidCrystal}. Together, these motivate a search for states with broken translational symmetry.

\begin{figure}[t!]
    \centering
    \includegraphics[width=\linewidth]{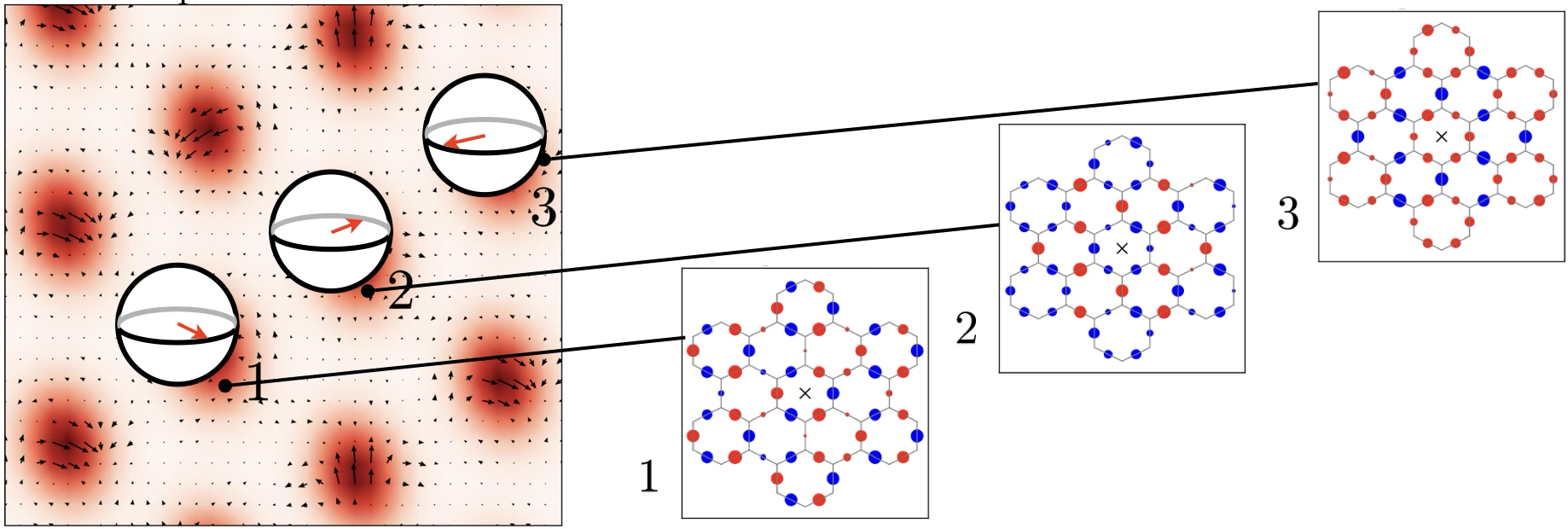}
    \caption{The real-space structure of an IKS with $\bm{q}_{\text{IVC}} = -\bm{G}_1/3$. The main plot shows charge density in color (dark spots correspond to AA regions) and complex IVC order parameter $\sim\braket{\tau_x\sigma_x} + i\braket{\tau_y\sigma_x}$ in arrows. Bloch spheres schematically illustrate the different IVC angles across three AA regions. Insets show the graphene scale charge patterns at three different AA regions. Blue (red) dots correspond to positive (negative) values of $\braket{\hat{c}^\dagger_A\hat{c}^{\phantom{\dagger}}_B} + \braket{\hat{c}^\dagger_B\hat{c}^{\phantom{\dagger}}_A}$, with the cross marking the center of each AA region. Adapted with permission from Kwan \textit{et al.}, Phys. Rev. X. 11, 041063 (2021)~\cite{Kwan2021IKS}. Copyright (2021) American Physical Society.}
    \label{fig:IKS_realspace}
\end{figure}

Ref.~\cite{Kwan2021IKS} performed extensive translational-symmetry breaking Hartree-Fock studies at all non-zero integer fillings, finding that the energetically favoured ground state with modest amounts of strain (with strain magnitude $\epsilon \gtrsim 0.15\%$) is the so-called {\it incommensurate Kekul\'e spiral} (IKS) order. This order breaks valley $U(1)$ symmetry in a subtle manner: it exhibits  coherence between a valley-$K$ electron at momentum $\bm{k}$, and a valley-$K'$ electron at momentum $\bm{k}+\bm{q}_{\text{IVC}}$ --- but, crucially, not to the valley-$K'$ electron at $\bm{k}-\bm{q}_{\text{IVC}}$ or $\bm{k}$. As we will see, this ``spiral'' structure enables to a simplified Hartree-Fock simulation.  (Readers familiar with spatially-modulated fermion-paired states might find it useful to make an analogy with the Fulde-Ferrell description of such states~\cite{fulde1964superconductivity}, rather than the Larkin-Ovchinnikov picture~\cite{larkin1965inhomogeneous}.)
However, the degree of intervalley coherence (IVC) is not spatially uniform across the mBZ: instead, at certain points  the  electronic state is fully polarized in either valley $K$ or valley $K'$, but these points are time-reversal conjugates such that the overall state preserves time-reversal symmetry.  We argue below that this is dictated by a combination of the  topology of the underlying bands and the energetic selection of states. Finally, depending on the detailed spin and  phase structure, an IVC order parameter necessarily involves  density/current order on the graphene scale in either the charge or spin channels. Thus,  a characteristic feature of the IKS state, that originates from its nonuniform IVC across the mBZ, is a moir\'e-scale modulation of a graphene-scale charge order across the sample, which is a sharp signature that can be detected by using scanning probes such as STM~\cite{NuckollsTextures,Kim2023STM}. The expected real-space structure, showing the moir\'e-scale rotation of the graphene-scale charge order, is schematically illustrated in Fig.~\ref{fig:IKS_realspace}.

\subsubsection{Accessing IKS Within Hartree-Fock}
We first summarize how we can access the  IKS state within the Hartree-Fock mean-field treatment. As noted above, while the IKS state breaks moir\'e translations, it does so in a ``spiral'' fashion that does not lead to a moir\'e-scale modulation of the electron density. As such,  rather than working with fully translational-breaking Hartree-Fock, we can instead work in a `boosted' coordinate system where the $\Gamma$-points of valleys $K$ and $K'$ are shifted by an amount $\pm \bm{q}_{\text{IVC}}/2$ with respect to each other, and treat $\bm{q}_{\text{IVC}}$ as a variational parameter\footnote{In a numerical calculation with a discrete momentum grid set by $N_1,N_2$, we typically require that $\bm{q}_\text{IVC}$ also lies on this grid. This is so that we can directly compare calculations performed with different values of $\bm{q}_\text{IVC}$. Another way of saying this is that we are simply selecting different submanifolds of the same full manifold of Slater determinants (that can break any symmetry). Hence by `incommensurate', we really mean that $\bm{q}_\text{IVC}$ is not tied to a specific high-symmetry momentum, and would vary continuously with parameters in the thermodynamic limit.}. In other words, we choose
\begin{equation}
P_{\tau'bs'\bm{k}^\beta, \tau a s \bm{k}^\alpha} \equiv \braket{\hat{c}^\dagger_{\tau a s}(\bm{k}^\alpha)\hat{c}_{\tau'bs'}(\bm{k}^\beta)} \sim \delta_{ss'}\delta_{\bm{k}^\alpha+\tau\bm{q}_{\text{IVC}}/2, \bm{k}^\beta+\tau'\bm{q}_{\text{IVC}}/2}
\end{equation}
where $\tau = \pm 1$ for $K$ and $K'$ valley respectively, and the Kronecker deltas on the RHS enforce spin-collinearity\footnote{The assumption of spin-collinearity will need to be relaxed when considering $SU(2)_K\times SU(2)_{K'}$ perturbations in Section~\ref{subsubsec:spin_structure_IKS}.}, and the constraint that any IVC must occur with spiral wavevector $\bm{q}_\text{IVC}$\footnote{The reader should be careful about the sign convention of $\bm{q}_\text{IVC}$ when comparing with Ref.~\cite{Kwan2021IKS}.}. For convenience and to make contact with our previous notation, we now define a new set of operators $\hat{\tilde{c}}^\dagger_{\tau a s}(\bm{k}) = \hat{c}^\dagger_{\tau a s}(\bm{k}-\tau\bm{q}_{\text{IVC}}/2)$, so that (throughout this subsection, we denote quantities with respect to this new set of momenta labels with a tilde)
\begin{equation}
    \tilde{P}_{\tau'bs'\bm{k}^\beta, \tau a s \bm{k}^\alpha} \equiv  \braket{\hat{\tilde{c}}^\dagger_{\tau a s}(\bm{k}^\alpha)\hat{\tilde{c}}_{\tau'bs'}(\bm{k}^\beta)} \propto \delta_{ss'}\delta_{\bm{k}^\alpha, \bm{k}^\beta}. 
\end{equation}
This block-diagonal structure allows us to write the one-particle density matrix exactly as before, but with $\tilde{c}$-operators replacing $c$-operators:
\begin{equation}
    \tilde{P}_{\tau'b,\tau a}(\bm{k},s) \equiv \braket{\hat{\tilde{c}}^\dagger_{\tau a s}(\bm{k})\hat{\tilde{c}}_{\tau'bs}(\bm{k})}.
\end{equation}
In terms of the new set of operators, the normal-ordered Coulomb interaction is given by (compare with Eq.~\ref{eq:Hint_n.o.})
\begin{align}
\begin{split}\label{eq:HCoul_prime}
    \hat{H}_{\text{int,\,n.o.}}
    &=\frac{1}{2A}\sum_{\mystackrel{ss'\tau\tau'}{abcd}}\sum_{\bm{q}}\sum_{\bm{k}^\alpha,\bm{k}^\beta}V(\bm{q})\tilde{\lambda}_{\tau,a,b}(\bm{k}^\alpha,\bm{q})\tilde{\lambda}^*_{\tau',d,c}(\bm{k}^\beta,\bm{q})\\
    &\quad\quad\quad\times \hat{\tilde{c}}^\dagger_{\tau as}(\bm{k}^\alpha)\hat{\tilde{c}}^\dagger_{\tau' cs'}(\bm{k}^\beta+\bm{q})\hat{\tilde{c}}_{\tau' ds'}(\bm{k}^\beta)\hat{\tilde{c}}_{\tau bs}(\bm{k}^\alpha+\bm{q}),
\end{split}
\end{align}
where $\tilde{\lambda}_{\tau,a,b}(\bm{k},\bm{q}) = \lambda_{\tau,a,b}(\bm{k} - \tau\bm{q}_{\text{IVC}}/2,\bm{q})$. It is then again easy to to show that the Hartree-Fock Hamiltonian  in terms of the $\tilde{c}$ operators is again diagonal in $\bm{k}$ and $s$, i.e.
\begin{equation}
    \tilde{H}^{\text{HF}}_{\tau a s \bm{k}^\alpha, \tau'bs'\bm{k}^\beta} = \delta_{ss'}\delta_{\bm{k}^\alpha,\bm{k}^\beta}\tilde{H}^{\text{HF}}_{\tau a, \tau' s}(\bm{k}^\alpha, s)
\end{equation}
for some $\tilde{H}^{\text{HF}}_{\tau a, \tau' s}(\bm{k}^\alpha, s)$, and with Hartree and Fock contributions  given by (compare with Eq.~\ref{eq:Hartree_and_Fock_term})
\begin{align}
\tilde{H}^\text{H}_{\tau a, \tau' b}(\bm{k}, s) &= 
\frac{\delta_{\tau \tau'}}{A}\sum_{s'\tau''cd}\sum_{\bm{G}}\sum_{\bm{k}'\in\text{ mBZ}}V(\bm{G})\tilde{\lambda}_{\tau,ab}(\bm{k},\bm{G})\tilde{\lambda}^*_{\tau'',dc}(\bm{k}',\bm{G})\tilde{P}_{\tau''d,\tau'' c}(\bm{k}',s') \nonumber \\
    \tilde{H}^{\text{F}}_{\tau a , \tau' b}(\bm{k},s)
    &=-\frac{1}{A}\sum_{cd}\sum_{\bm{q}\in\text{all}}V(\bm{q})\tilde{\lambda}_{\tau,ad}(\bm{k},\bm{q})\tilde{\lambda}^*_{\tau',bc}(\bm{k},\bm{q})\tilde{P}_{\tau d, \tau' c}(\bm{k}+\bm{q},s).
\end{align}
The mean-field Hamiltonian is obtained as the `tilded' version of the standard translationally-invariant HF Hamiltonian presented in Eq.~\ref{eq:HHF_T_H_F}. Hence, the HF machinery introduced in Section~\ref{subsec:HF_TBG} can be straightforwardly repurposed here, if we `boost' the $\bm{k}$ argument of the form factors and the one-body terms (which can either come from the kinetic energy or the interaction scheme). In the periodic gauge that we use here, this operation is straightforward, and can be separately implemented before the HF self-consistent procedure\footnote{See the companion code for an example implementation.}.

\subsubsection{Wavefunctions in the Boosted and Unboosted Frames}
The assumption of a single-$\bm{q}$ spiral order allowed us to go to a boosted frame in which the HF projector is diagonal in momentum space, viz. $\tilde{P}_{{\bm{k}}, {\bm{k}}'} \propto \delta_{{\bm{k}}, {\bm{k}}'}$. This compact notation hides a rather complicated structure when translated back to a projector $P$ in the unboosted frame: intravalley matrix elements of $P$ remain momentum-space diagonal, but intervalley terms are off-diagonal, linking valley-$K$ states at momentum $\bm{k}$ with valley-$K'$ states at momentum $\bm{k}+\bm{q}_{\rm IVC}$. When discussing the structure of the IKS state below, we will exclusively work in the boosted frame; one can obtain the projectors in the unboosted frame via
\begin{equation}\label{eq:PtotildeP}
    P_{\bm{k},\bm{k'}} = e^{\frac{\tau_z}{2}\bm{q}_{\rm IVC}\cdot\overrightarrow\nabla_{\bm{k}}}\tilde{P}_{{\bm{k}}, {\bm{k}}'}e^{\frac{\tau_z}{2}\bm{q}_{\text{IVC}}\cdot\overleftarrow\nabla_{\bm{k}'}},
\end{equation}
where $\bm{q}_{\rm IVC}$ is the IVC boost vector, and $\overleftarrow\nabla_{\bm{k}'}$ denotes gradient against $\bm{k}'$ acting on expression to the left. Note that although $\tilde{P}_{\bm{k}, \bm{k}'}$ is diagonal in $\bm{k}$-space, the $\bm{k}$-space boost encoded by $\overrightarrow\nabla_{\bm{k}}$ and $\overleftarrow\nabla_{\bm{k'}}$ renders $P_{\bm{k},\bm{k'}}$ off-diagonal.

\subsubsection{Initialization and Optimization}\label{subsubsec:IKS_initialization}

Since the one-particle density matrix in the boosted frame is again diagonal in $\bm{k}$ and $s$, we can use the same discussion as in Section~\ref{subsubsec:TBG_initialization} when constructing the initial density matrix. Furthermore, to optimize over possible IVC spirals, we treat $\bm{q}_{\text{IVC}}$ as a variational parameter. We perform a set of HF calculations (with multiple types of initial states if necessary) for each value of $\bm{q}_{\text{IVC}}$, and choose the value of $\bm{q}_{\text{IVC}}$ with the lowest energy if we are interested in the ground state. If  $\bm{q}_{\text{IVC}}$ is left completely unconstrained, this  adds another $O(N_k)$ factor to the time complexity of our computation; combining this with the $O(N_k)$ scaling of the translationally invariant HF, we see that the overall complexity is $O(N_k^3)$, which is the same as completely unrestricted HF. However, in practice, there are still significant advantages in using restricted Hartree-Fock. For example, the scan over all possible choices of $\bm{q}_{\textrm{IVC}}$ is easily parallelizable, and restricted Hartree-Fock usually requires fewer iterations to converge and fewer seeds to reach the ground state due to the restricted variational space. Furthermore, there is often a good guess for $\bm{q}_{\text{IVC}}$ that can be used to significant restrict the candidate values of $\bm{q}_\text{IVC}$~\cite{Kwan2021IKS}. 

\subsection{Properties of the IKS State}
On adopting the modifications to the Hartree-Fock procedure, we find IKS order at each nonzero integer filling of the central bands of realistic TBG (i.e., for experimentally relevant choices of the interaction parameters and the chiral ratio $w_\textrm{AA}/w_\textrm{AB}$) with modest amounts of strain~\cite{Kwan2021IKS}. In this subsection, we summarize various basic aspects of these IKS states, beginning with its structure, which we rationalize in terms of both topology and energetics. We will focus on  integer fillings, but will mention the metallic states at non-integer filling when discussing the phenomenology. Some further caveats are in order. We will invoke particle-hole symmetry, which is only an approximate symmetry of the BM model and its interacting descendants and is notably absent in experiments. However, modelling the experimentally-observed extent of particle-hole symmetry breaking likely requires inclusion of bands outside the central octet and additional terms in the continuum model, and would take us somewhat beyond the scope of this review.  We also note that the integer-filling numerics show that the IKS states at $|\nu|=1$ are either gapless or have a very small gap, and as noted strained TBG is a semimetal at neutrality. Nevertheless, for completeness we also provide an account of the $|\nu|=1$ IKS state.  
Finally, we also observe that if we neglect the electron-phonon coupling as well as the intervalley interaction terms generated by the short-range part of the interaction, the symmetry group of the model contains a $SU(2)_K\times SU(2)_{K'}$ subgroup corresponding to independent spin rotations in each valley. The manifold of IKS states inherits this symmetry, so we can always choose an appropriate spin orientation in each valley to make their spin structure especially simple, even with intervalley coherence. However, including these terms reduces the symmetry to that of the physical spin rotations; we briefly discuss how the experimental observability of the IKS state in STM allows us to deduce aspects of the resulting spin structure.

\subsubsection{IKS Wavefunctions}

\begin{figure}[t!]
    \centering
    \includegraphics[width=\linewidth]{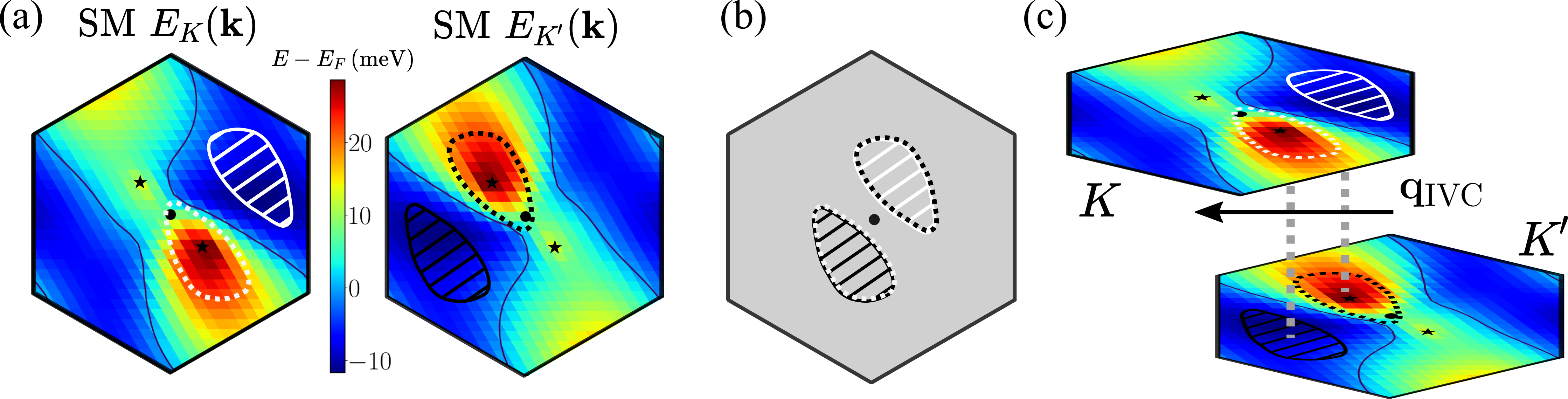}
    \caption{(a) Dispersions of the lower band for the symmetry-preserving self-consistent Hartree-Fock at $\nu = -2$. System size is $24 \times 24$, strain is 0.2\%, and $\bm{q}_{\rm IVC} = -\bm{G}_1/3$. Low- (high-) energy regions are marked with hatched (dotted) lobes. Fermi surfaces are indicates by black lines, and Dirac point locations are marked by black stars. (b),(c) Schematic illustration of the construction of the IKS states. The appropriate $\bm{q}_{\text{IVC}}$ should align the high-energy region in valley $K$ with the low energy region in valley $K'$, and vice versa. Adapted with permission from Kwan \textit{et al.}, Phys. Rev. X. 11, 041063 (2021)~\cite{Kwan2021IKS}. Copyright (2021) American Physical Society.}
    \label{fig:IKS_lobes}
\end{figure}

The ``primitive'' IKS state is that at $|\nu|=3$, since this corresponds to a filling of a single electron or hole band of the central octet. The Hartree-Fock projector of the $\nu=-3$ IKS state is parameterized in the boosted frame in the Chern basis (see Section~\ref{subsec:Chern_basis}) as 
\begin{equation}\label{eq:P-3boosted}
    \tilde{P}_{-3}({\bm{k}},{\bm{k}}
')=  \frac{1}{8} (1 + s_z) (1+ \bm{n}^\perp_{\bm{k}}\cdot\bm{\gamma}) (1+ \bm{m}_{\bm{k}}\cdot\bm{\eta})\delta_{{\bm{k}},{\bm{k}'}}
\end{equation}
where the inter- and intra-Chern Pauli triplets $\bm{\gamma},\bm{\eta}$ were introduced previously in \eqref{eq:Chern_triplet}.  Here, $\bm{n}, \bm{m}$ are unit three-vectors, with $\perp$ denoting components in the $xy$-plane, and $n^z=0$.  Several comments are in order. 
\begin{enumerate}
    \item It is easy to see that  $\tr  \tilde{P}_{-3}({\bm{k},\bm{k}})  = 1 = (\nu+4)$ as expected; the projector at $\nu=3$ is given by particle-hole conjugation.
    \item Note that we have partially fixed the gauge of the Chern basis by requiring $\hat C_{2z}$ to act as $\sigma_x\tau_x$ and $\mathcal{T}$ as $\tau_x \mathcal{K}$; this means that $\bm{n}^{\perp}_{\bm{k}}$ and the in-plane components of $\bm{m}_{\bm{k}}$ are formally gauge-variant. 
   \item Many key properties of the IKS state are encoded in the fact that $\bm{m}_{\bm{k}}$ has a nontrivial texture: across most of the mBZ, it lies in the $xy$-plane with a constant angle that can be changed by a global $U(1)_v$ rotation; over these regions of the mBZ, there is non-zero IVC. However, for  reasons discussed below, the Euler topology enforced by the unbroken $\hat{C}_{2z}\hat{\mathcal{T}}$ symmetry\footnote{Since valley-$U(1)$ rotation does not commute with $\hat{C}_{2z}$, what we really mean is that there exists a valley $U(1)$ rotation such that the IKS satisfies $\hat{C}_{2z}\hat{\mathcal{T}}$.} forbids a uniform IVC across the mBZ. Thus, there are two regions or ``lobes'' where $\bm{m}_{\bm{k}}$ orients towards the poles, reflecting valley polarization in these momentum regions. In contrast to $\bm{m}_{\bm{k}}$, the $\bm{n}^\perp_{\bm{k}}$ always lies in the $xy$-plane owing to $\hat{C}_2\hat{\mathcal{T}}$ symmetry.  
   \item While the fact that $\bm{m}_{\bm{k}}$ is textured for any fully gapped state is enforced by topology, choosing the optimum texture is set by energetics, and gives us a reasonable estimate for $\bm{q}_{\rm IVC}$, as follows. First, consider the dispersion of the Hartree-Fock renormalized dispersion of TBG, in the absence of any IVC or other spontaneous symmetry breaking, but including the explicit $\hat{C}_{3z}$-breaking due to strain. Once $\hat{C}_{3z}$ is broken, the dispersion in valley $K$ will generically have some energy maximum and energy minimum, at momenta $(\bm{q}_{\rm max}, \bm{q}_{\rm min})$; valley $K'$ has the time-reversed dispersion, with maximum and minimum at $(-\bm{q}_{\rm max}, -\bm{q}_{\rm min})$. Now, assuming that IVC is generically favored over valley polarization, if the system is forced (because of the nontrivial texture) to valley polarize at two points in the mBZ, it would be energetically optimal for $\bm{m}_{\bm{k}}$ to lie in valley $K$ for  momenta $\bm{k} \sim \bm{q}_{\rm min}$ where its dispersion has a minimum. A further optimization is achieved by introducing a relative boost $\bm{q}_\text{IVC}$ in the valleys such that this coincides with the {\it maximum} in valley $K'$ at $-\bm{q}_{\rm max}$, so that the loss of IVC occurs at a place where any rotation towards $K'$ would cost the most energy. Thanks to time reversal, exactly similar conditions arise near $-\bm{q}_{\rm min}$ with the roles of $K$ and $K'$ swapped. The necessary boost is given by setting $\bm{q}_{\rm IVC} = -\bm{q}_{\rm min} -\bm{q}_{\rm max}$. This heuristic ``lobe principle'', illustrated in Fig.~\ref{fig:IKS_lobes},  provides a rather good quantitative estimate of the optimal $\bm{q}_{\rm IVC}$ and using this as an initial guess and exploring in its vicinity can significantly speed up the HF numerics in some cases.
   \item As mentioned in the motivating discussion, the structure of the IVC projector has direct experimental consequences. Intervalley coherence involves a $\sqrt{3}\times \sqrt{3}$ tripling of the graphene-scale unit cell; owing to the texturing and the boost, this is non-uniform and modulated on the moir\'e scale at an incommensurate wavevector $\bm{q}_{\rm IVC}$. One can show in all cases, the $\sqrt{3}\times \sqrt{3}$ ``Kekul\'e'' order is a density-wave rather than a current-loop order on the graphene scale\footnote{This is to be contrasted with the KIVC state which has a  Kekul\'e current order, which for symmetry reasons does not have an associated charge or spin density and is hence invisible to STM~\cite{Calugaru2021KD,Hong2021KD}.}    but its precise decomposition  into spin and charge sectors depends on the relative alignment of the spins in the two valleys, as we discuss below. The corresponding graphene-scale Kekul\'e patterns can be detected by spin-unpolarized or spin-polarized STM, respectively, and by phase coherently comparing STM measurements in nearby moir\'e unit cells one can also detect the incommensurate modulation that is the hallmark of IKS. Pioneering experiments  have detected Kekul\'e charge order near $\nu=2$ in both MA-TBG \cite{NuckollsTextures} and its close cousin, twisted symmetric trilayer graphene \cite{Kim2023STM} with a $\bm{q}_{\rm IVC}$ consistent with both detailed HF energetics and the ``lobe principle''.
\end{enumerate}

We now briefly discuss the structure of the IKS states at $|\nu|=1,2$. At $\nu=-2$, we can construct a `spin singlet' IKS state by simply creating $\nu=-3$ IKS states for each spin species separately. Specifically, we have
\begin{align}\label{eq:P-2boosted}
    \tilde{P}_{-2}({\bm{k}},{\bm{k}}
')&= \frac{1}{4} (1+ \bm{n}^\perp_{\bm{k}}\cdot\bm{\gamma}) (1+ \bm{m}_{\bm{k}}\cdot\bm{\eta})\delta_{{\bm{k}},{\bm{k}'}}.
\end{align}
The $\nu=-1$ state is somewhat different in spirit. At this filling, we need to fill $3$ of the 8 central bands. We can construct the state as follows. In one spin species, we preserve all symmetries of the strained BM model and fill $2$ of the $4$ bands.  As noted previously, this leads to a nematic\footnote{`Nematic' is somewhat of a misnomer here since $\hat{C}_{3z}$ is broken explicitly. However, a similar semimetal has been proposed to be energetically competitive (but still above the strong-coupling states) at zero strain~\cite{Liu2021nematic}, in which case it is actually nematic.} semimetal. In the other spin sector, we simply create a primitive IKS state, filling one of the four bands. The combination of these states is then a gapless state that has nontrivial IKS order. As before, the states at $\nu=1,2$ can be obtained by particle-hole conjugating those at $\nu=-1,-2$.

\subsubsection{``Topological Frustration''}
As we have mentioned above, the IVC in the IKS is necessarily modulated in momentum space as a consequence of the Euler topology protected by the $\hat{C}_{2z}\hat{\mathcal{T}}$ symmetry of the single-valley band structure. Since this is a particularly subtle variant of band topology, we reason here by analogy with the simpler example of IVC between two bands with distinct Chern numbers $\pm C$ --- as emerges, for instance, in moir\'e graphene systems without $\hat{C}_{2z}$ symmetry, or TMD homobilayers~\cite{WangCTI}. 

As a model system, consider a pair of spinless weakly dispersive bands, one each in valley $\tau=\pm 1$. These bands have  Chern numbers $C\tau$ (where $C$ is an integer), and are interchanged by time-reversal symmetry. We wish to construct a mean-field insulating state at filling $\nu=1$. Evidently, one  option is to spontaneously break time reversal and valley polarize, resulting a state with a net Chern number. Now, imagine tuning the bandwidth or the competition between inter- and intravalley- interactions to destabilize this in favor of a state with non-zero IVC. Such a state would be characterized by a nonzero value of the ``excitonic'' order parameter\footnote{For convenience we discuss a $\bm{k}$-diagonal, unboosted state, but a similar argument holds for a boosted IVC state similar to the IKS.} $\Delta_{\bm{k}} \equiv \langle c^\dagger_{\bm{k},+} c^{\phantom\dagger}_{\bm{k},-}\rangle$. However, if we work in a smooth (but non-periodic) gauge, it is possible to show that the phase of this  order parameter has a net winding of $4\pi C$ in the Brillouin zone. As such, it must therefore vanish at some subset of ``IVC nodes'' around which the phase of $\Delta_{\bm {k}}$ winds by an integer multiple of $2\pi$, such that the net winding is $4\pi C$: the IVC is {\it topologically frustrated} from being uniform across the moir\'e BZ~\cite{Bultinck2020mechanism,KwanCTI}. 

Naively, it seems that this precludes a fully-gapped state~\cite{Bultinck2020mechanism}. However, there is one way out: the order parameter ``escapes into the third dimension'', by valley polarizing in the IVC nodes. By placing the nodes at time-reversal conjugate momenta and choosing to polarize in opposite valleys at each node, one arrives at a fully gapped time-reversal preserving insulator, dubbed the ``Chern texture insulator''. Evidently, the resulting state has a texture in the moir\'e Brillouin zone as can be captured by the (non-singular) ``bimeron'' configuration of the 3-component vector $\bm{n} = \frac{1}{2} \hat{c}^\dagger_{s}\bm{\tau}_{ss'} \hat{c}^{\phantom\dagger}_{s'} = ({\rm Re}\Delta_{\bm{k}},{\rm Im}\Delta_{\bm{k}},\frac{1}{2}(\hat{c}^\dagger_{\bm{k},+} \hat{c}^{\phantom\dagger}_{\bm{k},+} -\hat{c}^\dagger_{\bm{k},-} \hat{c}^{\phantom\dagger}_{ \bm{k},-}))$. When combined with a nonzero ``boost'' between the valleys, the state closely resembles the IKS, except that the underlying bands have Chern rather than Euler topology and it emerges from filling one out of two rather than four bands (ignoring spin). 

Generalizing these considerations to the MA-TBG setting is more subtle, since the constraint imposed on the IVC order parameter by the Euler topology does not lend itself to as simple an expression as the ``winding'' rule for the Chern case. However, one can show that the structure of the Bloch states continues to enforce a qualitatively similar ``Euler texture'' structure on the IVC, leading to the characteristic properties of the IKS state once one combines this with the lobe principle for energetics.

These and other topological aspects of mean-field IVC states in systems with nontrivial Euler or Chern topology were explored extensively in Ref.~\cite{KwanCTI}, which coined the collective term ``textured exciton insulators'' to describe such states.  An important aspects of the TRS-preserving states is that --- within the Hilbert space of the bands involved --- their valley order cannot be captured within a local moment description of the valley degree of freedom. The reader is referred to Ref.~\cite{KwanCTI} for a detailed discussion of this and other aspects of the topology of the IKS states and its close cousins in bands with Chern topology.

\subsubsection{Breaking $SU(2)_K\times SU(2)_{K'}$: Spin Structure and Observability of IKS}\label{subsubsec:spin_structure_IKS}
Above, we have assumed that $SU(2)_K\times SU(2)_{K'}$ spin symmetry is not broken explicitly, allowing us to  independently pick the spins in each valley. In other words, we `factored out' the spin part and discussed the different fillings by considering the physics in each spin sector separately. For example, in writing \eqref{eq:PtotildeP}, we simply chose a single spin sector (up) and then built in intervalley coherence within this sector. One can view the parts of the projector after the $(1+s_z)$ as defining a `spinless IKS' state in the four-band model of TBG for a single spin species; picking the spin to occupy then defines a `spin polarized' version of the $\nu=-3$ primitive IKS. This also allowed us to build the IKS states at other fillings in terms of stacking the ``primitive'' IKS. Realistic TBG explicitly breaks the $SU(2)_K\times SU(2)_{K'}$ symmetry. We can explore the states that result by  parametrizing the manifold of states obtained from the special choices listed above by performing $SU(2)_K\times SU(2)_{K'}$ rotations, and then computing the variational energy of these trial states in the presence of various terms that break the symmetry down to the physical $SU(2)_s$~\cite{wang_putting_2025}. This approach relies on the fact that the formation of the IKS at finite wavevector occurs on a much higher energy scale than the splitting of the  $SU(2)_K\times SU(2)_{K'}$ manifold by weaker terms that originate in either the electron-phonon interaction or subleading short-distance contributions from the Coulomb interaction. These effects are sometimes packaged together in terms of an effective ``intervalley Hund's coupling'', but it is important to recognize that their energetics receives both Hartree and Fock contributions in generic HF states with IVC.
The role of these terms is different at $\nu=-2, -3$, so we discuss them separately (the  generalization to the remaining fillings follow by stacking and/or particle-hole transformation.)

At $\nu=-3$,  the states obtained by acting with $SU(2)_K\times SU(2)_{K'}\times U(1)_v$ rotations on the parent state in \eqref{eq:P-3boosted} can be parameterized in terms of the global $U(1)_v$ angle, as well as the individual spin orientations in each valley, $\hat{\bm{n}}_{K}$ and $\hat{\bm{n}}_{K'}$. The  $SU(2)_K\times SU(2)_{K'}$ breaking terms then fix the relative angle between $\hat{\bm{n}}_{K}$ and $\hat{\bm{n}}_{K'}$\footnote{These are not to be confused with $\bm{n}$ in Eq.~\ref{eq:P-2boosted}.}. If these are antiferromagnetically aligned, we obtain a so-called ``spin-valley locked'' IKS state, which has IVC between opposite spin sectors, and as consequence exhibits a Kekul\'e spin texture rather than a Kekul\'e charge density.

At $\nu=-2$, the manifold generated by the $SU(2)_K\times SU(2)_{K'}\times U(1)_v$ action on the parent state in  \eqref{eq:P-2boosted} can be parametrized in terms of the global $U(1)_v$ phase, the {\it relative} IVC phase $2\gamma$ between the two spin sectors, as well as a unit vector $\hat{\bm{n}}$ that parametrizes the axis of spin quantization (note that $\hat{\bm{n}}$ is redundant for the specific choice $\gamma=0$ corresponding to a spin singlet). Physically, this has a natural interpretation in terms of ``stacking'' two $\nu=-3$ states to construct $\nu=-2$: we have one copy of IKS each in the sectors with spin parallel and anti-parallel to a chosen axis $\hat{\bm{n}}$, with a relative $U(1)_v$ phase difference of $2\gamma$ between the two, with the remaining freedom being the global IVC phase. In this case, it is possible to show that  $SU(2)_K\times SU(2)_{K'}\times U(1)_v$ symmetry eliminates any energetic dependence on $2\gamma$, whereas restoring the physical $SU(2)_s\times U(1)_v$ symmetry pins $\gamma$ to $0$ or $\pi$. The former ``ferro-IVC'' state corresponds to a Kekul\'e charge density, whereas the latter ``antiferro-IVC'' state has a Kekul\'e spin density.

At $\nu = -2$, electron-phonon interaction favors maximal Kekul\'e charge density pattern, i.e.~relative IVC angle of 0, while short-range Coulomb contribution favors vanishing Kekul\'e charge density pattern, i.e.~relative IVC angle of $\pi$. At $\nu = -3$, broadly speaking, electron-phonon coupling and short-range Coulomb contribution favor antiferromagnetic and ferromagnetic alignments respectively, though this can in principle depend on the competition between various terms in the Hartree-Fock energy functional. The observation of a charge Kekul\'e pattern for the IKS near $\nu=-2$~\cite{NuckollsTextures,Kim2023STM} indicates that electron-phonon coupling dominates, which leads us to expect the IKS at $\nu = -3$ to have antiferromagnetic alignment, with the additional caveat that the parameter regime for phonons to dominate has some weak dependence on the filling factor. 

\subsection{IKS: Phenomenology and Fermiology}
We close our discussion of the IKS state with a broad-brushstrokes summary of its phenomenology and that of the proximate metallic states accessed by doping the correlated insulators.

We have already mentioned that the spatially-modulated intervalley coherence of the IKS state has a direct signature in STM. This has already been observed in recent experiments \cite{NuckollsTextures,Kim2023STM}, and the fact that they involved spin-unpolarized measurements and hence interrogated the charge density rather than the spin density also gives insight into the spin structure of the state and hence the relative magnitude of the electron-phonon and short-range Coulomb interactions. The STM  measurements also allow us to track how $\bm{q}_\textrm{IVC}$  evolves on tuning various parameters, e.g.~on doping away from integer filling. As underscored by the ``lobe principle'', $\bm{q}_\textrm{IVC}$ is set by a delicate energetic competition between various factors that shape the renormalized symmetric dispersion out of which the IKS state evolves; as such, the robustness of the IKS state can be viewed as a consequence of its ability to respond to various changes by adjusting $\bm{q}_\textrm{IVC}$ while remaining in the same phase. Furthermore, the evolution of $\bm{q}_\textrm{IVC}$ with doping is nontrivial, and may provide one route to explaining rotation of a nematic superconducting order parameter with doping reported in Ref.~\cite{Cao2021nematicity} \footnote{Observe that even though IKS only emerges in the presence of explicit $\hat{C}_{3z}$-breaking by strain, $\bm{q}_\textrm{IVC}$ does not directly track the strain and even relatively modest strains produce a substantial $\bm{q}_{\text{IVC}}$, so it has some features of a true nematic even though it is not one in a strict sense.}.

Turning to thermodynamics, we observe that the compressibility computed within Hartree-Fock for strained TBG in the regime where IKS order is prevalent is broadly compatible with the reported ``cascade'' features seen in experiments \cite{Wong2020cascade,Zondiner2020cascade}. That said, it is by now reasonably widely believed that the cascades  primarily give evidence for the formation of local moments, rather than serve as a sensitive test of the precise nature of the low-temperature state \cite{Datta2023heavy}. Evidence in favour of this view comes from both experiment --- which see cascades at a far higher scale than the onset of correlated insulating behaviour --- and from theoretical modelling in the heavy fermion picture, where this separation of scales is especially clear. Finally, a detailed analysis of the spin response of the IKS states provides a possible explanation of why  the  ``isospin Pomeranchuk effect'' --- wherein fluctuations stabilize an ordered state on {\it increasing} temperature from a low-temperature disordered state --- is quenched by an in-plane magnetic field \cite{Saito2021pomeranchuk,Rozen2021pomeranchuk}.

A complete description of transport  is much more challenging, and remains in many ways an open problem. We remark that at least one key {\it qualitative} feature, namely the sequence of correlated gaps, and their absence in favour of a semimetallic state at charge neutrality, appears to be reasonably well-captured by the mean-field studies of strained TBG, in the regime where IKS order is present at integer filling. The situation in  the (most likely strongly correlated) normal state away from the correlated insulators and above the superconducting transition remains murky, with a notable absence  of  a fully plausible model for the reported linear-$T$ resistivity~\cite{Cao2020strange,Polshyn2019linear,Jaoui2021critical}. However, one feature of these normal states, namely the magnetotransport, {\it is} captured quite well by the mean-field modelling, as we now summarize.

Recall that  magnetotransport in a 2D system is often most easily understood by tracking ``Landau fans'': lines in the density-magnetic field plane corresponding to the densities $n(B)$ at which an integer number of Landau levels are fully filled for magnetic field $B$. Under the assumption that the parent state has zero Chern number, and that the  effective carrier density $n$ (and hence the Fermi surface volume) is not altered by the  field~\footnote{And hence also requires that the spin-splitting of the bands in a magnetic field is negligible.}, a standard semiclassical calculation shows that these densities satisfy
\begin{eqnarray}
    n(B) =  \frac{g eB}{h}N
\end{eqnarray}
where $g$ is the effective degeneracy, i.e.~the number of flavours, and $N$ is an integer. The traces for different values of $N$ form a fan that emanates from some insulating or semimetallic filling (for TBG, usually an integer), with the interpretation that the oscillations come from the Fermi surfaces that emerge on doping the corresponding, likely reconstructed, bandstructure. If we then measure $n$ with respect to this insulating filling,  $g$ provides a measure of the degeneracy of the corresponding Fermi surfaces\footnote{Note that this implicitly assumes that each of the degenerate Fermi surfaces has the same area, or equivalently that the quantum oscillations have a single frequency.}. 

\begin{figure}[t!]
    \centering
    \includegraphics[width=\linewidth]{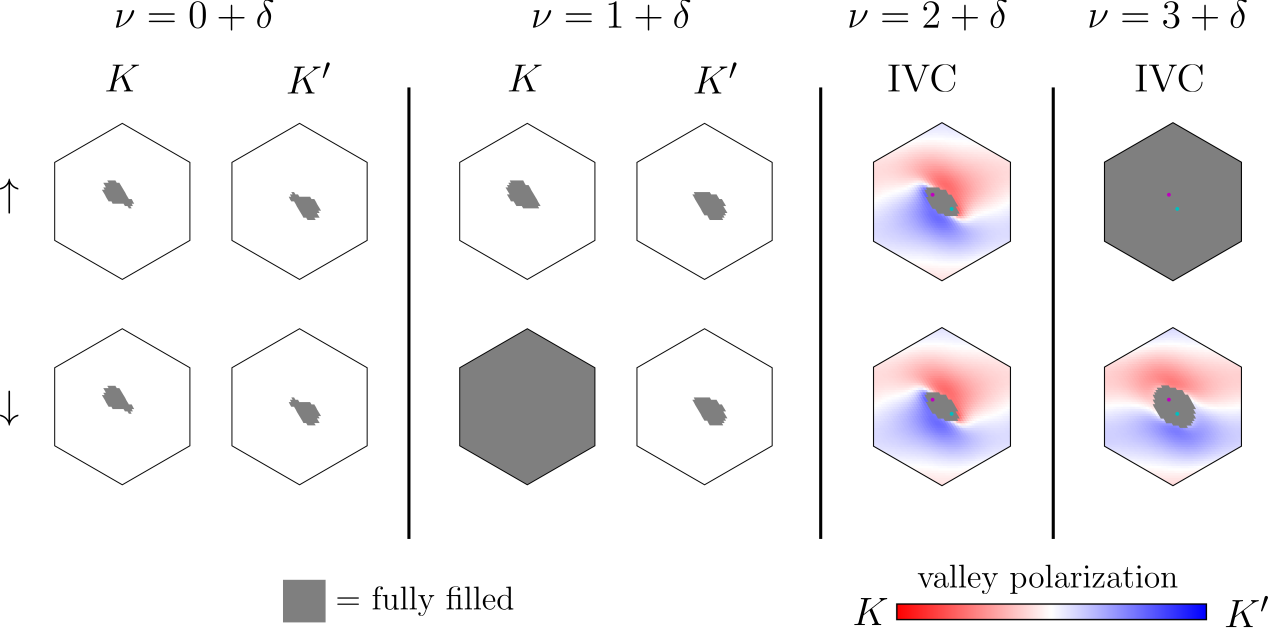}
    \caption{Fermi surfaces at filling factors $\nu = \mathrm{integer} + \delta$, obtained with $10 \times 10$ HF interpolated to $50 \times 50$, with $\nu = 0.12, 1.12, 2.08, 3.08$ and $\epsilon=0.25\%$ strain along $\hat{x}$-direction. For $\nu = 0 + \delta, 1 + \delta$, the ground states preserve $U(1)_v$ symmetry for these parameters, and Fermi surfaces are shown for each spin-valley sector. For $\nu = 2 + \delta, 3 + \delta$, ground states spontaneously break $U(1)_v$ symmetry. The Fermi surfaces are shown for each spin sector, and cyan and magenta dots indicate the $\Gamma_M$-points of valley $K$ and $K'$ respectively. Gray areas denote fully occupied sectors. For IVC states, color plots show the valley polarization of occupied orbitals. For $\nu = 0+\delta, 2 + \delta, 3 + \delta$, we obtain Fermi surface degeneracies $4, 2, 1$, consistent with experiments. We find the $\nu = 1$ state to be metallic. As such, we do not expect Landau fans to be visible for $\nu = 1 + \delta$. The calculations are done with with $\theta = 1.05^\circ$, $w_\textrm{AA} = 80$~meV, $w_\textrm{AB} = 110$~meV, and $\epsilon_r = 10$.}
    \label{fig:fermi_surfaces}
\end{figure}

Two features of the Landau fan data in TBG are particularly noteworthy. First, they only appear when doping a correlated state {\it away} from charge neutrality; second, the states at $|\nu|=0, 2,\,{\rm and}\,3$ respectively have fan degeneracies of $4, 2,\,{\rm and}\,1$ when the corresponding states are insulating or semimetallic in a given sample (there is usually no clear fan at $|\nu|=1$ as with rare exceptions there is never a strong insulator here). Both these features naturally emerge in the mean-field modelling of the IKS state~\cite{Wagner2022global}. First, we find that the dispersion of the doped electrons is significantly flatter (more massive) when doping towards charge neutrality than when doping away from it. While the physical origins of this feature are probably better understood from the heavy-fermion perspective of TBG, it is  nevertheless present in the Hartree-Fock mean-field results on the continuum model. Second, studying the doping of the correlated IKS states at  $|\nu|=0, 2,\,{\rm and}\,3$ indeed reproduces the fan degeneracy, as illustrated in Fig.~\ref{fig:fermi_surfaces}. This is in contrast to the strong-coupling states once the Hartree renormalization of the dispersion is included, and adds additional credence to the IKS picture.

Of course, the most striking experimental finding in TBG is that of gate-tunable superconductivity, particularly on doping the correlated insulator near $\nu=2$. As such, mean field modelling of the normal state cannot provide direct insight into this problem, but can yield important clues. Most notably, a mean-field study of the nature of the doped Fermi surfaces both for realistic TBG and its close cousin, symmetric trilayer graphene (which has an added tunability, via displacement field) gives us some insight into possible pair order parameters, assuming that the normal state from which superconductivity emerges is the doped IKS state~\cite{wang_putting_2025}. Some of the features of the fermiology are also {\it prima facie} consistent with experiments  that track a ``double dome'' feature in the superconducting state~\cite{Zhou2025},

Finally, experiments are beginning to probe properties of the collective modes of TBG~\cite{hesp2021observation,Morissette2023,xie2024long} and other moir\'e materials. Here, time-dependent HF can offer at least a preliminary model for the modes that arise as a consequence of broken symmetry in the correlated states. We defer a discussion of these points to our next case study.

\section{Case Study II: Collective Modes}
\label{sec:collective}
\subsection{Background}\label{subsec:collective_bg}
Goldstone's theorem \cite{Goldstone1961field,Goldstone1962broken} states that the spontaneous breaking of a continuous symmetry gives rise to gapless collective modes. With the complex array of symmetry-broken phases in MA-TBG, as discussed above, we expect a similarly rich spectrum of Goldstone modes. In order to organize these modes, we must consider two key aspects of the problem.

The first of these is {\it mode counting}. For Lorentz-invariant systems, every broken symmetry gives rise to exactly one linear gapless Goldstone mode. However, for condensed matter systems, it is well known that this is not the case. This is readily illustrated by the Heisenberg ferromagnet, where the two spin rotations along axes orthogonal to the ferromagnetic axis are broken, but there is only a single quadratic mode. We briefly summarize the systematic approach to count  Goldstone modes in non-relativistic systems~\cite{watanabe_counting_2020}.  Consider the set of linearly independent broken generators $\hat{Q}_i$ for $1 \leq i \leq n_{\text{BG}}$, and define
\begin{equation}
    \rho_{ij} \equiv -i\lim_{A \rightarrow \infty} \frac{1}{A}\braket{[\hat{Q}_i, \hat{Q}_j]},
\end{equation}
where $A$ is the area of the system. The set of Goldstone modes can be classified into Type-A and Type-B; excluding fine-tuned parameters, these correspond respectively to linear and quadratic modes.  The numbers of the two types of Goldstone modes are respectively given by~\cite{watanabe_counting_2020}
\begin{align}\label{eq:goldstone_counting}
    n_{\text{A}} &= n_{\text{BG}} - \rank\rho, \\
    n_{\text{B}} &= \frac{1}{2}\rank\rho.
\end{align}
We will use this approach to organize the numerical study of collective mode spectrum of the broken-symmetry states using time-dependent Hartree-Fock (TDHF), as discussed in Sec.~\ref{subsec:TDHF}.

The second consideration is that of  {\it which symmetries are present to be broken}, which can be subtle in the MA-TBG context. Formally, the only continuous symmetry present in MA-TBG is that of $U(1)_c$ charge conservation, since all the remaining symmetries are discrete. Of course, the smallness of the spin-orbit coupling means that we can promote spin to a full $SU(2)_s$ symmetry, and as we have argued previously we can safely promote the valley degree of freedom to a $U(1)_v$ conservation symmetry, at least at the moir\'e scale. Thus, we take the ``physical'' symmetry group to be $SU(2)_s\times U(1)_c\times U(1)_v$.

However, there is an additional hierarchy of scales present, once again stemming from  the separation of the moir\'e and microscopic length scales. To wit, the magnitude of the inter-valley Hund's couplings, which are of the order of $0.1$~meV, is  far below the scale of the interaction terms responsible the symmetry-breaking that drives the formation of either the IKS or strong-coupling states. As such, at  energy scales above this splitting but below, e.g., the gap to remote bands or the scale of the interactions, it is reasonable to enhance the symmetry group to the $U(2)\times U(2)$ of the interacting BM Hamiltonian introduced in Sec.~\ref{sec:basics} (and preserved in its strained counterpart in Sec.~\ref{sec:IKS}).  We shall term this symmetry group, corresponding to that of independent spin rotations and charge conservation  in each  valley, the ``BM symmetry group''. For realistic TBG this symmetry is broken by intervalley Hund's couplings to the physical $SU(2)_s\times U(1)_v \times U(1)_c $ symmetry. Finally, there is yet another symmetry enhancement by tuning to the chiral limit and ignoring the dispersion: the  $U(4)\times U(4)$ ``chiral-flat''  symmetry of rotations within each Chern quartet (see Section~\ref{sec:strong}), which reduces to $U(2)\times U(2)$ when moving away from this limit back to the BM model.
These three scales are illustrated below
\begin{equation}
\underset{(\text{chiral-flat})}{U(4)_{C=1}\times U(4)_{C=-1}} \rightarrow  \underset{(\text{BM})}{U(2)_K\times U(2)_{K'}} \rightarrow \underset{(\text{physical MA-TBG})}{SU(2)_s\times U(1)_c\times U(1)_v}.
\end{equation}
Of course, on a formal level the last symmetry is further explicitly broken to the exact microscopic symmetries. However, these are irrelevant to the experimentally accessible energy scales, and  we therefore do not discuss this step further.

While Goldstone's theorem technically only applies to spontaneously broken continuous symmetries, given this hierarchy of scales it allows us to organize the soft collective mode spectrum even when the symmetries are explicitly broken into what are sometimes termed ``pseudo-Goldstone modes'', such that modes within each hierarchy gap at distinct energy scales. As we have argued, it is reasonable to consider the ``physical'' scale to be $SU(2)_s\times U(1)_c\times U(1)_v$, but from the preceding sections we have seen that the  state selection in this limit is complex and depends on fine details that remain experimentally debated. Therefore, we focus primarily in this section on the ``chiral-flat'' and ``BM'' symmetries.

\subsection{Collective Modes of Strong Coupling states}\label{subsec:collective_sc}

The collective modes of MA-TBG in the strong coupling regime were systematically explored in Refs.~\cite{Khalaf2020soft,Bernevig2021TBG5,Khalaf2021charged,kumar2021lattice,Vafek2020RG}.  From the discussion above, we expect that we will find a large number of modes in the $U(4)\times U(4)$ chiral-flat limit (termed ``approximate Goldstone modes'' in Ref.~\cite{Khalaf2020soft}) with a corresponding reduction to a subset of these modes (termed ``exact'' in Ref.~\cite{Khalaf2020soft}) on moving to the $U(2)\times U(2)$ BM scale. We discuss the counting and properties of these in turn.

\subsubsection{Goldstone Spectra at the Chiral-Flat and BM Scales}\label{subsubsec:AG}

In the $U(4)\times U(4)$ limit, we can derive a rule for counting  chiral-flat Goldstone modes with the following intuitive arguments. Such a Goldstone mode can be understood as being built out of particle-hole excitations within the same Chern sector $\gamma = \pm$ (in this section, $\gamma$ denotes band Chern numbers while $C$ denotes the total Chern number of a many-body state), i.e.
\begin{equation}
    \hat{\varphi}_{\bm{q},\gamma\gamma,\alpha\beta}(\bm{k}) = \sum_{\bm{k}}\varphi_{\bm{q},\gamma\gamma,\alpha\beta}(\bm{k})\hat{c}^\dagger_{\gamma\alpha}(\bm{k})\hat{c}^{\phantom\dagger}_{\gamma\beta}(\bm{k}+\bm{q}).
\end{equation}
Here, $\alpha$ runs over the empty bands while $\beta$ runs over the filled bands within the same Chern sectors. Let us suppose the number of filled bands in the $\gamma = \pm$ sectors are $n_\pm$ respectively, with total filling factor $\nu = n_+ + n_--4$ and Chern number $C = n_+ - n_-$. A guess for the number of chiral-flat Goldstone modes can be made by taking the sum of the numbers of particle-hole excitation types in each Chern sector 
\begin{equation}
    n_{\text{AG}} = n_+(4 - n_+) + n_-(4 - n_-) = \frac{16 - \nu^2 - C^2}{2}.
\end{equation}

It is possible to formulate the above counting more rigorously using the theory discussed in Sec.~\ref{subsec:collective_bg}. Ref.~\cite{Khalaf2020soft} finds the number of broken generators to be $16 - \nu^2 - C^2$, while $\rho$ is a full rank matrix. This gives a total number of $ \frac{16 - \nu^2 - C^2}{2}$ Goldstone modes, all of which are quadratic. Once we move away from the chiral-flat limit, some of the Goldstone modes will become gapped at the scale of terms that break $U(4)\times U(4)$ down to $U(2)\times U(2)$ (i.e.,  they are pseudo-Goldstone modes\footnote{Observe that any mode identified as ``pseudo-Goldstone'' in Ref.~\cite{Khalaf2020soft} remains so here, but a subset of the $U(2)\times U(2)$ Goldstone modes are additional pseudo-Goldstones on reducing the symmetry further to the physical $SU(2)_s\times U(1)_c\times U(1)_v$.}) of the BM symmetries, while the remaining gapless modes do not necessarily remain quadratic in the $U(2)\times U(2)$ theory.

We can use the same approach as above, but now with only $U(2) \times U(2)$ symmetry, to calculate the number and type of the Goldstone modes at this new scale. Since the  details are state and filling-dependent,  we  illustrate this with a single example, that of the KIVC state at $\nu = 0$.

For the spinless model, according to Eq.~\ref{eq:QKIVCspinless}, KIVC is captured by the order parameter  $Q = \sigma_y\tau_{x,y}$. Here, $\tau_z$ is the only broken generator, leading to 1 linear Goldstone mode. Upon incorporating spin, at $\nu = 0$ the set of broken generators depend on which state is selected from the degenerate manifold generated by $U(2) \times U(2)$ symmetry. Since they all have identical excitation spectra, one can choose the spin singlet KIVC state, given by $Q = \sigma_y\tau_{x,y}s_0$. The broken symmetry generators are $\tau_z$ and $s_{x,y,z}\tau_z$. It can be shown that $\rho$ vanishes identically, leading to 4 linearly dispersing Goldstone modes.

\subsubsection{Nematic Modes}

Besides the approximate Goldstone modes, Ref.~\cite{Khalaf2020soft} also studied a set of gapped low-energy collective modes that correspond to particle-hole excitations between bands with opposite Chern numbers. These modes are termed ``nematic modes" because they carry non-zero angular momentum under $\hat{C}_{3z}$, and the condensation of these modes yields a nematic semimetal.

Using a similar argument as in Sec.~\ref{subsubsec:AG}, the number of nematic modes is given by
\begin{equation}
    n_{\text{N}} = n_-(4 - n_+) + n_+(4 - n_-) = \frac{16 - \nu^2 + C^2}{2}.
\end{equation}
The total number of chiral-flat Goldstone modes and nematic modes is given by $16 - \nu^2$, which only depends on the filling factor. They can be understood as the ``Goldstone modes" of filling $4 + \nu$ out of 8 bands with $U(8)$ symmetry. To understand this $U(8)$ symmetry, one can imagine concentrating the Berry curvature within each Chern sector to a point, which can then be removed by a singular gauge transformation so that the distinction between Chern sectors is lost. As such, the spread of Berry curvature is responsible for breaking the $U(8)$ symmetry. Moving away from the chiral limit, the $U(4) \times U(4)$ symmetry gets broken to $U(2) \times U(2)$, but the Berry curvature becomes more concentrated. Interestingly, at least at the level of TDHF, Ref.~\cite{Khalaf2020soft} finds that the gaps of the pseudo-Goldstone modes and the nematic modes are of a similar magnitude at realistic values of the chiral ratio.

\subsection{Collective Modes of IKS}

The collective modes of IKS were investigated by the present authors in Ref.~\cite{wang_putting_2025}, which studied the effects of $U(2) \times U(2)$-breaking terms in the Hamiltonian. Here, we briefly summarize the results in the $U(2) \times U(2)$-symmetric limit. (Recall that the chiral-flat limit precludes the presence of the IKS state as an energetically competitive ground state, so the additional scale hierarchy present at strong coupling is absent.)

\subsubsection{IKS at $|\nu| = 2$}
From the degenerate manifold of $|\nu|=2$ IKS states, we can pick the spin singlet state, where the same `primitive' IKS state is occupied in both spin sectors. The arguments for the Goldstone mode counting turn out to be very similar to KIVC state at $\nu = 0$, namely, the broken generators are $\tau_z$ and $s_{x,y,z}\tau_z$. It can again be shown that $\rho$ vanishes identically, leading to 4 linearly dispersing Goldstone modes.

\subsubsection{IKS at $|\nu| = 3$}
The IKS state at $|\nu| = 3$ has finite spin polarization in each valley. Since an $SU(2)$ ferromagnet has one quadratic Goldstone mode, we expect  $|\nu| = 3$ to have two quadratic Goldstone modes from the spin polarization of each valley. There is an additional linear Goldstone mode coming from the broken valley $U(1)$ symmetry, giving a total of 1 linear and 2 quadratic Goldstone modes.

We can also obtain the same result more rigorously using the method discussed in Sec.~\ref{subsec:collective_bg}. For definiteness, we consider the case of $\nu = -3$ with electrons fully polarized into the $s_z = +1$ sector. There are a total of $n_{\text{BG}} = 5$ broken symmetry generators, given by $\tau_z, s_x, s_y, \tau_zs_x, \tau_zs_y$. In this order, the matrix $\rho$ is given by
\begin{equation}
    \rho \propto \begin{pmatrix}
        0 & 0 & 0& 0& 0 \\
        0& 0& 1 & 0 & 0 \\
        0& -1& 0 & 0& 0 \\
        0& 0& 0& 0 & 1 \\
         0& 0& 0& -1 & 0
    \end{pmatrix},
\end{equation}
with $\rank \rho = 4$. Using Eq.~\ref{eq:goldstone_counting}, we find $n_{\text{A}} = 1$ and $n_{\text{B}} = 2$, i.e., 1 linear Goldstone mode and 2 quadratic Goldstone modes, consistent with the intuitive argument.

\subsection{Time-dependent Hartree-Fock Results}

As we have outlined in Sec.~\ref{subsec:TDHF}, performing time-dependent Hartree-Fock simulations with reference to a broken symmetry state gives insight into the corresponding spectrum of collective modes (at the RPA level), which should fall into the pattern captured by the general considerations built by combining Goldstone's theorem with the MA-TBG energy scale hierarchy. 
\begin{figure}[t!]
    \centering
    \includegraphics[width=\linewidth]{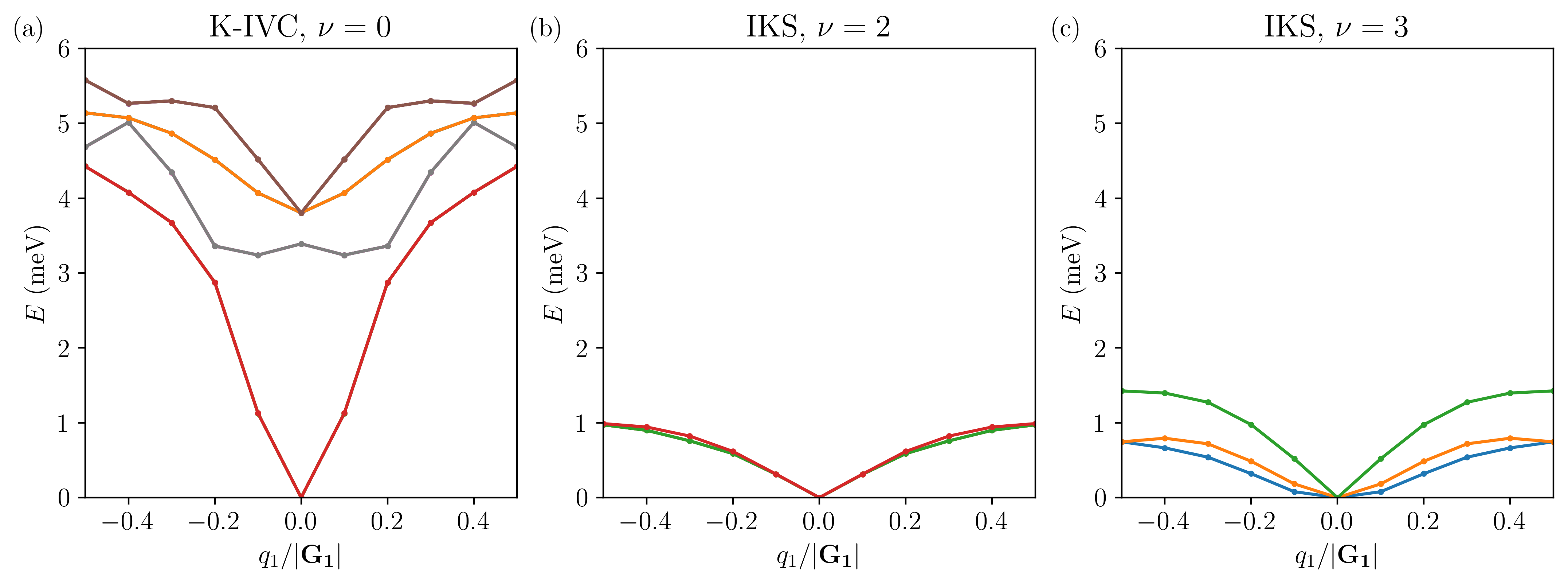}
    \caption{Neutral excitation spectra for various correlated insulators: (a) KIVC state at $\nu = 0$ without strain. All the data points are 4-fold degenerate. (b) IKS state at $\nu = 2$ with $0.2\%$ heterostrain. The green curve is a triplet while the red curve is a singlet. (c) IKS state at $\nu = 3$ with $0.2\%$ heterostrain. For all of the plots, we used $10 \times 10$ TDHF, with $\theta = 1.05^\circ$, $w_\textrm{AA} = 80$~meV, $w_\textrm{AB} = 110$~meV, and $\epsilon_r = 10$. The direction of heterostrain, if non-zero, is along $\hat{x}$-direction.}
    \label{fig:TDHF}
\end{figure}

In Fig.~\ref{fig:TDHF} (a), we show the neutral excitation spectra for the KIVC at $\nu = 0$ obtained from numerical TDHF calculations. We have zoomed in on the low-energy excitations, so that the particle-hole continuum is not shown. All the modes are 4-fold degenerate. We observe 4 gapless Goldstone modes, and 12 gapped modes. From the discussion in Sec.~\ref{subsec:collective_sc}, we expect 4 true Goldstone modes, 4 pseudo-Goldstone modes and 8 nematic modes. As we have already remarked, at realistic values of the chiral ratio, pseudo-Goldstone modes and nematic modes have similar energy gaps.

In Fig.~\ref{fig:TDHF} (b) and (c), we show the neutral excitation spectra for IKS at $\nu = 2$ and $3$ respectively. We observe 4 Goldstone modes (forming a triplet and a singlet) for $\nu = 2$ IKS and 3 Goldstone modes for $\nu = 3$ IKS. A set of gapped collective modes, similar to nematic modes of the strong coupling states, are also present in IKS (see the discussion in Ref.~\cite{wang_putting_2025}), though they are outside the range shown in Fig.~\ref{fig:TDHF}.

Finally, we comment that the collective mode spectra for IKS states are calculated within the ``boosted frame", as explained in Sec.~\ref{sec:IKS}. While the excitation energy is obviously independent of the chosen ``frame", in order to label the excitations with definite crystal momenta $\bm{q}$, we require the HF state from which we perform TDHF to preserve (generalized) translation symmetry, as is manifest in Eq.~\ref{collmode}. Generally, for states that break translation symmetry, the collective modes are only labeled with momenta defined within the reduced Brillouin zone of the translation-symmetry-breaking states.

\section{Case Study III: Domain-wall Energetics in Orbital Chern Insulators}\label{sec:domain_wall}

In our final case study, following Ref.~\cite{Kwan2021domain}, we change emphasis slightly from aspects of global ordering properties, and instead turn to considering effects in the mesoscopic regime, triggered by sample imhomogeneities. Specifically, we consider an example where the strong-coupling prediction of an orbital Chern insulator holds for a pristine sample, but where local imhomogeneities lead to domain walls between different phases distinguished by their valley order, their local Chern topology, or both. In doing so, we showcase how the Hartree-Fock methods introduced in Section~\ref{sec:Hartree-Fock} can be generalized to work with spatially inhomogeneous Hamiltonians and mean-field configurations. 

\subsection{Motivations}

As we discussed in Section~\ref{sec:strong}, the hBN substrate can have a significant impact on TBG. In particular, the hBN substrate affect the orbital magnetism in TBG \cite{Liu2021orbital}. Samples in which the hBN is aligned with the TBG can exhibit an  anomalous Hall effect at $\nu=+3$ \cite{Sharpe2019ferromagnetism}, an observation sharpened by later experiments that revealed a quantized Hall response at this filling~\cite{Serlin2020QAH}. In such an aligned configuration, the coupling between the substrate and the TBG is enhanced. The substrate breaks $\hat{C}_{2z}$ symmetry, and leads to the formation of Chern bands with $C=\pm1$ if the coupling is modelled by a sublattice mass $\Delta \sigma_z$. In this regime, in consonance with the strong-coupling analysis in the absence of a substrate, interactions spontaneously break time-reversal symmetry, leading to a spin and valley polarized QAH state \cite{Bultinck2020mechanism,Liu2021theories,Zhang2019hbn}. In this sense, one can view substrate alignment as `biasing' the system towards strong coupling. However, due to the $\sim 2$\% mismatch of the lattice constants of graphene and hBN, even in the case of perfect alignment the hBN will slip in and out of registry with the neighbouring graphene layer, which can lead to domain formation. SQUID-on-tip magnetic imaging has confirmed the presence of domains \cite{Tschirhart2021imaging, Grover2022mosaic}.

\subsection{Domain Wall Types: Chern Wall vs.~Valley Wall}
In this section we consider the domain walls (DWs) arising for the orbital Chern insulator at $\nu=+3$ in the presence of a spatially varying sublattice potential $\Delta(\mathbf r)$. For simplicity, we assume that the sublattice potential acts on just one layer, and strain is negligible. The sublattice potential breaks the $\hat{C}_{2z}$ symmetry and gaps out the Dirac points of the BM band structure, leading to Chern bands with opposite Chern number in  opposite valleys. The conduction bands have Chern number $C=\textrm{sgn}(\Delta)\tau_z$\footnote{This can be motivated from the discussion of the strong-coupling Chern basis in Section~\ref{subsec:Chern_basis}, where $C=\tilde{\sigma}\tau$. A positive sublattice potential will tend to raise the energy of the $\tilde{\sigma}=A$ bands relative to the $\tilde{\sigma}=B$ bands, so the kinetic conduction bands are expected to have $C=\tau$.}.  

Breaking time-reversal spontaneously by spin-valley polarizing the electrons leads to the orbital Chern insulator. For example at $\nu=+3$, if we fill all central bands except the conduction band for valley $K'$ and spin $\downarrow$, then the resulting many-body state has Chern number $C=+1$ and positive spin and valley polarization.
We focus on a region where the sublattice potential locally changes sign along some line in real space. For concreteness, we orient this line (we will also refer to this as a DW) along the $\bm{a}_2$ direction, and let the sublattice potential vary along the $\bm{a}_1$ direction such that $\textrm{sgn}(\Delta(\bm{r}))=-\textrm{sgn}(x_1)$, where $x_1$ is the spatial coordinate along $\bm{a}_1$. We label the region $x_1<0$ to the left of the DW as L, and label the region on the other side as R.  
The sublattice potential is taken to be $+\Delta_\textrm{max}$ deep inside region L, and $-\Delta_\text{max}$ deep inside region R. An example choice of sublattice potential profile consists of linearly interpolating between these values over a length scale $2w$. This leads to $\Delta(x_1)=-\Delta_\textrm{max}x_1/w$ for small $x_1$. 

Within the `bulk' of each region $|x_1|\gg w$, we anticipate that the system will locally form a spin-valley polarized QAH state. We assume that the spin is polarized along $\uparrow$ everywhere due to spin exchange physics. Hence, it remains to determine the spontaneous valley polarization within each domain. Let us consider the scenario where region L is polarized in valley $K$, and locally has Chern number $C=+1$. For region R, we have two options for the local valley polarization: 
\begin{enumerate}
    \item The valley switches across the DW, such that region R is polarized in valley $K'$ and has Chern number $C=+1$. We dub this the \emph{valley wall}. Since the Chern numbers $C_L=C_R$ are unchanged, the gapless chiral edge modes of the two domains are counter-propagating and carry opposite valley character. If intervalley hybridization occurs\footnote{This could occur due to explicitly $U(1)_V$-breaking terms such as short-range scatterers, or spontaneous IVC. However, the latter is not expected to be present beyond mean-field theory due to the Mermin-Wagner theorem applied to the 1D DW.}, then these edge modes can gap each other out.
    \item The valley remains the same across the DW, such that region R is polarized in valley $K$ and has Chern number $C=-1$. We dub this the \emph{Chern} wall. Since the Chern numbers $C_L=-C_R$ are opposite, the gapless chiral edge modes of the two domains are co-propagating. If this occurs in a patchwork fashion across a mesoscopic sample, this is referred to as a \emph{Chern mosaic} \cite{Grover2022mosaic}.
\end{enumerate} 
Neglecting the DW energy, these two solutions would be energetically degenerate. However, as well will see, the delicate energetics at the DW can tilt the system towards either solution, depending on the continuum model parameters and the gradient of the sublattice potential (set by $w$). The choice of DW can affect global transport characteristics, as well as the Luttinger parameters of the 1D chiral/helical modes (if they remain ungapped).

\subsection{Numerical Particularities}

\begin{figure}[t!]
    \centering
    \includegraphics[width=\linewidth]{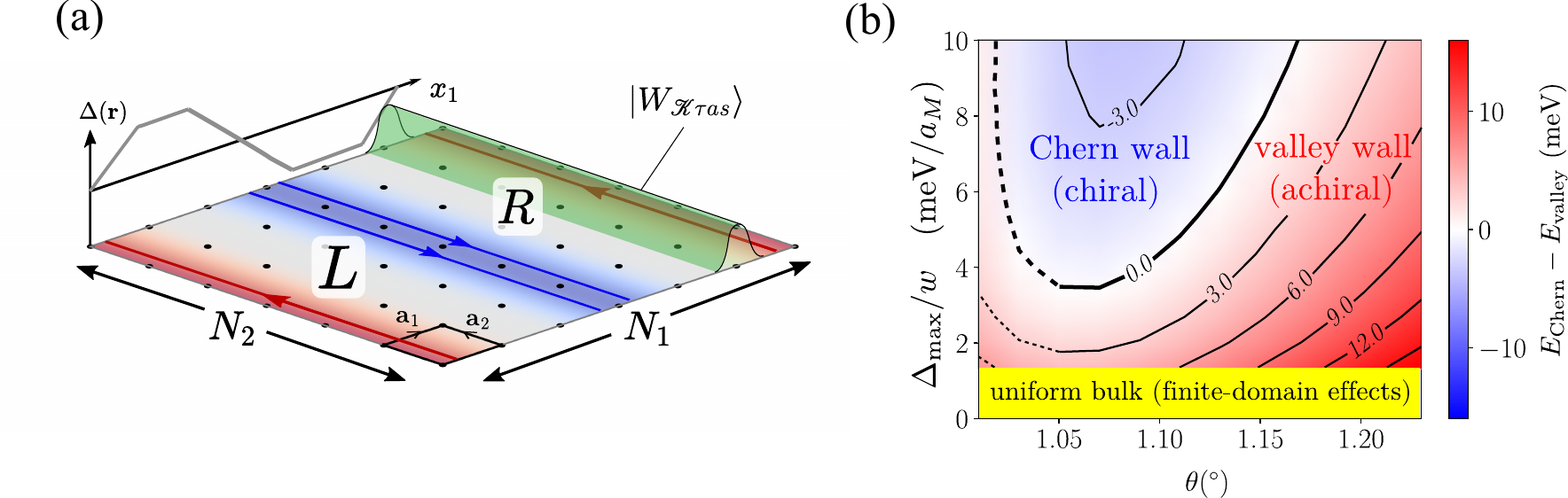}
    \caption{(a) Perspective view of numerical setup for studying DWs. The system contains $N_1N_2$ moir\'e unit cells (illustration shows $N_1=N_2=6$). Black dots indicate the $AA$-stacking moir\'e sites, which are connected by the moir\'e lattice vectors $\bm{a}_{1,2}$. The substrate potential $\Delta(\bm{r})$ varies with $x_1$ along the $\bm{a}_1$ direction, but is uniform along the $\bm{a}_2$ direction. The sign-changing potential divides the system into $L$ and $R$ regions, and can generate two sets of Chern or valley DWs due to the periodic boundary conditions. We schematically illustrate the case of Chern DWs which host co-propagating chiral modes. Green envelope represents a hybrid Wannier state $\ket{W_{\mathscr{K}\tau a s}}$ which is exponentially localized along $\bm{a}_1$ but extended along $\bm{a}_2$ (note that the actual wavefunction amplitude will vary along $\bm{a}_2$ according to the moir\'e periodicity). (b) Phase diagram of the energy competition between Chern DWs and valley DWs as a function of twist angle $\theta$ and substrate potential gradient $\Delta_\text{max}/w$. Contours for $\theta<1.05^\circ$ are dashed because the gap to the remote bands is sufficiently small that treating the remote bands as inert is less good of an approximation. Adapted with permission from Kwan \textit{et al.}, Phys. Rev. B. 104, 115404 (2021)~\cite{Kwan2021domain}. Copyright (2021) American Physical Society.}
    \label{fig:DW}
\end{figure}

\subsubsection{One-dimensional Translational Symmetry Breaking}
For the numerical calculations (see Fig.~\ref{fig:DW}a for an illustration of the setup), we assume the usual periodic boundary conditions with $N_1,N_2$ moir\'e unit cells along the $\bm{a}_1,\bm{a}_2$ directions respectively. For computational ease, we restrict to sublattice potentials that vary only along the $\bm{a}_1$ direction. This implies we must have have an even number of DWs where $\Delta(x_1)$ changes sign. We place DWs at $x_1=0$ and $x_1=N_1|\bm{a}_1|/2$ so that the system is partitioned into two regions L and R of equal size. At each DW, the sublattice potential varies linearly over a length $2w$, and reaches $+\Delta_\text{max}$ $(-\Delta_\text{max})$ in the `bulk' of region L (R).

To treat this spatially inhomogeneous system, we have to generalize the HF algorithm to allow for translational symmetry breaking. The system can fully break translation symmetry along the $\bm{a}_1$ direction, but we enforce moir\'e translation symmetry along the $\bm{a}_2$ direction. As a result, the HF projector takes the form
\begin{equation}
\langle \hat c^\dagger_{\tau as}(k_1,k_2)\hat c^{\phantom{\dagger}}_{\tau'a's'}(k_1',k_2')\rangle=\delta_{k_2,k_2'}P_{k_1'\tau'a's';k_1\tau as}(k_2).
\end{equation}

\subsubsection{Fixed Hilbert Space Method}
The presence of an explicit translation symmetry-breaking potential $\Delta(\bm{r})$ suggests that we need to rediagonalize the BM model with this term. The large gap between the central bands and remote bands enables an alternative and simpler approach called the ``Fixed Hilbert space" method. Here, we start with the central band subspace of the BM model in the absence of any substrate potential, and then project the spatially-varying sublattice potential $\Delta(\bm{r})$ into this subspace. Numerically it can be shown that the overlap between the fixed Hilbert space and the Hilbert space for a given constant sublattice potential $\Delta_0$ is very high in the range $\Delta_0\lesssim30$meV, justifying this approach \cite{Kwan2021domain}. The substrate potential operator takes the form
\begin{equation}
    \hat{\Delta}(\bm{r})=
    \begin{pmatrix}
    \Delta_1(\bm{r})\sigma_z&0\\
    0&\Delta_2(\bm{r})\sigma_z
    \end{pmatrix},
\end{equation}
where $\Delta_l(\bm{r})$ is the substrate potential on layer $l$. We typically consider cases where only one of the layers is aligned with the hBN, i.e.~$\Delta_1(\bm{r})=\Delta(\bm{r}),\ \Delta_2(\bm{r})=0$. We project this sublattice potential directly onto the central bands $\ket{\psi_{\tau a}(\bm{k} )}$ evaluated at $\Delta=0$, i.e.
\begin{equation}
    \bra{\psi_{\tau a}(\bm{k})}\hat{\Delta}\ket{\psi_{\tau'a'}(\bm{k}')}=\delta_{\tau,\tau'}\delta_{k_2,k_2'}\Delta_{k_1a;k'_1a'}(k_2,\tau),
\end{equation}
where we assumed that the substrate potential is only a function of $x_1$. This leads to a term in the projected Hamiltonian
\begin{equation}
    \hat H_\textrm{substrate}=\sum_{\tau aa's k_1k_1'k_2}\Delta_{k_1a;k'_1a'}(k_2,\tau)\hat{c}^\dagger_{\tau a s}(k_1,k_2)\hat{c}_{\tau a' s}(k_1',k_2).
\end{equation}
This method can be generalized to other spatially-varying potentials. 

\subsubsection{Hybrid Wannier Orbitals}
In order to initialize the HF state to target particular types of DWs, as well as visualize and interpret the converged HF solutions, we utilize hybrid Wannier orbitals~\cite{Hejazi2021hybrid,Kang2020nonabelian}. As is well known, Chern bands have a  topological obstruction that rules out using them to build exponentially-localized Wannier orbitals. However, it is possible to construct hybrid Wannier orbitals which are only localized in one direction \cite{Qi2011Wannier,Wu2012Wannier,Scaffidi2012adiabatic,Barkeshli2012nematic,Liu2019pseudo}. In our case, it will prove most useful to orient the hybrid Wannier wavefunctions  such that they are localized perpendicular to the DW (and hence delocalized parallel to it), and build them using the HF eigenstates for a uniform substrate potential. This construction leads to Wannier states that not only approximate well the wavefunction within the `bulk' of each region, but are also capable of capturing a change in the order parameter in the $x_1$ direction.

On a discrete lattice, the Berry connection of the HF Bloch functions takes the form
\begin{equation}
    \mathcal{A}_{\tau a}^{(1)}(n_1,n_2)=\mathrm{Im} \log \sum_{\bm{G}l\sigma} u^\textrm{HF*}_{\tau a;l\sigma}(n_1,n_2;\bm{G})u^\textrm{HF}_{\tau a;l\sigma}(n_1+1,n_2;\bm{G}),
\end{equation}
where $n_1,n_2$ parameterize the momentum $\bm{k}$ according to Eq.~\ref{eq:n1n2mtm}, and $u^\textrm{HF}_{\tau a;l\sigma}(n_1,n_2)$ are Bloch coefficients of the HF Bloch wavefunctions (obtained for a uniform substrate potential) in a similar notation to Eq.~\ref{eq:plane_wave_expansion_k}. In terms of the HF Bloch states $\ket{\psi^\textrm{HF}_{\tau as}(n_1,n_2)}$, the corresponding hybrid Wannier orbitals take the form \cite{Scaffidi2012adiabatic,Qi2011Wannier}
\begin{equation}
\label{eq:Bloch_to_WQ_SM}
        \ket{W_{\mathscr K\tau as}}=\frac{1}{\sqrt{N_1}}\sum_{n_1=0}^{N_1-1} e^{-\mathrm{i}\sum_{n_1'=0}^{n_1} \mathcal{A}^{(1)}_{\tau a}\left(n_1',n_2\right)-\mathrm{i} \frac{2\pi n_1}{N_1}\left(\tilde n_1-\frac{\theta_{\tau a}\left(n_2\right)}{2 \pi}\right)}\ket{\psi^\textrm{HF}_{\tau as}(n_1,n_2)},
\end{equation}
where $\mathscr{K}=\frac{n_2}{N_2}+\tilde n_1$, $\tilde n_1=0,1,\dots,N_1-1$ and $\theta_{\tau a}\left(n_2\right)=(\sum_{n_1=0}^{N_1-1} \mathcal{A}^{(1)}_{\tau a}\left(n_1,n_2\right))\mod 2\pi$. The quasimomentum $\mathscr K$ is related to the average position of a given hybrid Wannier orbital in the $x_1$ direction (see Fig.~\ref{fig:DW}a).

We compute one set of hybrid Wannier orbitals for each choice of valley polarization and uniform substrate potential $\pm\Delta_\text{max}$, leading to four sets in total. From these, we can design initial states that are specifically tailored to converge to a Chern wall or a valley wall. For example, consider the situation where we want to obtain a Chern wall solution that is polarized to valley $K$ everywhere. We construct the initial density matrix by specifying the Wannier states that we occupy. For the set corresponding to valley $K$ and sublattice potential $+\Delta_\text{max}$, we occupy quasimomenta $\mathscr K$ corresponding to orbitals whose average $x_1$ position lies in region L. Moreover, for the set corresponding to valley $K$ and sublattice potential $-\Delta_\text{max}$, we occupy quasimomenta $\mathscr K$ corresponding to orbitals whose average $x_1$ position lies in region R. In a similar vein, we can design an initial state that converges to the valley wall.

\subsection{Energetic Competition Between Chern and Valley Walls}

There will be a loss of (negative) exchange energy at the DWs, since the system is forced to change its local valley polarization or topology\footnote{Loosely, since the Chern number changes at a Chern wall, then the Bloch functions that make up the occupied bands have to change in a significant manner.}. The system  can however smoothly interpolate or `texture' its local wavefunctions between region L and region R, and thereby reduce the amount of exchange energy lost. For the valley wall, we thus expect a smoothly textured DW, with the width of the DW set by the valley stiffness and the gradient of the sublattice potential, $\Delta_\textrm{max}/w$. On the other hand, there is a topological obstruction towards forming inter-Chern coherence~\cite{Bultinck2020mechanism} (see discussion in Section~\ref{sec:strong}) and thus the Chern wall is sharp. This leads to the following energetic competition reflected in the numerics in Fig.~\ref{fig:DW}b: For small $\Delta_\textrm{max}/w$, the valley wall can texture and thereby reduce the exchange energy loss compared to the Chern wall, so that the valley wall has lower energy. For large $\Delta_\textrm{max}/w$, the valley wall becomes sharp and does not benefit from any texturing. In this regime, the Chern DW is lower in energy. We emphasize the unusual nature of this competition, where the global 2D pattern of symmetry-breaking is determined by the physics occurring at domain walls which only make up a 1D subregion.

\subsection{Intertwined Walls}

It it possible to have DWs even in the presence of a uniform substrate potential. For instance for a uniform $\Delta>0$, we could have one domain be polarized in valley $K$ with $C=+1$, and an adjacent domain be polarized in valley $K'$ with $C=-1$. We call this type of DW an intertwined wall, since both the Chern number and the valley flip across the DW. In this case, these DW solutions are meta-stable, since they have a higher energy than the state with uniform valley polarization. Numerically, we find that the intertwined DW is pinned by the moiré potential. In our HF calculations, the intertwined DW does not relax to a uniform QAH state even after many HF iterations, demonstrating the strong pinning of the state. At finite temperature, these intertwined DWs can proliferate due to the entropy associated with domain formation.

\section{Concluding Remarks}\label{sec:conclusions}

In this review, we have attempted to give both an introduction to the broad field of moir\'e systems, with a focus on magic-angle twisted bilayer graphene, as well as the role of Hartree-Fock mean-field studies in theoretically modelling some of their rich physics. In doing so, we have also attempted to provide some rationale for what might, {\it prima facie}, seem like the ``unreasonable effectiveness'' of these techniques in what is putatively a strongly correlated system. Our choice of illustrative case studies, while to an extent  influenced by our own involvement, was also made in order to highlight various modes in which the Hartree-Fock method can be deployed. There are of course other applications of HF in the moir\'e setting. For instance, a subset of us~\cite{Kwan2021skyrmion} have used such methods to explore the emergence of skyrmion excitations of the strong-coupling states~\cite{Khalaf2021charged,schindler2022trions,Khalaf2021polaron}, and explore their stability, and the possibility that they bind into charge-$2e$ bosons that can then condense, fueling superconductivity \cite{Khalaf2021charged,Chatterjee2020DMRG}, or crystallize into a variety of translationally-breaking states when doped into correlated insulators \cite{Bomerich2020tetarton}. As another example, studying the evolution of the Bethe-Salpeter solutions for  one-particle-one-hole bound states across the mBZ can be used to extract the topology of exciton bands in orbital Chern insulators~\cite{Kwan2021exciton}.

In concluding this review, we give a brief account of some of the advantages and failings of Hartree-Fock, before closing with a summary of some open problems that might stimulate further work.

\subsection{Advantages and Shortcomings of Hartree-Fock}

Hartree-Fock is a computationally inexpensive method that is not as severely plagued by finite size effects as are other methods such as exact diagonalization or density matrix renormalization group (DMRG). Hartree-Fock calculations on the central bands of TBG can be routinely performed on $25\times25$ momentum space meshes and larger, and with the interpolation scheme discussed above, the resulting bandstructures can be plotted on much finer grids. Due to the low computational cost, large parameter sweeps are possible, which is especially relevant for TBG where some parameters in the Hamiltonian, such as the chiral ratio, are not accurately known and so a more global picture is necessary. 
Furthermore, Hartree-Fock allows for any type of symmetry breaking order. If the initial Hartree-Fock projector breaks all symmetries, then the self-consistent solution may break any number of symmetries. By imposing symmetries on the initial Hartree-Fock projector, one can obtain the Hartree-Fock ground states which respect those symmetries. This can be useful to diagnose whether there are closely competing orders. 
Finally, unlike quantum Monte Carlo whose computational effectiveness is only maximal at certain sign-problem free fillings and limits, HF is straightforwardly applicable at any filling factor. 

However, as a mean-field method, there are also certain limitations to Hartree-Fock. For instance, it is known to overestimate the stability of symmetry breaking order due to its neglect of quantum fluctuations and in TBG this reflects in the overestimate of the band gaps. The band gaps found in central band Hartree-Fock calculations are typically on the order of 20\,meV, which is an order of magnitude larger than the transport gaps measured in experiment. Strongly correlated states such as fractional Chern insulators are entirely beyond mean-field theory and require other methods such as exact diagonalization and DMRG\footnote{Although fractional Chern insulators cannot be described by a mean field theory of electrons, certain filling fractions can be captured by a mean-field theory of {\it composite fermions}. Note also that in many moir\'e systems the onset of valley polarization is necessary in order to break time-reversal symmetry, a prerequisite for nonzero Chern number; this sort of valley ferromagnetism is often  reasonably well-captured by Hartree-Fock.}.

As we have seen, Hartree-Fock can capture the zero-temperature normal-state phase diagram of TBG well at integer fillings.  While  we presented the zero-temperature version of the method in this review, it can be generalized to finite temperature, although its neglect of fluctuations will lead to an overestimate of critical temperatures for the various ordered states. However, the finite-temperature physics in TBG appears to have various features, such as (non-)local moment physics, that might be better captured by DMFT. We therefore omitted any discussion of finite-temperature Hartree-Fock from this review, since in our view the method is better deployed as a guide to the physics at low temperatures.

\subsection{Open Questions}

Two features of the phase diagram of twisted bilayer graphene have attracted attention from the early days of the field in 2018: The correlated insulators and the superconducting domes. While the similarity to the cuprate phase diagram is striking, we now know thanks to the success of Hartree-Fock calculations that the nature of the correlated insulators in twisted bilayer graphene is very different to the Mott insulator in the cuprates. Two examples of correlated insulators in TBG that have been experimentally confirmed are (1) the IKS state at $|\nu|=2$ confirmed via the characteristic modulated Kekulé pattern STM experiments \cite{NuckollsTextures,Kim2023STM} and (2) the quantized anomalous Hall insulator at $\nu=3$ in hBN aligned samples confirmed via transport measurements \cite{Serlin2020QAH,Sharpe2019ferromagnetism}.  Both of these states can be well captured by the Hartree-Fock method outlined in this review. 

On the other hand, the pairing symmetry and the mechanism behind the superconductivity remain poorly understood. Progress has been made towards constraining the pairing symmetry: experimental probes such as tunneling spectroscopy~\cite{Oh2021unconventional,park2025simultaneoustransporttunnelingspectroscopy} and superfluid stiffness measurements \cite{banerjee2024superfluidstiffnesstwistedmultilayer,tanaka2024superfluid} suggest the intriguing possibility of a nodal order parameter. Furthermore, the observation of nematicity in the superconductivity \cite{Cao2021nematicity} coupled with a Ginzburg-Landau analysis \cite{lake2022pairing} also suggests a nodal order parameter. 
However, while many putative mechanisms for have been proposed \cite{Khalaf2021charged,Cea2021Coulomb,Po2018mott,Chatterjee2020skyrmionic,Kennes2018,Roy2019,Sharma2020,Isobe2018,Khalaf2020unitary,Chichinadze2020,Wang2020,Gonzales2019,Liu2018Chiral,Kozii2019,chou2019versus,Fidrysiak2018,Lewandowski2021,Wu2018theory,Lian2019SC,Wu2019phonon,Peltonen2018MFT,Shavit2021theory,Schrodi2020,Blason2022Kekule,Yu2022euler,Choi2018SC,Wu2019chiral,Wu2019identification,Lake2021reentrant,lake2022pairing,huang2022pseudospin,gonzalez2023universalmechanismisingsuperconductivity,ingham2023quadraticdiracfermionscompetition,Christos:2023aa,Lewandowski:2021aa,You2019SC,Christos2020SC,HUANG2019310}, there is no consensus on the superconducting mechanism. At the mean-field level, there must be a bare attractive term in the interaction for superconductivity to be possible \cite{Back1994HF}; beyond the mean-field level, of course, there can be a variety of sources of pairing `glue', ranging from fluctuations of spin or other collective modes, to `Kohn-Luttinger' superconductivity from the ``overscreening'' of purely repulsive interactions \cite{Kohn1965}. Identifying which pairing mechanism is appropriate is a question where guidance from experiment is crucial. Still, understanding the parent normal state is often a key initial step to solving this puzzle, and one where mean-field methods such as those discussed in this review has had and will likely continue to have important applications.

\section*{Acknowledgements}
We are grateful to numerous sources of support for pieces of research reported in this review, including a Leverhulme Trust International Professorship (Grant Number LIP-202-014, Z.W.), a  Swiss National Science Foundation (SNSF)  Ambizione Grant  (Number PZ00P2-216183, G.W.), the  NCCR
MARVEL, a National Centre of Competence in Research,
funded by the Swiss National Science Foundation (Grant
No. 182892, G.W.), a postdoctoral research fellowship at the Princeton Center for Theoretical Science (Y.H.K.), the European Research Council under the European Union Horizon 2020 Research and Innovation Programme  (Grant Agreements No. 804213-TMCS, S.A.P., and 101076597-SIESS, N.B.), the UK Engineering and Physical Sciences Research Council EPSRC (Grants EP/S020527/1
 and EP/X030881/1, S.H.S.), and a  UKRI Horizon Europe Guarantee (Grant No. EP/Z002419/1, S.A.P.).

We note that several portions of our discussion  are adapted from sections of the doctoral thesis of one of the present authors (YHK), that have not previously appeared in stand-alone publications.

\section{Appendices}

\noindent\textbf{Appendix A. Derivation of the Bistritzer-MacDonald Model}\medskip

In this appendix section, we provide a clean and direct derivation of the BM model for TBG~\cite{Bistritzer2011BM} to demonstrate the simplicity of the approach and highlight where the approximations enter. We also describe how to incorporate uniaxial heterostrain into the theory.

Recall that in the low-energy limit, we can focus on plane wave momenta $\bm{p}\simeq \bm{K}$ near valley $K$. The corresponding Hamiltonian for valley $K'$ is obtained using spinless TRS $\hat{\mathcal{T}}=\tau_x\mathcal{K}$. We consider four species of fermions (two layers $l=1,2$ and two sublattices $\sigma=A,B$). The top layer ($l=1$) is rotated by $\theta/2$, while the bottom layer ($l=2$) is rotated by $-\theta/2$. We consider a vanishing interlayer shift $\bm{d}=0$ between the layers (a non-zero $\bm{d}$ can be gauged away for TBG~\cite{Bistritzer2011BM}). The intralayer physics is modelled by rotated versions of the Dirac Hamiltonian. The interlayer coupling leads to a spatially-modulating hopping amplitude between the layers, whose functional form is to be determined below. 

We work in the basis of plane waves: for an electron residing in layer $l$ and sublattice $\sigma$, we define the Bloch states
	\begin{equation}
		\ket{\bm{p},l\sigma}=\frac{1}{\sqrt{N}}\sum_{\bm{R}}e^{i\bm{p}\cdot(\bm{R}^{l}+\bm{\tau}_\sigma^l)}\ket{\bm{R}^{l},l\sigma}
	\end{equation}
where $\bm{R}$ is an untwisted monolayer lattice vector, the superscript $l=1,2$ on a vector denotes rotation by $R_{\pm\theta/2}$, $\bm{\tau}_\sigma$ are the intra-unit-cell sublattice coordinates, and $N$ is the number of graphene unit cells per layer. Therefore $\bm{K}^l$ correctly identifies the new positions of valley $K$ in each layer. $\ket{\bm{R}^{l},l\sigma}$ represents an atomic orbital belonging to layer $l$ and sublattice $\sigma$ in the unit cell centered at $\bm{R}^l$. 

Consider the following interlayer matrix element of the valley-$K$ BM Hamiltonian $H_\text{BM}$
	\begin{equation}
		T^{\sigma,\sigma'}_{\bm{p},\bm{p}'}=\bra{\bm{p},1\sigma}H_\text{BM}\ket{\bm{p}',2\sigma'}
	\end{equation}
which corresponds to an electron on sublattice $\sigma'$ on the bottom layer with momentum $\bm{p}'$ hopping to sublattice $\sigma$ on the top layer with momentum $\bm{p}$. Now compute
\begin{align}
\begin{split}
	T^{\sigma,\sigma'}_{\bm{p},\bm{p}'}&=
	\frac{1}{N}\sum_{\bm{R},\bm{R}'}e^{-i\bm{p}\cdot(\bm{R}^1+\bm{\tau}_\sigma^1)+i\bm{p}'\cdot(\bm{R}^{\prime2}+\bm{\tau}_{\sigma'}^2)}\bra{\bm{R}^1,1\sigma}H_\text{BM}\ket{\bm{R}^{\prime2},2\sigma'}\\
	&=\frac{1}{N}\sum_{\bm{R},\bm{R}'}e^{-i\bm{p}\cdot(\bm{R}^1+\bm{\tau}_\sigma^1)+i\bm{p}'\cdot(\bm{R}^{\prime2}+\bm{\tau}_{\sigma'}^2)}t(\bm{R}^1-\bm{R}^{\prime2}+\bm{\tau}_\sigma^1-\bm{\tau}_{\sigma'}^2)\\
	&=\frac{1}{N^2A_\text{UC}}\sum_{\bm{R},\bm{R}'}\sum_{\bm{q}}
	e^{i(\bm{q}-\bm{p})\cdot\bm{R}^1+i(\bm{p}'-\bm{q})\cdot\bm{R}^{\prime2}}
	e^{i(\bm{q}-\bm{p})\cdot\bm{\tau}_\sigma^1}
	e^{i(\bm{p}'-\bm{q})\cdot\bm{\tau}_{\sigma'}^2}t(\bm{q})\\
	&=\sum_{\bm{G}_\text{G},\bm{G}_\text{G}'}\frac{t(\bm{p}+\bm{G}^1_\text{G})}{A_{\text{UC}}}e^{i(\bm{G}_\text{G}\cdot\bm{\tau}_\sigma-\bm{G}_\text{G}'\cdot\bm{\tau_{\sigma'}})}\delta_{\bm{p}+\bm{G}^1_\text{G},\bm{p}'+\bm{G}^{\prime2}_\text{G}}\label{eqn:BM_derivation}.
\end{split}
\end{align}
In the second line, we introduced the interlayer hopping matrix element $t(\bm{r})$ which depends only on the distance between the sites (two-centre approximation). In the third line, we inserted the Fourier transform of the hopping element $t(\bm{r})=\frac{1}{NA_{\text{UC}}}\sum_{\bm{q}}t(\bm{q})e^{i\bm{q}\cdot\bm{r}}$ where $A_\text{UC}$ is the area of the graphene unit cell. In the fourth line, the sum over lattice sites is used to enforce \mbox{$\bm{q}=\bm{p}+\bm{G}^1_\text{G}=\bm{p}'+\bm{G}^{\prime2}_\text{G}$} for some $\bm{G}_\text{G},\bm{G}'_\text{G}$ (in-plane Bragg scattering) belonging to the unrotated monolayer reciprocal lattice. 

At this stage, no additional approximations have been made. Now we exploit the fact that we are interested in momenta close to the monolayer Dirac points, so $|\bm{p}-\bm{K}^1|,|\bm{p}'-\bm{K}^2|\ll|\bm{K}|$. Since the the orbital overlap is smooth as a function of inter-orbital distance, the hopping element $t(\bm{q})$ is expected to fall rapidly with $|\bm{q}|$. Therefore in the sum over $\bm{G}_\text{G}$ in the last line of Eq.~\ref{eqn:BM_derivation}, we only keep the terms $\bm{G}_\text{G}=\bm{0},-\bm{b}_{\text{G}1},-\bm{b}_{\text{G}2}$, where the $\bm{b}_{\text{G}i}$ are the monolayer reciprocal lattice vectors shown in Fig.~\ref{fig:superlattice}b. These are the three terms for which $t(\bm{p}+\bm{G}^1_\text{G})$ has the smallest argument---effectively we have kept the lowest harmonic of the moir\'e potential (dominant harmonic approximation). Its value is approximated as $t(|\bm{K}|)$. The delta function constrains $\bm{G}'_\text{G}$ to take the same three reciprocal lattice vectors if we demand that $|\bm{p}'-\bm{K}^2|$ remains small\footnote{The Bloch waves $\ket{\bm{p},l\sigma}$ and $\ket{\bm{p}+\bm{G}^l_\text{G},l\sigma}$ are not independent and only differ up to a phase. In other words, $\bm{k}$ is a crystal momentum. For simplicity, we therefore always pick the representative whose momentum label lies closest to $\bm{K}^l$.}. 
Aggregating these observations and performing some straightforward algebra, we obtain the BM model presented in Sec.~\ref{subsec:BM_model}
\begin{gather}
	\bra{\bm{p},1}H_\text{BM}\ket{\bm{p}',1} = \hbar v_F \bm{\sigma}^*_{\theta/2}\cdot(\bm{p}-\bm{K}^1)\,\delta_{\bm{p},\bm{p}'}\label{eqn:BM_start_SM}\\
	\bra{\bm{p},2}H_\text{BM}\ket{\bm{p}',2} = \hbar v_F \bm{\sigma}^*_{-\theta/2}\cdot(\bm{p}-\bm{K}^2)\,\delta_{\bm{p},\bm{p}'}\\
	\bra{\bm{p},1}H_\text{BM}\ket{\bm{p}',2} = 
	T_1\delta_{\bm{p}-\bm{p}',\bm{0}} + T_2\delta_{\bm{p}-\bm{p}',\bm{b}_{1}+\bm{b}_2} + T_3\delta_{\bm{p}-\bm{p}',\bm{b}_{2}}\label{eqn:BM_inter_SM}\\
	\bm{\sigma}^*_{\theta/2}=e^{-(i\theta/4)\sigma_z}(\sigma_x,\sigma_y^*)e^{(i\theta/4)\sigma_z}\label{eqn:pauli_twist_SM}\\
	T_1 = \begin{pmatrix}w_\textrm{AA}&w_\textrm{AB}\\w_\textrm{AB}&w_\textrm{AA}\end{pmatrix}\\
	T_2 = \begin{pmatrix}w_\textrm{AA}&w_\textrm{AB}e^{i\phi}\\w_\textrm{AB}e^{-i\phi}&w_\textrm{AA}\end{pmatrix}\\
	T_3 = \begin{pmatrix}w_\textrm{AA}&w_\textrm{AB}e^{-i\phi}\\w_\textrm{AB}e^{i\phi}&w_\textrm{AA}\end{pmatrix}\\
	\phi=\frac{2\pi}{3}\label{eqn:BM_end_SM}.
\end{gather}
Some comments to help decipher the above equations:
\begin{enumerate}
	\item The sublattice degree of freedom has been absorbed into the matrix structure.
	\item The intralayer kinetic terms pick up a twist factor due to the rotation of the layers. If we neglect this Pauli twist, which works well in the small-angle limit, the model gains an exact particle-hole symmetry.
	\item In the interlayer hopping terms, we have defined the basis moir\'e reciprocal lattice vectors (Fig.~\ref{fig:superlattice}b)
		\begin{equation}
		 \begin{gathered}\label{eqn:moire_b1_SM}
		\bm{b}_{1}=(R_{\theta/2}-R_{-\theta/2})(\bm{b}_{\text{G}2}-\bm{b}_{\text{G}1})=\sqrt{3}k_{\theta}(1,0)\\
		\bm{b}_{2}=(R_{\theta/2}-R_{-\theta/2})\bm{b}_{\text{G}2}=\sqrt{3}k_{\theta}(-\frac{1}{2},\frac{\sqrt{3}}{2}),
		\end{gathered}
		\end{equation}
	where $k_\theta=2k_D\sin\theta/2$ is the moir\'e wavevector and $k_D=\frac{4\pi}{\sqrt{3}a_\text{CC}}$ is the monolayer Dirac wavevector, with $a_\text{CC}$ the C$-$C bond length.
	\item There are now two interlayer hopping strengths $w_\textrm{AA}$ and $w_\textrm{AB}$, while we initially had a single interlayer hopping amplitude $t(|\bm{p}|)$. This generalization is allowed by the symmetries. 
	\item $H_\text{BM}$ is an infinite-dimensional matrix in the momentum basis, similar to the situation in the nearly free electron model. To solve for the energies numerically, one has to impose a plane wave cutoff that ideally preserves the point group symmetries.
\end{enumerate}	

Because the interlayer terms only allow momentum changes by integer combinations of $\bm{b}_{1}$ and $\bm{b}_{2}$, the resulting theory is periodic. Inspection of the new RLVs reveals that the new mBZ is the monolayer BZ rotated by $90^\circ$ and shrunk by the moir\'e scale factor $2\sin(\theta/2)$. Therefore, the real-space superlattice is similarly rotated and expanded by $1/(2\sin(\theta/2))$. By taking the inverse Fourier transform, we can interpret the continuum model as Dirac fermions on two layers with an interlayer hopping that is local\footnote{The locality of the hopping derives from the fact that we neglected the $\bm{p}$ dependence in the hopping amplitude in the last line of Eq.~\ref{eqn:BM_derivation}. Retaining the momentum-dependence leads to corrections known as non-local tunneling or momentum-dependent tunneling.} and spatially modulated with wavevector $\sim k_\theta$.

How did a periodic Hamiltonian emerge from generically incommensurate twist angles? 
Due to the intralayer kinetic penalty, we imagined restricting to plane wave momenta $\bm{p}$ near $\bm{K}$. However, an electron with starting momentum $\bm{p}$ in principle may scatter to $\bm{p}+\Delta \bm{p}$ that is far from $\bm{K}$ through multiple interlayer tunneling processes that change the momentum by integer multiples of $\bm{b}_j$. In particular, if through this multistep scattering process, it ends up on layer $l$ and gains momentum $\Delta \bm{p}=\sum_j n_j\bm{b}_j\simeq \bm{G}^l_\text{G}$ that is close to a monolayer RLV, it can undergo `Bragg scattering' by $-\bm{G}^l_\text{G}$ to return near $\bm{K}$. The incommensurability manifests in the fact that $\bm{G}^l_\text{G}$ is not expressible as an integer sum of moir\'e RLVs for general twist angles. However, such effects are exponentially suppressed at small $\theta$ due the high order of perturbation theory required. A similar argument explains the emergent $U_V(1)$ valley conservation symmetry---a high order of perturbation theory is required to scatter an electron from the neighbourhood of valley $K$ to that of valley $K'$.

We now turn to the alterations required to account for uniaxial heterostrain (henceforth referred to as just strain), in which the two layers are strained oppositely. Homostrain is typically not included, since to first order it does contribute to the distortion of the moir\'e lattice vectors, and it has a substantially smaller impact on the electronic structure \cite{Huder2018,Mesple2021heterostrain}. These strains, which have a finite average over mesoscopic lengthscales, are to be contrasted with intra moir\'e cell strains and relaxations which preserve the spatial symmetries. 

The moir\'e geometry is deformed depending on the value of the strain ratio $\epsilon$ and strain angle $\varphi$ with respect to the $x$-axis. The orthogonal direction is also stretched/compressed due to the Poisson ratio $\nu_\text{P}\simeq 0.16$ \cite{Cao2014poisson}. Following Refs.~\cite{Bi2019strain,Parker2021strain}, the effect of uniaxial strain on the unstrained graphene lattice vectors $\bm{r}$ and RLVs $\bm{g}$ is expressed through layer-dependent transformation matrices $M_l$
\begin{equation}
    \begin{gathered}
    \bm{r}_l=M_l^T\bm{r},\quad \bm{g}_l=M_l^{-1}\bm{g}\\
    M_l\simeq 1+\mathcal{E}_l^T\\
    \mathcal{E}\simeq R(\theta)-1+S(\epsilon,\varphi)\simeq \begin{pmatrix}
    \epsilon_{xx} & \epsilon_{xy}-\theta\\
    \epsilon_{xy}+\theta & \epsilon_{yy}
    \end{pmatrix}\\
    S(\epsilon,\varphi)=R^{-1}(\varphi)\begin{pmatrix}
    -\epsilon & 0\\
    0 & \nu_\text{P}\epsilon
    \end{pmatrix}
    R(\varphi)
    \end{gathered},
\end{equation}
where $R(\theta)$ is a rotation matrix. Heterostrain constrains the parameters to satisfy $\theta_1=-\theta_2=\theta/2$, $\varphi_1=\varphi_2=\varphi$, and $\epsilon_1=-\epsilon_2=\epsilon/2$. To first order in $\epsilon$ and $\theta$, the twist angle is unaffected. In analogy with the unstrained case, the new basis moir\'e RLVs are, in terms of the basis monolayer RLVs,
\begin{gather}
    \bm{b}_1=(M^{-1}_1-M^{-1}_2)(\bm{b}_{G2}-\bm{b}_{G1}),\quad \bm{b}_2=(M^{-1}_1-M^{-1}_2)\bm{b}_{G1}.
\end{gather}
This will affect the sampling of the momentum grid. 

The strained and rotated intralayer kinetic term in valley $\tau$ becomes
\begin{gather}
    \bra{\bm{p},l}H_\textrm{BM}\ket{\bm{p}',l}=\hbar v_F\left[M_l(\bm{p}-\tau \bm{A}_l)-\bm{K}_\tau\right]\cdot
    \begin{pmatrix}
    \tau\sigma_x\\
    -\sigma_y
    \end{pmatrix}\delta_{\bm{p},\bm{p}'}\\
    \bm{A}=\frac{\beta}{2a}(\epsilon_{xx}-\epsilon_{yy},-2\epsilon_{xy})
\end{gather}
where $\beta=3.14$, and $a_\text{CC}$ is the C$-$C bond length. Note that $\bm{p}$ above is still measured with respect to the global momentum origin, and $\bm{K}_\tau=\tau \bm{K}_D$ is the original Dirac wavevector in valley $\tau$. $\bm{A}_l$ acts as an effective layer-dependent vector potential in a similar fashion to the orbital effect of an in-plane magnetic field~\cite{Kwan2020parallel,antebi2022inplane} (except for the contrasting valley dependence).  
\\\\
\noindent\textbf{Appendix B. Specifications of Symmetry Operations}\medskip

In this section, we specify the actions of various symmetry operations mentioned in the main text on the plane wave basis. We organize the creation operators at a single momentum as a (row) vector $\hat{\bm{d}}^\dagger_{\bm{k}}$, with layer, sublattice, valley and spin indices. In this appendix section, $\bm{k}$ {\it is measured from the Dirac point of the relevant layer and valley} for convenience. The action of a symmetry operator will be written in the matrix form as
\begin{equation}
    \hat{O}\hat{\bm{d}}^\dagger_{\bm{k}}\hat{O}^{-1} = \hat{\bm{d}}^\dagger_{\hat{O}\bm{k}}U^{\hat{O}}
\end{equation}
for some unitary or anti-unitary operator $\hat{O}$. Matrix multiplication is implied on the right hand side. We use $\bm{\sigma}$, $\bm{\tau}$ and $\bm{\mu}$ to denote Pauli matrices in sublattice, valley and layer indices respectively.

The anti-unitary (spinless) time-reversal operator exchanges the two valleys, flips momentum from $\bm{k}$ to $-\bm{k}$, and is given by
\begin{equation}
    \hat{\mathcal{T}}\hat{\bm{d}}^\dagger_{\bm{k}}\hat{\mathcal{T}}^{-1} = \hat{\bm{d}}^\dagger_{-\bm{k}}\tau_x
\end{equation}
The two-fold rotation around $z$-aixs, i.e. $\hat{C}_{2z}$, is unitary, exchanges valley and sublattice indices, and flips momentum. Its action is given by 
\begin{equation}
    \hat{C}_{2z}\hat{\bm{d}}^\dagger_{\bm{k}}\hat{C}_{2z}^{-1} = \hat{\bm{d}}^\dagger_{-\bm{k}}\tau_x\sigma_x
\end{equation}
The combined $\hat{C}_{2z}\hat{\mathcal{T}}$ symmetry is anti-unitary and local in $\bm{k}$, given by
\begin{equation}
    (\hat{C}_{2z}\hat{\mathcal{T}})\hat{\bm{d}}^\dagger_{\bm{k}}(\hat{C}_{2z}\hat{\mathcal{T}})^{-1} = \hat{\bm{d}}^\dagger_{\bm{k}}\sigma_x.
\end{equation}
The $\hat{C}_{2x}$ symmetry exchanges the two layers, and its action is given by
\begin{equation}
    \hat{C}_{2x}\hat{\bm{d}}^\dagger_{\bm{k}}\hat{C}^{-1}_{2x} = \hat{\bm{d}}^\dagger_{\hat{C}_{2x}\bm{k}}\sigma_x\mu_x,
\end{equation}
where $\hat{C}_{2x}(k_x, k_y)^T = (k_x, -k_y)^T$.

The $\hat{C}_{3z}$ symmetry (counter-clockwise rotation by $2\pi/3$) preserves valley and sublattice index, and it is given by
\begin{equation}
\hat{C}_{3z}\hat{\bm{d}}^\dagger_{\bm{k}}\hat{C}_{3z}^{-1} = \hat{\bm{d}}^\dagger_{\hat{C}_{3z}\bm{k}}e^{i\frac{2\pi}{3}\sigma_z\tau_z}.
\end{equation}
To derive the non-trivial phase factor, one can consider a monolayer graphene and consider the action of $\hat{C}_{3z}$ on $\hat{c}^\dagger_{\bm{p}\sigma}$, where $\bm{p}$ is momentum measured from graphene $\Gamma$-point, and $\sigma =A/B$ is the sublattice index. For concreteness, consider a monolayer Bloch wave near the Dirac point $K$, with $\bm{p} = \bm{K} + \bm{k}$. Then, 
\begin{equation}\label{eq:C3z}
    \hat{C}_{3z}\hat{c}^\dagger_{\bm{K} + \bm{k},\sigma}\hat{C}_{3z}^{-1} = \hat{c}^\dagger_{\hat{C}_{3z}\bm{K} + \hat{C}_{3z}\bm{k},\sigma},
\end{equation}
where $c^\dagger_{\bm{p}\sigma}$ creates a Bloch wave on sublattice $\sigma$, and $\bm{p}$ is the momentum measured with respect to the origin. By definition, 
\begin{equation}
    \hat{c}^\dagger_{\bm{p}\sigma} \sim \sum_{\bm{R}}e^{-i\bm{p}\cdot(\bm{R} + \bm{r}_\sigma)}\hat{c}^\dagger_{\bm{R}\sigma}
\end{equation}
where $\bm{R}$ is unit cell vector and $\bm{r}_\sigma$ is the intra-cell atomic position vector, as denoted in Fig.~\ref{fig:superlattice}a. As $\hat{C}_{3z}\bm{K} = \bm{K} - \bm{b}_{G2}$, we have
\begin{equation}
    \hat{c}^\dagger_{\hat{C}_{3z}\bm{K} + \hat{C}_{3z}\bm{k},\sigma} = \hat{c}^\dagger_{\bm{K} + \hat{C}_{3z}\bm{k},\sigma}e^{i\bm{b}_{G2}\cdot \bm{r}_\sigma} = \hat{c}^\dagger_{\bm{K} + \hat{C}_{3z}\bm{k},\sigma}e^{\pm i\frac{2\pi}{3}}
\end{equation}
for $\sigma = A/B$ respectively. It can be shown that the sign is reversed for valley $K'$, confirming Eq.~\ref{eq:C3z}.

The unitary particle-hole symmetry, which is exact for the BM model if the small ``Pauli twist" in the kinetic terms are ignored, is given by\cite{Hejazi2019multiple,Song2019topological}
\begin{equation}
    \hat{\mathcal{P}}\hat{\bm{d}}^\dagger_{\bm{k}}\hat{\mathcal{P}}^{-1} = (i\sigma_x\mu_y)\hat{\bm{d}}_{-\bm{k}}.
\end{equation}

At the ``chiral-limit", i.e. $\kappa = 0$, there is an additional anti-unitary chiral symmetry, given by~\cite{Tarnopolsky2019chiral}
\begin{equation}    
\hat{\mathcal{S}}\hat{\bm{d}}^\dagger_{\bm{k}}\hat{\mathcal{S}}^{-1} = \sigma_z\hat{\bm{d}}_{\bm{k}}
\end{equation}

\noindent\textbf{Appendix C. Transformation Properties under Symmetries}\medskip

In this appendix we illustrate the derivation of the transformation matrices that act on the one body density matrix, as used in Sec.~\ref{sec:observables}. As an example, we will derive the transformation matrix for  $\hat{C}_{2z}\hat{\mathcal{T}}$. The transformation matrices for the other symmetry operations can be derived similarly. From Eq.~\eqref{eq:plane_wave_expansion_k} we have the moir\'e Bloch wavefunctions 
\begin{equation}
    \ket{\psi_{\tau a}(\bm{k})}=\sum_{\bm{G},l,\sigma}u_{\tau a;l\sigma}(\bm{k},\bm{G})\ket{\bm{k},\bm{G},\tau l\sigma},
\end{equation}
with corresponding operator
\begin{equation}
        c^\dagger_{\tau a}(\bm{k})=\sum_{\bm{G},l,\sigma}u_{\tau a;l\sigma}(\bm{k},\bm{G})d_{\bm{k}+\bm{G},\tau l\sigma}^\dagger,
\end{equation}
where $d_{\bm{k}+\bm{G},\tau l\sigma}^\dagger$ is an operator in the plane wave basis. We then have the transformation
\begin{align}
    (\hat{C}_{2z}\hat{\mathcal{T}})c^\dagger_{\tau a}(\bm{k})(\hat{C}_{2z}\hat{\mathcal{T}})^{-1}&=\sum_{\bm{G},l,\sigma}u^*_{\tau a;l\sigma}(\bm{k},\bm{G})d_{\bm{k}+\bm{G},\tau l\bar\sigma}^\dagger\\&=\sum_{\bm{G},l,\sigma, a'}u^*_{\tau a;l\sigma}(\bm{k},\bm{G})u^*_{\tau a';l\bar\sigma}(\bm{k},\bm{G})c^\dagger_{\tau a'}(\bm{k})\\&\equiv \sum_{a'\tau'}U^{\hat C_{2z}\hat{\mathcal{T}}}_{\tau' a',\tau a}(\bm{k},s)c^\dagger_{\tau' a'}(\bm{k}),
\end{align}
where
\begin{equation}
    U^{\hat C_{2z}\hat{\mathcal{T}}}_{\tau' b,\tau a}(\bm{k},s)=\delta_{\tau,\tau'}\sum_{\bm{G}, l, \sigma}u^*_{\tau a;l\sigma}(\bm{k},\bm{G})u^*_{\tau b;l\bar\sigma}(\bm{k},\bm{G}).
\end{equation}
Defining the one-body density matrix 
\begin{equation}
    {P}_{\tau'b,\tau a}(\bm{k},s) \equiv \braket{\hat{{c}}^\dagger_{\tau a s}(\bm{k})\hat{{c}}_{\tau'bs}(\bm{k})},
\end{equation}
we deduce the transformation property
\begin{gather}
    P^{\hat C_{2z}\hat{\mathcal{T}}}_{\tau a,\tau'a'}(\bm{k},s)=\sum_{\tau''\tau'''bb'}U^{\hat C_{2z}\hat{\mathcal{T}}}_{\tau a,\tau''b}(\bm{k},s)\left[U^{\hat C_{2z}\hat{\mathcal{T}}}_{\tau'a',\tau'''b'}(\bm{k},s)\right]^*P^*_{\tau''b,\tau'''b'}(\bm{k},s).
\end{gather}

\bibliographystyle{tfq}
\bibliography{refs.bib}

\end{document}